# ABSTRACT

Title of Dissertation: NONLINEAR INSTABILITIES IN COMPUTER NETWORK DYNAMICS

Priya Ranjan, Doctor of Philosophy, 2003

Dissertation directed by: Professor Eyad H. Abed
Department of Electrical and Computer Engineering

This work studies two types of computer networking models. The primary focus is to understand the different dynamical phenomena observed in practice due to the presence of severe nonlinearities, delays and widely varying operating conditions.

The first models considered are of senders running TCP (Transmission Control Protocol) and traffic passing through RED (Random Early Detection) gateways. Building on earlier work, a first order nonlinear discrete-time model is developed for the interaction scenario between transport protocols like TCP and UDP (User Datagram Protocol) and Active Queuing Management schemes like RED. It is shown that the dynamics resulting from the interaction with TCP is consistent with various dynamical behaviors and parameter sensitivities observed in practice. Using bifurcation-theoretic ideas it is shown that TCP-RED type networks may lose their stability through a period doubling bifurcation followed by border collision bifurcations. The nonlinear dependence of the throughput function of TCP-type flows on drop probability is found to be responsible for the period doubling bifurcation, whereas limited buffer space and lack of sufficient damping results in border collision bifurcations. Using the "period three implies chaos" theorem of Li and Yorke the existence of chaos is proved and the phenomenon is shown to occur as different parameters are varied.

Observation of period doubling bifurcation in the model leads to ideas for using feedback control to extend the region of stable operation in the parameter space. Using earlier work on control of period doubling bifurcations, a washout filter based control is proposed.

A second class of models studied in this work deals with optimal rate control in networks and are based on the rate-control framework proposed by Kelly. Using the results on delay-differential equation stability, the stability and its lack thereof is studied through an underlying map which arises naturally in time delay systems. An invariance property of this map is used to prove delay-independent stability and to compute bounds on periodic oscillations. In particular, it is shown that the sender's utility and the receiver's price functions, both are responsible for the dynamical behavior of the system. Instability results indicate that the system again loses stability through a period doubling bifurcation resulting in oscillations. These oscillations are well understood in the Mathematics community and are known as "slowly oscillating periodic (SOP)" orbits.

It is also argued that these periodic orbits will be observed in networking and network-based control systems more often.

# NONLINEAR INSTABILITIES IN COMPUTER NETWORK DYNAMICS

by

Priya Ranjan

Dissertation submitted to the Faculty of the Graduate School of the
University of Maryland, College Park in partial fulfillment
of the requirements for the degree of
Doctor of Philosophy
2003

Advisory Committee:

      Professor Eyad H. Abed , Chairman
      Assistant Professor Richard J. La (Co-advisor)
      Professor P. S. Krishnaprasad
      Professor Armond M. Makowski
      Professor Mark I. Freidlin

# PREFACE

क्षणैः क्षणैः यत् नवताम् उपैति इति सौन्दर्यम्
-भरत मुनि प्रणीत नाट्यशास्त्र से (200 ईसा पूर्व)

Beauty is something that renews itself every moment.
-Bharat Muni, *NatyaShastra*, c. 200 B.C.

Hence, chaos is beautiful!

# DEDICATION

This work is dedicted to the public schooling system of the Republic of India and the United States of America, and to the Indian and the American taxpayers.

# ACKNOWLEDGEMENTS

I would like to take this opportunity to thank my advisor Prof. Abed, who has been a great friend over the last five years of my stay at College Park. I have immensely benefited in terms of research, writing and thinking from his company.

I would like to acknowledge the great help from my advisory committee: Profs. A. M. Makowski, P. S. Krishnaprasad, R. La and M. Freidlin for their reading of the dissertation and providing fruitful comments. I also appreciate Prof. Krishnaprasad's 'small pokings' in the corridor and suggestions to improve my work.

I would also like to thank Prof. R. La for constant encouragement and numerous discussions on and beyond networking. His friendship is very much appreciated. I would also like to thank Prof. D. L. Elliott for many enlightening conversations and helpful suggestions. Thanks are also due to Prof. M. Freidlin for introducing me to a number of exciting mathematical ideas.

I also want to thank Prof. André Tits for being a great teacher and a nice friend to me. I have learned a lot from him in the due course of time. He has always been a great inspiration.

I would like to thank my friends who have provided the much needed life support in different ways. I would also like to thank my family who have kept their patience for a long time.

Many thanks are also due to "AID" (Association for India's Development) which has taught me a whole lot about life, life-style and living.

It will be unfair to close this without thanking the local restaurants like Woodlands and Food Factory who have provided excellent food (for thought).

Finally, I would like to extend my thanks to the faculty, staff and fellow graduate students at the ISR for providing an excellent and stimulating environment for study and research.

This research has been supported in part by the NSF under grants ECS-01-15160 and ANI-02-19162.

# TABLE OF CONTENTS

# LIST OF TABLES

# LIST OF FIGURES

# Chapter 1

# Introduction

## 1.1 Introduction

The Internet is a phenomenon which can be said to have been developing for the last thirty years if we think of it as a computer networking or simply making two computers talk to each other [11]. Internetworking is the art and science of making computers talk. It has been a very challenging problem over decades and still remains a rather difficult problem for modeling and analysis due to its complex nature and enormity. Extremely wide variations in parameters for a given technology and widely varying technology both on the networking and application levels are hallmark of any internet. This kind of heterogeneity and parametric variation lead to a very rich world of dynamical behaviors ranging from smooth, stable operation to rather complicated irregular patterns. The rich variety of phenomenology should not come as a surprise, considering that networks and network elements have varying levels of different adaptation techniques and they often try to optimize mutually conflicting objectives. They are also very nonlinear by design due to their built-in fault tolerance, which responds completely differently in different regimes.

In this work we are concerned with a rather simple version of this problem where we look at the interaction between network transport and feedback protocols. The basic motive is to build on earlier work and provide low-dimensional models of the observed phenomena, study the dynamics of the models and finally develop new control algorithms. The modeling depends heavily on the recently developed throughput functions for network transport protocols in the presence of drop probability. This comes naturally as discrete event modeling of these protocols leads to the explosion in the state space and significance of different events becomes harder to understand. This work also relies heavily on simulations in network simulator by Lawrence Berkeley Laboratory *ns-2* [15]. The typical approach will be to develop the models, analyze the dynamics as a low dimensional simulation models in MATLAB® [14] and in DYNAMICS® [64] software packages and then get back to *ns-2* to understand the significance of different dynamical phenomena for the actual networking environment. We begin with developing the terminology and required groundwork.

### 1.1.1 Formal Definition of Internet

The document "A Brief History of The Internet" describes the formal definition of internet and its evolution [52].

*On October 24, 1995, the Federal Networking Council (FNC) unanimously passed a resolution defining the term Internet. This definition was developed in consultation with members of the internet and intellectual property rights communities. RESOLUTION: The FNC agrees that the following language reflects our definition of the term "Internet". "Internet" refers to the global information*

*system that –*

(i) is logically linked together by a globally unique address space based on the Internet Protocol (IP) or its subsequent extensions / follow-ons;

(ii) is able to support communications using the Transmission Control Protocol / Internet Protocol (TCP/IP) suite or its subsequent extensions / follow-ons, and/or other IP-compatible protocols; and

(iii) provides, uses or makes accessible, either publicly or privately, high level services layered on the communications and related infrastructure described herein [52].

Simply speaking, internet is an interconnected structure of computing machines which may exhibit an emergent behavior due to their underlying connectivity. In many situations this emergent behavior can be much more powerful than plain summation of its components and hence the underlying mechanism of internetworking is of crucial importance. In the above definition of the term "internet" refererence is made to Internet Protocols (IP) and Transmission Control Protocol (TCP). According to the Miriam and Webster Dictionary (www.m-w.com) the word protocol is defined as

1 : an original draft, minute, or record of a document or transaction

2 a : a preliminary memorandum often formulated and signed by diplomatic negotiators as a basis for a final convention or treaty b : the records or minutes of a diplomatic conference or congress that show officially the agreements arrived at by the negotiators

3 a : a code prescribing strict adherence to correct etiquette and precedence (as in diplomatic exchange and in the military services) b : a set of conventions governing the treatment and especially the formatting of data in an electronic communications system 4 : a detailed plan of a scientific or medical experiment, treatment, or procedure. For two engineering systems using information technology for interconnection, a protocol is a predefined set of rules which two or more communicating nodes follow to achieve certain communication or connectivity objectives. These protocols can be defined differently for different technologies, logical layers and the objectives to be achieved.

### 1.1.2 Introduction to TCP-IP

TCP/IP is a set of protocols developed to allow computers to share communication/computation/control resources across a network in a cooperative manner. It was originally developed by a community of researchers at the ARPAnet. However, there is a whole spectrum of products that support TCP/IP, and thousands of networks of all kinds including sensor and wireless use it.

TCP provides a connection oriented, reliable, byte stream type service. The term connection-oriented means the two applications using TCP must establish a TCP connection with each other before they can exchange data. It is a full duplex protocol, meaning that each TCP connection supports a pair of byte streams, one flowing in each direction. TCP includes a flow-control mechanism for each of these byte streams that allows the receiver to limit how much data the sender can transmit. TCP also implements a congestion-control mechanism to keep the network running with acceptable performance. Some of the underlying features of TCP are discussed next:

**Adaptive Retransmission**

TCP guarantees reliable delivery and so it retransmits each segment if an ACK is not received in a certain period of time. TCP sets this timeout as a function of the RTT (Round Trip Time) it expects between the two ends of the connection. Unfortunately, given the range of possible RTT's between any pair of hosts in the Internet, as well as the variation in RTT between the same two hosts over time, choosing an appropriate timeout value is not that easy. To address this problem, TCP uses an adaptive retransmission mechanism. We describe this mechanism and how it has evolved over time. Most of the material presented here can be found in [81].

**Original Algorithm**

Measure Sample RTT for each segment/ACK pair
    Compute weighted average of RTT
    EstimatedRTT = a*EstimatedRTT + b*SampleRTT, where a+b = 1
    a between 0.8 and 0.9
    b between 0.1 and 0.2
    Set timeout based on EstimatedRTT
    TimeOut = 2 * EstimatedRTT

**Karn/Partridge Algorithm**

Do not sample RTT when retransmitting
    Double timeout after each retransmission

**Jacobson/Karels Algorithm**

This mechanism can be found in Jacobson seminal 1988 paper [38].

New calculation for average RTT

Difference = SampleRTT - EstimatedRTT

EstimatedRTT = EstimatedRTT + ( d * Difference)

Deviation = Deviation + d ( |Difference| - Deviation)), where d is a fraction between 0 and 1

Consider variance when setting timeout value

Timeout = u * EstimatedRTT + q * Deviation, where u = 1 and q = 4

### 1.1.3 Congestion Control Mechanism of TCP

In the earlier development of Internet, the idea was to keep the network as simple as possible and place most of the intelligence required to regulate the communication on the end nodes [76]. Keeping this paradigm in mind, Jacobson developed new algorithms for TCP arguing the stability as the most important concern. He used the Lyapunov function approach to reason that the network should be selfclocked to avoid any clogging in the network [38]. In this section we review some of the TCP features and operating regimes which are designed to keep the system stable.

**Slow Start**

This phase of TCP aims to share the network resources in a cooperative manner and without overwhelming the existing resources. It is designed observing that the rate at which new packets should be injected into the network is the rate at which the acknowledgments are returned by the other end.

Slow start adds another window to the sender's TCP: the congestion window,

called "cwnd". When a new connection is established with a host on another network, the congestion window is initialized to one segment (i.e., the segment size announced by the other end, or the default, typically 536 or 512). Each time an ACK is received, the congestion window is increased by one segment essentially as a multiplicative increase mechanism. The sender can transmit up to the minimum of the congestion window and the advertised window. The congestion window is flow control imposed by the sender, while the advertised window is flow control imposed by the receiver. The former is based on the sender's assessment of perceived network congestion; the latter is related to the amount of available buffer space at the receiver for this connection.

The sender starts by transmitting one segment and waiting for its ACK. When that ACK is received, the congestion window is incremented from one to two, and two segments can be sent. When each of those two segments is acknowledged, the congestion window is increased to four. This provides an exponential growth, although it is not exactly exponential because the receiver may delay its ACKs, typically sending one ACK for every two segments that it receives.

At some point the capacity of the internet can be reached, and an intermediate router will start discarding packets. This tells the sender that its congestion window has gotten too large.

Early implementations performed slow start only if the other end was on a different network. Current implementations always perform slow start.

**Congestion Avoidance**

Congestion can occur when data arrives on a big pipe (a fast Local Area Network or LAN) and gets sent out a smaller pipe (a slower Wide Area Network or WAN).

Congestion can also occur when multiple input streams arrive at a router whose output capacity is less than the sum of the inputs. Congestion avoidance is a way to deal with lost packets.

The assumption of the algorithm is that packet loss caused by damage is very small (much less than 1%), therefore the loss of a packet signals congestion somewhere in the network between the source and destination. There are two indications of packet loss: a timeout occurring and the receipt of duplicate ACKs.

Congestion avoidance and slow start are independent algorithms with different objectives. But when congestion occurs TCP must slow down its transmission rate of packets into the network, and then invoke slow start to get things going again. In practice they are implemented together.

Congestion avoidance and slow start require that two variables be maintained for each connection: a congestion window, cwnd, and a slow start threshold size, ssthresh. The combined algorithm operates as follows:

1. Initialization for a given connection sets cwnd to one segment and ssthresh to 65535 bytes.

2. The TCP output routine never sends more than the minimum of cwnd and the receiver's advertised window.

3. When congestion occurs (indicated by a timeout or the reception of duplicate ACKs), one-half of the current window size (the minimum of cwnd and the receiver's advertised window, but at least two segments) is saved in ssthresh. Additionally, if the congestion is indicated by a timeout, cwnd is set to one segment (i.e., slow start).

4. When new data is acknowledged by the other end, increase cwnd, but the way it increases depends on whether TCP is performing slow start or congestion avoidance.

If cwnd is less than or equal to ssthresh, TCP is in slow start; otherwise TCP is performing congestion avoidance. Slow start continues until TCP is halfway to where it was when congestion occurred (since it recorded half of the window size that caused the problem in step 2), and then congestion avoidance takes over.

Slow start has cwnd begin at one segment, and be incremented by one segment every time an ACK is received. As mentioned earlier, this opens the window exponentially: send one segment, then two, then four, and so on. Congestion avoidance dictates that cwnd be incremented by segsize*segsize/cwnd each time an ACK is received, where segsize is the segment size and cwnd is maintained in bytes. This is a linear growth of cwnd, compared to slow start's exponential growth. The increase in cwnd should be at most one segment each round-trip time (regardless how many ACKs are received in that RTT), whereas slow start increments cwnd by the number of ACKs received in a round-trip time.

**Fast Retransmit**

TCP may generate an immediate acknowledgment (a duplicate ACK) when an out- of-order segment is received. This duplicate ACK should not be delayed. The purpose of this duplicate ACK is to let the other end know that a segment was received out of order, and to tell it what sequence number is expected.

Since TCP does not know whether a duplicate ACK is caused by a lost segment or just a reordering of segments, it waits for a small number of duplicate ACKs to be received. It is assumed that if there is just a reordering of the segments, there will be only one or two duplicate ACKs before the reordered segment is processed, which will then generate a new ACK. If three or more duplicate ACKs are received in a row, it is a strong indication that a segment has been lost. TCP then performs a retransmission of what appears to be the missing segment, without waiting for a retransmission timer to expire.

**Fast Recovery**

After fast retransmit sends what appears to be the missing segment, congestion avoidance, but no slow start is performed. This is the fast recovery algorithm. It is an improvement that allows high throughput under moderate congestion, especially for large windows.

The reason for not performing slow start in this case is that the receipt of the duplicate ACKs tells TCP more than just a packet has been lost. Since the receiver can only generate the duplicate ACK when another segment is received, that segment has left the network and is in the receiver's buffer. That is, there is still data flowing between the two ends, and TCP does not want to reduce the flow abruptly by going into slow start.

The fast retransmit and fast recovery algorithms are usually implemented together as follows.

1. When the third duplicate ACK in a row is received, set ssthresh to one-half the current congestion window, cwnd, but no less than two segments. Retransmit the missing segment. Set cwnd to ssthresh plus 3 times the segment size. This inflates the congestion window by the number of segments that have left the network and which the other end has cached .

2. Each time another duplicate ACK arrives, increment cwnd by the segment size. This inflates the congestion window for the additional segment that has left the network. Transmit a packet, if allowed by the new value of cwnd.

3. When the next ACK arrives that acknowledges new data, set cwnd to ssthresh (the value set in step 1). This ACK should be the acknowledgment of the retransmission from step 1, one round-trip time after the retransmission. Additionally, this ACK should acknowledge all the intermediate segments sent between the lost packet and the receipt of the first duplicate ACK. This step is congestion avoidance, since TCP is down to one-half the rate it was at when the packet was lost.

### 1.1.4 More general TCP-type algorithms

The general class of TCP-type congestion control algorithms known as *Binomial Congestion Control algorithms* [7] can also be considered in our framework as shown in the next chapter. These algorithms have been developed to design TCP-friendly multimedia streams that cannot sustain window halving during congestion as prescribed in the multiplicative decrease phase of TCP. A theoretical derivation and empirical validation of throughput function for this class of algorithms can be found in [7]. Indeed, the empirical validation has been performed in the presence of RED gateways, which makes this a good candidate for analysis using the basic model proposed in this work. Briefly, the binomial congestion control algorithms can be described as follows:

$$I \;:\; w_{t+R} \leftarrow w_t + \frac{\alpha}{w_t^k} \;;\; \alpha > 0$$
$$D \;:\; w_{t+\delta t} \leftarrow w_t - \beta w_t^l \;;\; 0 < \beta < 1$$

where $I$ denotes the increase in the window as a result of the receipt of one window of acknowledgments in a round-trip time (RTT) and $D$ denotes the decrease in window size on detection of congestion by the sender, $w_t$ denotes the window size at time t, $R$ is the $RTT$ of the flow, and $\alpha$ and $\beta$ are constants. Parameters $l$ and $k$ describe the functional dependence of increase and decrease respectively, on the current window size $w_t$. Binomial congestion control algorithms are characterized by $l + k = 1$. Classical TCP is characterized by $k = 0$, $l = 1$. Other TCP-type algorithms can be considered for $k = -1$, $l = 1$ which is MIMD (Multiplicative increase/ Multiplicative decrease) used by slow start in TCP. For $k = -1$, $l = 0$, we get MIAD (Multiplicative Increase and Additive Decrease) and for $k = 0, l = 0$ we get AIAD (Additive Increase and Additive Decrease), which completes the classification of all affine TCP-type algorithms.

## 1.2 Throughput Function of TCP and TCP-type Algorithms

These algorithms are shown to possess the following approximate throughput profile [7]:

$$T(p, R) = \frac{M}{R} \frac{K}{p^{\frac{1}{k+l+1}}} \tag{1.1}$$

where

$$K = \frac{\alpha}{\beta}^{\frac{1}{k+l+1}}$$

This class of congestion control algorithms are important due to their potential use for multimedia streams applications and TCP friendliness. This is to stress that nonlinear stability analysis framework proposed in [71] remains valid even for quickly growing presence of TCP-like non-TCP traffic as long as they react to the packets drops and have throughput functions similar to that of TCP. It can also be shown that these assumptions will remain valid under the limited presence of UDP traffic along with Binomial traffic.

The assumptions for existence of an inverse can again be verified in parameter regions of interest $(p > 0, T > 0)$ for a general class of TCP and TCP-type binomial congestion control algorithms with timeouts considered in an average sense. Next, we analyze a more detailed TCP-throughput function [58, 7, 89] than in the examples above.

$$T(p,R) = \frac{M}{R\sqrt{\frac{\beta p}{\alpha}} + T_0 \min\left(1, 3\sqrt{\frac{\beta p}{\alpha}}\right)p(1 + 32p^2)} \qquad (1.2)$$

where

$$T_0 = \text{timeout period}$$

It is clear the average effect of timeout is to decrease the throughput. If we again consider this average effect of timeout opposed to the instantaneous effect, we can show that our modeling framework holds. First, analyzing the function $\min\left(1, 3\sqrt{\frac{\beta p}{\alpha}}\right)$ tells us that for drop probability $p < \frac{\alpha}{9\beta} = 0.222$ for $\alpha = 1$ and $\beta = 0.5$ (typical TCP implementations) which is reasonably high drop rate given that generally $p_{max}$ in RED approximately 0.1. Hence, throughput function can be reduced to:

$$T(p,R) = \frac{M}{R\sqrt{\frac{\beta p}{\alpha}} + 3T_0\sqrt{\frac{\beta p}{\alpha}}p(1 + 32p^2)} \qquad (1.3)$$

Finally, even if the $p < \frac{\alpha}{9\beta}$ condition is not valid uniformly, we can show that throughput function is a piecewise smooth function and the inverse exists for separate segments. We have already shown that for small $p$ segment it satisfies the required properties to compute an inverse. For $p$ sufficiently large, throughput and corresponding derivatives can be written as:

$$T(p, R) = \frac{M}{R\sqrt{\frac{\beta p}{\alpha}} + T_0 p(1 + 32p^2)}$$

## 1.3 Congestion Control

With the growing size and popularity of the Internet, the problem of congestion control is emerging as a more crucial problem. Poor management of congestion can render one part of a network inaccessible to the rest and significantly degrade the performance of networking applications. Researchers have proposed various approaches to address this issue. One approach is to keep the network dumb and place most of required intelligence at the end hosts by implementing a more sophisticated end-user rate control allocation [10, 61, 76, 27]. Another approach is to control congestion level at each router through an Active Queue Management (AQM) mechanism, *e.g.,* Random Early Detection (RED) [24], Random Early Marking (REM) [3], Virtual Queue (VQ) [26], and Adaptive Virtual Queue (AVQ) [46]. A common goal for these AQM mechanisms is to detect early signs of congestion and provide feedback to the adaptive sources so that congestion can be avoided without causing a significant degradation in network performance.

Hollot *et al.* [32, 31] have developed a linearized model of a RED gateway with TCP connections to characterize the stability region of the system. Based on their single node analysis they have provided a guideline for selection of the RED parameters. They observed that exponential averaging introduces an unnecessary feedback delay in addition to the built-in averaging effect of the queue size, and proposed Proportional and Proportional-Integral controllers to improve the responsiveness and stability of the RED mechanism, using a similar linearized model. Low *et al.* [56] have argued that the dynamic behavior of queue is governed mostly by the stability of TCP-RED. This argument forms the basic ground multi-link multi-source utility-based model for studying the interaction of TCP with RED gateways and also for a local stability condition in a single link case.

### 1.3.1  Random Early Detection (RED)

The RED mechanism, proposed by Floyd and Jacobson [24], attempts to control the congestion level at a bottleneck by monitoring and updating the average queue size. The basic idea of RED is to sense impending congestion before it happens and provide feedback to the sources by either dropping or marking their packets. The packet marking/dropping probability is the control administered by the RED gateways when they detect queue build-up beyond a certain threshold. Although the RED mechanism is conceptually very simple and easy to understand, its interaction with Transmission Control Protocol (TCP) connections has been shown to be rather complex and is poorly understood. Most of the rules for setting the parameters of the RED mechanism are based on limited empirical data and come from networking experience. These rules have been evolving, as the effects of controller parameters are not very clearly understood. There are reports that discourage wide deployment of RED (*e.g.,* [59]), arguing that there is insufficient consensus on how to select controller parameter values, and that the RED does not provide a drastic improvement in performance. Furthermore, researchers have observed that the RED mechanism is sensitive to parametric variations, which is undesirable given the limited understanding of the effects of various system parameters.

### 1.3.2 Adaptive Random Early Detection (ARED)

As the name suggest Adaptive Random Early Detection (ARED) is adaptive version of RED [23]. Adaptation has been proposed in parameter $p_{max}$ to keep the queue size within a prescribed range. An additive increase/multiplicative decrease algorithm modifies the value of $p_{max}$ based on average queue occupancy computed from the RED's averaging scheme. This adaptive version has shown considerable promise and is still under investigation [50].

## 1.4 Optimal Rate Control

In this section we describe the optimal rate-control model proposed by Kelly [41]. This material forms the basis for stability results described in chapter 5.

We consider a network with a set $\mathcal{J}$ of resources or links and a set $\mathcal{I}$ of users. Let $C_j$ denote the finite capacity of link $j \in \mathcal{J}$. Each user has a fixed route $\mathcal{J}_i$, which is a non-empty subset of $\mathcal{J}$.[1] We define a zero-one matrix $A$, where $A_{i,j} = 1$ if link $j$ is in user $i$'s route $\mathcal{J}_i$ and $A_{i,j} = 0$ otherwise. When the throughput of user $i$ is $x_i$, user $i$ receives utility $U_i(x_i)$. For example, suppose that a user is transferring a file. The per-transfer delay is inversely proportional to the rate it receives. Hence, the utility of the user may be measured as a function of its rate. We assume that the utility $U_i(x_i)$ is an increasing, strictly concave and continuously differentiable function of $x_i$ over the range $x_i \geq 0$.[2] Furthermore, we assume that the utilities are additive so that the aggregate utility of rate allocation $x = (x_i, i \in \mathcal{I})$ is $\sum_{i \in \mathcal{I}} U_i(x_i)$. This is a reasonable assumption since these utilities are those of independent network users. Let $U = (U_i(\cdot), i \in \mathcal{I})$ and $C = (C_j, j \in \mathcal{J})$. In this work we study the feasibility of achieving the maximum total utility of the users in a distributed environment with non-zero delay, by characterizing the stability under certain genera classes of utility and price functions. Under our model this problem can be formulated as follows.

---

[1] We assume that there is a routing mechanism and do not discuss the issue of routing in this paper.

[2] Such a user is said to have elastic traffic.

*SYSTEM(U,A,C):*

$$\begin{align}
maximize \quad & \sum_{i \in \mathcal{I}} U_i(x_i) \tag{1.4}\\
subject\ to \quad & A^T x \leq C \\
over \quad & x \geq 0
\end{align}$$

The first constraint in the problem says that the total rate through a resource cannot be larger than the capacity of the resource. Given that the system knows the utility functions $U$ of the users, this optimization problem may be mathematically tractable. However, in practice not only is the system not likely to know $U$, but also it is impractical for a centralized system to compute and allocate the users' rates due to the large scale of the network. Hence, Kelly in [41] has proposed to consider the following two simpler problems.

Suppose that each user $i$ is given the price per unit flow $\lambda_i$. Given $\lambda_i$, user $i$ selects an amount to pay per unit time, $p_i$, and receives a flow $x_i = \frac{p_i}{\lambda_i}$.[3] Then the user's optimization problem becomes the following [41].

*$USER_i(U_i; \lambda_i)$ :*

$$maximize \quad U_i(\frac{p_i}{\lambda_i}) - p_i \tag{1.5}$$

---
[3]This is equivalent to selecting its rate $x_i$ and agreeing to pay $p_i = x_i \cdot \lambda_i$.

$$\text{over} \quad p_i \geq 0$$

The network, on the other hand, given the amounts the users are willing to pay, $p = (p_i, i \in \mathcal{I})$, attempts to maximize the sum of weighted log functions $\sum_{i \in \mathcal{I}} p_i \log(x_i)$. Then the network's optimization problem can be written as follows [41].

NETWORK(A,C;p) :

$$\begin{align}
\text{maximize} \quad & \sum_{i \in \mathcal{I}} p_i \log(x_i) \tag{1.6}\\
\text{subject to} \quad & A^T x \leq C \\
\text{over} \quad & x \geq 0
\end{align}$$

Note that the network does not require the true utility functions $(U_i(\cdot), i \in \mathcal{I})$, and pretends that user $i$'s utility function is $p_i \cdot \log(x_i)$ to carry out the computation. It is shown in [41] that one can always find vectors $\lambda^* = (\lambda_i^*, i \in \mathcal{I}), p^* = (p_i^*, i \in \mathcal{I})$, and $x^* = (x_i^*, i \in \mathcal{I})$ such that $p_i^*$ solves $USER_i(U_i; \lambda_i^*)$ for all $i \in \mathcal{I}$, $x^*$ solves $NETWORK(A, C; p^*)$, and $p_i^* = x_i^* \cdot \lambda_i^*$ for all $i \in \mathcal{I}$. Further, the rate allocation $x^*$ is also the unique solution to $SYSTEM(U, A, C)$. This suggests that the problem of solving $SYSTEM(U, A, C)$ can be achieved by an algorithm that solves $NETWORK(A, C; p(t))$ for a given $p(t)$ at time $t$ on a smaller time scale, and drives $p(t)$ to $p^*$ on a larger time scale.

Our work focuses on determining the stability of user's problem in the presence of arbitrary return trip delay. In particular we are also interested in understanding the interaction between the utility and price functions to determine the shape of periodic solution in the case of instability. As we show in chapter 5 that these problems are directly related to some of the standard delay-differential equations and the results for delay-differential equations provide some interesting insight into the design of utility and price functions.

# Chapter 2

# Mathematical Preliminaries

## 2.1 Introduction

Periodicities are fundamental to the behavior of dynamical and physical systems. They constitute the next level of complexity after equilibrium points. Most of the systems observed in nature or built for engineering purposes either approach an equilibrium point or oscillate. The origin of periodicities in a dynamical system is one of the most complicated and oldest area of research. The latest understanding that chaotic systems are nothing but a hierarchical organization of unstable periodic orbits (UPO) has made periodicities more relevant than ever before [25].

Periodicity in a system sometimes may be unwanted since it leads to performance degradation and inefficient use of the available resources. For this reason, it will be useful to know if a system is on the verge of performing a periodic motion so that, if possible some kind of control to suppress the periodicities can be administered. This also raises the question of what parameter leads the system to the periodic behavior and how it can be mitigated. This is very relevant now with the advent of very sophisticated computer control systems which are basically discrete or switched type in nature, while the systems they control are continuous.

Control of continuous systems with discrete elements is an area of very active research. From power systems, power electronics to communication and computer networks, there is a whole class of engineering systems which use digital control but their dynamics is essentially modeled as continuous. These systems make use of switching type of control where switching is based on a particular value of state variable, known as 'threshold'. Threshold based control laws have been shown to exhibit anomalous behavior under parameter drift or stress. The latest example in this category is that of the internet where TCP-RED systems behave in strange ways under certain parameter settings. Power electronic systems have been reported to show similar behavior which can be explained based on "border collision bifurcation" theory proposed recently.

This chapter is organized as follows: we illustrate the reduction of differential equations to maps using Poincaré and stroboscopic mapping techniques in section 2.2. In section 2.3, we discuss the theory behind basic smooth bifurcations. Section 2.4 deals with the background material on nonsmooth bifurcations which is a developing field by itself.

## 2.2 Discrete Time Modeling

A very general class of system can be be modeled by a set of ordinary differential equations(ODE) of the form

$$\dot{x} = f(x, t, \mu) \quad (2.1)$$

with $f : R^{n+1+p} \to R^n$ being a piecewise smooth function, $\mu \in R^p$ a parameter vector and $x \in R^n$. It is clear that everywhere smooth systems are a special case of eq.(2.1). The phase space of a general piecewise smooth system, such as eq.(2.1) with $f$ being piece wise smooth, can be divided into countably many regions; in each region the system having a different smooth functional description. At the border of these regions $f$ may either be discontinuous or have discontinuous first derivative(non-smooth). The piecewise affine system makes a good example for this kind of system description. These are very important from the point of view of both local analysis of bifurcations and in applications. Power electronic converters can be taken as an engineering example for this class of systems where the evolution of the state variables $i$ and $v_c$ during the *on* and *off* periods are described by an affine differential equations [4]:

$$\text{`On' period:} \quad \frac{di}{dt} = \frac{V_{in}}{L}, \quad \frac{dv_c}{dt} = -\frac{v_c}{CR} \qquad (2.2)$$

$$\text{`Off' period:} \quad \frac{di}{dt} = \frac{V_{in}}{L} - \frac{v_c}{L}, \quad \frac{dv_c}{dt} = \frac{i}{C} - \frac{v_c}{CR} \qquad (2.3)$$

where the *on* and *off* periods depends on the discrete state of a switch in the system. Thus, phase space of the power converter can be augmented with two discrete states and the borders can be drawn depending on the value of states for which the switch changes its state. The precise notion of this kind of border will be made clear when we model this system as a stroboscopic map.

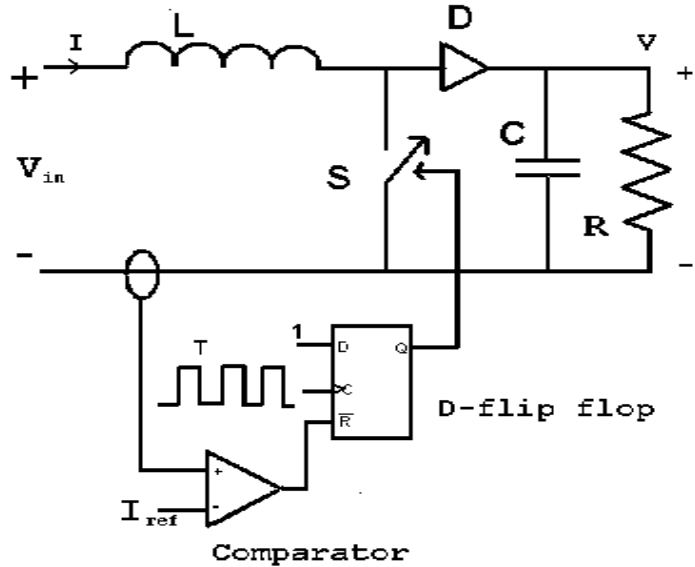

Figure 2.1: Circuit diagram for a current controlled boost converter

In general, dynamical behavior of a non-smooth system can change in two ways as the system parameters are varied. Border collision bifurcation is a phenomenon that can occur if the flow is continuous but non-smooth with respect to the state variables across the boundary as in the system shown above [4]. The other type of bifurcation, known as grazing, may be observed if the system is discontinuous at the phase space boundary. This kind of behavior is reported in power electronic systems in [18]. These kind of bifurcations are observed when the system parameter is such that the trajectory (e.g. periodic orbits) becomes tangent to the border separating the different regions in the state space where vector field in eq.(2.1) have different functional descriptions.

Lets assume that domain $D$ of a piecewise smooth system described by eq.(2.1) can be split into finitely many sub domains $D_1, D_2, ..., D_r$ in which the differential eq.(2.1) is continuous and has a sufficiently smooth right hand side. Now we can have two classes of bifurcations: 1) Trajectories stay in one of the domains $D_k$ and there is a qualitative change due to the inherent nonlinearity of the system as system parameter varies. This kind of bifurcation will include classical bifurcations like saddle-node, pitchfork, Hopf etc. 2) This class of bifurcation includes border collision or grazing as described above. Systems undergo this kind of bifurcation, when upon parameter variation a periodic orbit, $P_0$, existing locally in a region $D_j$ of phase space, becomes tangential to the hyper-surface, $\Phi_j$, bordering a different region, $D_i$. Border collision has been an area of active interest from an engineering point of view and system in power electronics [4, 5], computer networks [19], tandem queues have been shown to have this kind of bifurcations. Theoretical results on this kind of bifurcations can be found in [5, 4, 8, 66, 65].

## 2.3 Local Analysis and Smooth Bifurcations

We will now describe a general framework to analyze bifurcations and motivate the use of Poincaré map for periodic orbits of eq.(2.1). Deriving a local map to describe the dynamics of eq.(2.1) is a basic technique to reduce the system to a normal form. The eigenvalues of this local map will provide the information about the different dynamical behaviors corresponding to the general system.

### 2.3.1 Poincaré map

Poincaré map or first return map is the most basic tool for studying the stability and bifurcations of periodic orbits. It was defined by Henri Poincaré in 1881. The idea of Poincaré map is very simple: If $\Gamma$ is a periodic orbit of the system

$$\dot{x} = f(x) \tag{2.4}$$

through the point $x_0$ and $\Sigma$ is a hyperplane transversal to $\Gamma$ at $x_0$, then for any point $x \in \Sigma$ sufficiently near $x_0$, the solution of eq.(2.4) through $x$ at $t = 0, \phi_t(x)$, will cross $\Sigma$ again at a point $P(x)$ near $x_0$. The mapping $x \to P(x)$ is called a Poincaré map [68].

### 2.3.2 Stroboscopic map

Since Poincaré map is only defined for autonomous systems something similar can be defined for systems with periodic forcing. The basic idea of a stroboscopic map is to sample the state once every clock period in the system. This technique also gives a similar map as Poincaré map and has been used in power electronics, networks etc. [4, 5, 19].

In this work most of the models will be analyzed in their reduced form as a map. The benefit of having a system as a map is the ease of analysis, simulation and control, although it sometimes does obscure the interesting things one can do with the original differential equations. From here on we will define the systems as a map:

$$x_{n+1} = f(x_n, \mu) \tag{2.5}$$

where $x_n \in R^n$, $f : R^{n+p} \to R^n$ and parameter vector $\mu \in R^p$. For simplicity lets assume that p=1.

### 2.3.3 Classical co-dimension one bifurcations of maps

Classical bifurcation analysis looks for ways in which a fixed point $x^0(\mu) = f(x^0, \mu)$ of map(2.5) can lose hyperbolicty. This can happen in a map in three ways:

1) $Df(x^0, \mu)$ may have an eigenvalue $+1$,

2) An eigenvalue $-1$, or

3) a pair of complex eigenvalues $\lambda, \overline{\lambda}$ with $|\lambda| = 1$.

The bifurcation scenario for fixed points with $\lambda = +1$ is completely analogous to the bifurcation scenario for equilibria with eigenvalue 0 in flows. The following theorem from [33] describes it precisely.

**Theorem 1** *Let $(\mu, x) \to f_\mu(x) : R^2 \to R$ be of class $C^k, k \geq 2$, near 0 and*

satisfy $f_\mu(0) = 0, Df_\mu(0) = \lambda(\mu)$ with $\lambda(0) = 1, \lambda' > 0$. Then there exists a unique bifurcated branch of fixed points $(\mu(s), x(s))$, for $s$ in some interval $(-\epsilon, \epsilon)$, for $f_\mu : f_{\mu(s)}(x(s)) = x(s)$, such that $\mu(0) = x(0) = 0, x'(0) \neq 0$ and the functions $\mu$ and $x$ are $C^{k-1}$ near 0.

The fixed point zero is unstable for $\mu > 0$ and stable for $\mu < 0$.

For fixed points with eigenvalue $-1$ with flip bifurcations, also known as *period−doubling* or *subharmonic* bifurcation. This does not have an analogue for equilibria in flows. This bifurcation is most important from the application point of view since it has been observed in variety of engineering systems. Again we quote following theorem from [33] to describe period doubling or flip bifurcation mathematically:

**Theorem 2** *Let $(\mu, x) \to f_\mu(x) : R^2 \to R$ be of class $C^k, k \geq 2$, near 0 and satisfy $f_\mu(0) = 0, Df_\mu(0) = \lambda(\mu)$ with $\lambda(0) = -1, \lambda' < 0$. Then there exists a unique <u>onesided</u> bifurcated branch of fixed points of order 2, $(\mu(s), x_j(s), j = 1, 2)$ for $f_\mu$, such that $\mu(0) = 0, \mu(-s) = \mu(s), x_1(-s) = x_2(s), \frac{dx_1}{ds}(0) = 1, x_j(0) = 0, f_\mu(x_j) = x_{j'}, j \neq j'$. The functions $\mu$ and $x_j$ are $c^{k-1}$ near 0. The fixed point 0 is stable for $\mu < 0$, unstable for $\mu > 0$. If $\mu'(s)$ keeps a constant sign for $s \in (0, \epsilon)$ or for $s \in (-\epsilon, 0)$, then the bifurcated fixed points are stable if $\mu > 0$ and unstable if $\mu < 0$.*

The third bifurcation scenario where a pair of eigenvalue become imaginary is analogous to the Hopf bifurcation for flows and can only be observed in maps of dimension 2 or higher. Since this is not directly relevant to our subject we will not be describing it in any detail.

## 2.4 Nonsmooth Bifurcations

As the name suggests, a border-collision bifurcation happens when the system trajectory crosses one of the borders between different phase space regions. After a border-collision bifurcation a whole range of different dynamical scenarios can be observed [4]. For example: (1) Period-1 to period-2 (2) Period-1 to chaos (3) Period-1 to period-1 or periodicity remains unchanged even if the orbits collide with the border etc. All these bifurcation scenarios have been observed in piecewise smooth systems in different applications like power electronics, impact oscillators etc. Next, we describe the setup for applying the border-collision bifurcation theory in smooth systems with discrete components.

Let $L_0$ be a periodic orbit of eq.(2.1) which is tangential to one of the borders $\Phi_0$ for $\mu = \mu_0$ which can be assumed to be 0 without loss of generality. Let $\Sigma$ be some appropriately chosen Poincaré plane transversal to the system flow on one side of the border $\Phi_0$. Let $M_0$ be the fixed point on $\Sigma$, corresponding to the grazing limit cycle $L_0$. Also, assume $S$ to be a line on $\Sigma$, such that trajectories starting on it stay tangential to $\Phi_0$. Clearly, the line $S$ is continuously mapped by the system flow into a line $l$ lying on the border $\Phi_0$. The line $S$ splits the plane $\Sigma$ in the two half planes $\Sigma^+$ and $\Sigma^-$. Trajectories starting in $\Sigma^-$ will eventually cross the border $\Phi_0$ while trajectories passing through $\Sigma^+$ will not. Let $\Pi$ be the mapping generated by the system flow from $\Sigma$ to itself. Lets also assume $\Pi$ to be continuous in the neighborhood of the fixed point $M_0$ and and such that its dependence on the parameter variation is sufficiently smooth. As trajectories may either cross the border $\Phi_0$ or not, $\Pi$ is accordingly a suitable composition of two sub-mappings $\Pi^- : \Sigma^- \to \Sigma$ and $\Pi^+ : \Sigma^+ \to \Sigma$ describing the system motion in the case depending on its crossing of the border($\Pi^-$) or ($\Pi^+$).

Now a system of coordinates on $\Sigma$ can be introduced with the origin located at the fixed point $M_0$, such that the sign of the $n$-th coordinate determines whether a given point, $x^* = (x_1^*, ...., x_n^*)$, is in the sub-plane $\Sigma^-$ or $\Sigma^-$(i.e. sign of $x_n^*$ determines if $x^*$ is in $\Sigma^-$ or $\Sigma^+$ $\forall$ $x^* \in \Sigma$).

Linearizing the system with respect to $x_i$ and $\mu$ in the neighborhood of $M_0$, the equation for Poincaré map can be derived as:

$$x^{k+1} = \begin{cases} A_1 x^k + c\mu & , \quad x_n^k > 0 \\ A_2 x^k + c\mu & , \quad x_n^k < 0 \end{cases} \tag{2.6}$$

where

$$A_1 = \left.\frac{\partial \Pi^+}{\partial x}\right|_{x=0}, \quad A_2 = \left.\frac{\partial \Pi^-}{\partial x}\right|_{x=0}$$

$$c = \left.\frac{\partial \Pi^+}{\partial \mu}\right|_{x=0} = \left.\frac{\partial \Pi^-}{\partial \mu}\right|_{x=0}.$$

Since we have assumed that mapping is continuous when $x_n^k = 0, \mu = 0$ for all $k$ and smooth for $x_n^k = 0, \mu = 0, k = 1, ..., n-1$, the matrices $A_1 = [a_{ij}^1]$ and $A_2 = [a_{ij}^2]$ are such that $a_{ij}^1 = a_{ij}^2$ if $j \neq n$.

### 2.4.1 Classification of border collision bifurcations for multidimensional systems

Once the first return map has been derived as in eq.(2.6) the condition for different border collision bifurcations can be given in terms of the eigenvalues of the first return map. Let $M^*$ and $M^{**}$ be two fixed points of the sub-mappings $\Pi^+$ and $\Pi^-$ respectively, and let $p^*(\lambda)$ and $p^{**}(\lambda)$ be the corresponding characteristic polynomials(i.e. $p^*(\lambda) = |A_2 - \lambda I|, p^{**}(\lambda) = |A_1 - \lambda I|$). Lets define the following quantities which are useful to express the results in term of the eigenvalues of the map.

$$\sigma_1^+ = \text{number of real eigenvalues of } A_1 \text{ greater than } +1$$
$$\sigma_2^+ = \text{number of real eigenvalues of } A_2 \text{ greater than } +1$$
$$\sigma_1^- = \text{number of real eigenvalues of } A_1 \text{ less than -1}$$
$$\sigma_2^- = \text{number of real eigenvalues of } A_2 \text{ less than -1}$$

The three basic bifurcations phenomena can be described as follows:

1) A periodic orbit exists on one side of the boundary, (corresponding to the fixed point $M^*$) and is smoothly converted at the border-collision bifurcation point($\mu = 0$) into a solution existing on the other side of the boundary (corresponding to $M^{**}$), if:

$$p^*(1)p^{**}(1) > 0 \quad \text{or } \sigma_1^+ + \sigma_2^+ \text{ is even.} \tag{2.7}$$

2) The two periodic orbits corresponding to the fixed points $M^*$ and $M^{**}$, both exist on one side of the boundary, collide on the border at a C-bifurcation, $\mu = 0$, and then vanish on the other side, if:

$$p^*(1)p^{**}(1) < 0 \quad \text{or } \sigma_1^+ + \sigma_2^+ \text{ is odd.} \tag{2.8}$$

3) A new two-periodic solution originates at the border collision bifurcation point if:

$$p^*(-1)p^{**}(-1) < 0 \quad \text{or } \sigma_1^- + \sigma_2^- \text{ is odd.} \tag{2.9}$$

The details of this derivation can be found in [8]. Special cases for one dimensional maps and their application in power electronic circuits can be found in [5] and for two dimensional maps in [4].

## 2.5 Border Collision Bifurcation for One- Dimensional Maps

In this section we list the Border Collision bifurcation results for continuous piecewise-smooth one-dimensional maps. The border collision bifurcation phenomena is a rich class of bifurcation phenomena and has been observed in a variety of applications.

Assume that $G : \quad \times \quad \to \quad$ is piecewise smooth and continuous where $\quad \subset \quad$ is an open interval. Suppose that for some $x_0 \in R$ and $\mu_0 \in \quad$, $G(x_0, \mu_0) = x_0$ and $(x_0, \mu_0)$ is a point at which $G$ is not differentiable with respect to $x$. Let $a \geq 0$ be the left partial derivative with respect to $x$ of the map $G$ at $(x_0, \mu_0)$ and let $b \leq 0$ be the right partial derivative with respect to $x$ of the map $G$ at $(x_0, \mu_0)$. Then for many choices of $a$ and $b$ the map has a bifurcation as $\mu$ increases through $\mu_0$. These bifurcations are closely connected with that of skew tent maps [66].

Suppose for $\mu \in (\mu_0 - \epsilon, \mu_0)$ the map $G$ has an attracting fixed point $x(\mu)$, where $x(\mu_0) = x_0$ and hence $0 \leq a \leq 1$. If the attracting fixed point can not be continued as an attracting fixed point beyond $\mu_0$, then $b \leq -1$. Assuming $0 < a < 1$, the attracting fixed point bifurcates at $\mu = \mu_0$ to an attractor when $b < -1$, but that new emerging attractor is either periodic or chaotic rather than a fixed point. This is known as border collision bifurcation.

Fig. 2.2 shows for which $(a, b)$ there are periodic attractors and for which there are chaotic attractors.

The next result is mentioned in [66] but is based on [83] and others. First we define $P_m$ [66] :

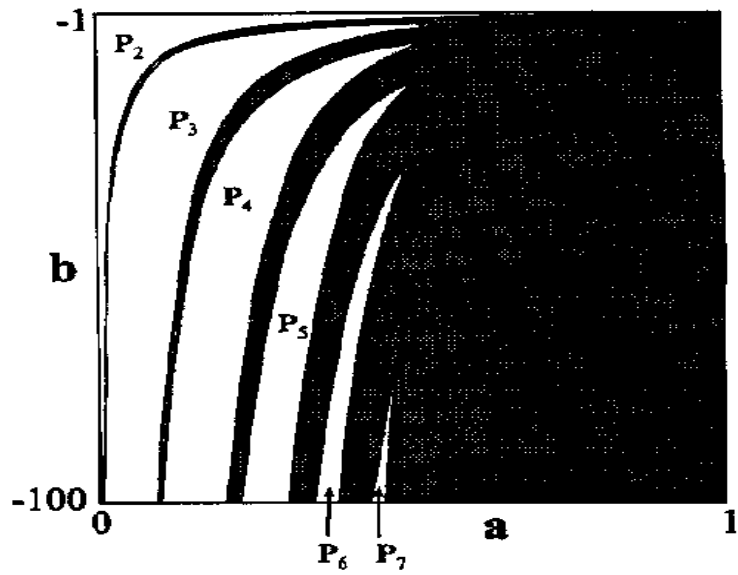

Figure 2.2: This figure shows two types of region in the (a,b) space. If (a,b) is in the white region, there exists a positive integer m ≥ 2 indicating the presence of an attracting period-m orbit. If (a,b) is in the dark region then there is a chaotic attractor.

$$P_m = (a,b) \in \quad : -a^{1-m} < b < \frac{a}{1-a}(1-a^{1-m}) \tag{2.10}$$

**Theorem 3** *Let $G: \quad \times \quad \to \quad$ be a piecewise smooth map. Assume that $G$ has isolated fixed point property at $(x_0, \mu_0)$. Let $m \geq 2$ and $k \geq 1$ be any integers. Write $a = \lim_{x \uparrow x_0} \frac{\partial G}{\partial x}(x, \mu_0)$ and $b = \lim_{x \downarrow x_0} \frac{\partial G}{\partial x}(x, \mu_0)$. Then*

*(1) if $(a,b) \in P_m$ as defined in eq. 2.10, then $G$ exhibits a border collision bifurcation from a fixed point attractor to a period-m attractor at $(x, \mu) = (x_0, \mu_0)$;*

*(2) if $(a,b) \notin P_m$, then $G$ exhibits a border-collision bifurcation from a fixed point attractor to a one or more piece chaotic attractor at $(x, \mu) = (x_0, \mu_0)$.*

A map $G$ as defined in theorem 3 has the *isolated crossing fixed point property* at $(x_0, \mu_0)$ if there exists an open neighborhood $N_0$ of $(x_0, \mu_0)$ and an open interval $_0 \subset J$ containing $\mu_0$ such that $G$ has the following properties:

(1) There exists a curve $K_c \subset \quad ^2$ such that $K_c = \{(c(\mu), \mu) : \mu \in J_0\}$, and for every $\mu \in J_0$, the map $G$ is not differentiable at the point $(c(\mu), \mu)$.

(2) The set $K_1 = \{(x(\mu), \mu) : G(x(\mu), \mu) = x(\mu) \text{ and } \mu \in \quad _0\}$ is a continuous curve and intersects the curve $K_c$ transversally at $(x_0(\mu_0), \mu_0)$, so $x_0 = c(\mu_0)$.

(3) For $\mu \in J_0$, write $L_\mu = \{x \in \quad : (x, \mu) \in N_0, x < c(\mu)\}$ and $R_\mu = \{x \in \quad : (x, \mu) \in N_0, x > c(\mu)\}$. For all $\mu \in \quad _0$, the map $G$ restricted to both $L_\mu$ and $R_\mu$ is a $C^1$-map, $\frac{\partial G}{\partial x}(x, \mu_0) > 0$ for $x \in L_\mu$ and $\frac{\partial G}{\partial x}(x, \mu_0) < 0$ for $x \in R_\mu$, and both the left and right limit exist;

(4) $G(c(\mu), \mu)$ is a smooth function of $\mu$, and $\frac{\partial G}{\partial \mu}(c(\mu_0), \mu_0) > 0$ for $\mu \in $ .

Although, the above theorem looks very complicated with its all technical conditions, its assumptions are not hard to verify for a map either analytically or numerically. This theorem will be applied in the analysis of TCP-RED map to prove the existence and border collision routes to chaos. We also need another lemma from [4] to show border collision in the event of two consecutive period doublings for a piecewise monotonic map. Actually, the result holds for weaker conditions but for our purpose this lemma will suffice. Suppose the map is given by

$$f(x;\mu) = \begin{cases} g(x;\mu) & \text{if } x \leq x_b \\ f(x;\mu) & \text{if } x \geq x_b \end{cases} \quad (2.11)$$

**Lemma 1** *[4] Assume g(x) and h(x) are each monotonic functions. Then a fixed point undergoes a period doubling bifurcation at $\mu_1$ and the double-period orbit goes through a period doubling bifurcation at $\mu_2 > \mu_1$ then the parameter range $[\mu_1, \mu_2]$ contains a border collision.*

We also need the Intermediate Value Theorem (IVT) and the Inverse Function Theorem (IFT) to prove some of the results in this work.

**Theorem 4 (IVT (see e.g. [22]))** *Let, the function $f : [a, b] \to $ be continuous, and c be a number strictly between $f(a)$ and $f(b)$; that is*

$$f(a) > c > f(b) \text{ or } f(a) < c < f(b).$$

*Then there exists a point $x_0 \in (a, b)$ with $f(x_0) = c$.*

Next is the Inverse Function Theorem:

**Theorem 5 (IFT (see e.g. [22]))** *Suppose $U$ is open in and suppose that the mapping $f : U \to$ is continuously differentiable. Let $x_0 \in U$ be a point at which the derivative matrix $df_{x_0}$ is invertible ($df_x \neq 0$ for $= 1$). Then there exist open neighborhoods $V \subset U$ containing $x$ and $W \subset f(U)$ contain $f(x_0)$ such that $f : V \to W$ is a diffeomorphism. That is, there exists a smooth inverse $f^{-1} : V \to W$, and for a point $y \in W$, if $x$ is a point in $V$ at which $f(x) = y$, then*

$$df^{-1}(y) = [df(x)]^{-1}.$$

# Chapter 3

# TCP-RED

## 3.1 Introduction

Computer networks are highly complicated systems, both in their temporal and spatial behavior [44]. Although they have traditionally been modeled and analyzed using stochastic methods, there have recently been several papers that use deterministic nonlinear modeling and analysis (e.g., [60, 31, 48, 20, 86, 21]).

In this chapter, we study a deterministic dynamical model of a simple computer network with Transmission Control Protocol (TCP) connections and implementing RED at the router end. The basic model that we consider was proposed recently by Firoiu and Borden [21]. We first modify their model with a simpler TCP throughput function [58, 30] to facilitate analysis. The calculations we give show that the model exhibits a rich variety of bifurcations leading to chaotic behavior of the computer network. The bifurcations occur as control parameters are slowly varied, moving the dynamics from a stable fixed point to oscillatory behavior and finally to a chaotic state. Then, the model is studied with a generic detailed throughput function [58, 30] as well as for a class of traffic using binomial congestion control [7].

A glimpse at the history of network congestion control reveals significant attempts to control congestion in the general network and telephony literature. Congestion and synchronization in tandem telephone queues have been studied in [19] using a piecewise affine model. A similar model has been applied to the dynamics of choke packets in a LAN to explain synchronization and sustained oscillations [20]. These models indeed explain qualitative changes in the operation of a network or that of a network component as parameters cross critical values. In contrast to the deterministic setting of [19] and [20], multi-stability or emergence of pseudo-stable states has been reported in a stochastic setting in [86]. The paper discusses the qualitative changes in the stochastic behavior of the network due to parameter change, which may lead to degradation in network performance.

There have been several attempts to address the issue of congestion control with TCP connections, which is the most popular network mechanism for data transfer. The most important scheme to avoid impending congestion was published in [24] and is known as random early drop, or RED. The basic idea of RED is to sense impending congestion before it happens and try to give feedback to the senders by either dropping or marking their packets.[1] The dropping probability is the control administered by the gateways once they detect queue build-up beyond a certain threshold. This scheme involves three parameters: 1) $p_{max}$, 2) $q_{max}$, and 3) $q_{min}$ that need to be selected. (The meanings of these parameters will be identified in the next section.) Most of the rules for setting these parameters are empirical, and come from networking experience. These rules have been evolving as the effects of controller parameters are not very clearly understood. There are papers discouraging implementation of RED (e.g., [59]), arguing that there is insufficient consensus on how to select controller parameter values, and that RED does not provide a drastic improvement in performance.

---

[1] Without loss of generality we assume that packets are dropped in the rest of the paper.

Initially, there was very little in the way of mathematical modeling of TCP-RED. However, with the recent efforts toward modeling TCP throughput for a transmission line with a packet drop probability [21, 60, 31, 48, 30], several papers have discussed TCP-RED in the framework of feedback control systems. Most of the models used are continuous-time and the analysis uses basic control theoretic results. The biggest problem with the continuous-time models is their inability to reflect delay, which is prominent in networks and can be very significant for large trunks [31, 48]. Third order, continuous-time, nonlinear models with variable (state dependent) delay are hard to analyze [48]. Even in the mathamatics literature, most of the state-dependent delay models are first order. The analysis reported on these models deals mostly with the stability of fixed points and limit cycles under different parameter settings. For the first time, chaotic behavior of TCP has been reported in [28]. The evidence for this irregularity is mostly explored by simulations. Some theoretical work on flow synchronization in TCP has been reported in [30, 39], but one of the very important issues which currently is not well understood is how a smoothly operating network transitions into chaos. To borrow dynamical systems terminology, the route to chaos starting from a stable fixed point is not well-studied.

In this work, a discrete-time map will be used to model the TCP-RED interaction. A dynamical systems approach will be used to explain the loss of stability, bifurcation behavior, and routes to chaos in TCP-RED networks. We will use bifurcation-theoretic ideas to explain nontrivial periodic behavior of the system.

The appearance of bifurcation and chaos should not be surprising, considering that the system response is nonlinear especially during heavy load conditions. We will show the performance of the system as a function of various control and system parameters in general and try to explain these irregular behaviors with the help of bifurcation diagrams.

Our work begins by realizing that the model proposed in [21] can be viewed as a first-order (rather than third-order) discrete nonlinear model. Our replacement of the TCP throughput function of [21] with a simpler version makes the analysis feasible. However, symbolic calculations could be used to allow treatment of the more complex throughput function of [21] as shown in section 3.8. The advantage of the current work is that the calculations are simple enough that the results are easily understood.

We borrow the model proposed by Firoiu and Borden [21] and use the well known formula for TCP throughput proposed by many others including [58, 30]. The motivation behind not using Firoiu and Borden's formula for TCP throughput is its complexity. Complex operations like inverse of a function in different parameters, which are needed to connect the TCP to the control mechanism RED, demand simplicity in the TCP throughput formulation. Although this TCP-RED formulation may not be the exact representation of the complicated mechanism, it does give a qualitative handle on its dynamics and enhances our understanding of chaos and other instabilities. We hope that this understanding will lead to better monitoring of the network congestion and help us in formulating robust but simplified control mechanisms.

This chapter is organized as follows. In Section 3.2, we describe the TCP-RED mechanism in control system framework. Section 3.3 contains the discrete map of TCP-RED mechanism. Section 3.4 deals with the stability and normalization scheme. Section 3.5 explains the different nonlinear phenomena we have observed in our models. Numerical results are shown in Section 3.6. In section 3.7 we discuss the results in networking context and show the results from NS-2 [15] simulation. Finally, section 3.8 shows that this modeling formalism is valid for almost all the throughput functions for various traffic streams.

## 3.2 TCP-RED: Feedback System Modeling

A computer network implementing TCP-RED is essentially a feedback loop where senders adjust their transmission rates based on the feedback they receive from the routers in the form of dropped packets. Routers on the other hand implement a control policy which can be either drop tail or RED [24]. There have been different approaches to model the dynamics of TCP-RED and various control schemes have been proposed [21, 60, 31, 48, 30] not only to control the system but to also enhance its dynamic performance. We closely follow the approach taken in [21] with a modified TCP throughput formula.

Each flow at a router sends packets with rate $r_{s,i}$. The sending rates of all $N$ flows combine at the buffer of link $l$ and generate a queue of size $q$ which is limited by its buffer size $B$. The controller at the router drops packets with a probability $p$ which is a function of average queue size $\overline{q}$. For $i$th flow let the forwarding rate at the router be $r_{t,i}$ which is the same as $r_{s,i}$ sans dropped packets. When a sender notices that its packets are being dropped, it adjusts its sending rate based on the drop probability $p$ it observes.

This makes a control system with sender's rate as control variable with the controller sitting at the router which issues the feedback signal in the form of a drop probability. The aim of this control system is to keep the cumulative throughput below or equal to the link's capacity $C$:

$$\Sigma_{j=1}^{n} r_{t,j} \leq C$$

We assume that TCP flows are long-lived connections and that the set of connections remains the same, then the throughput of each TCP flow follows the steady state model derived in [58, 30].

$$r_{t,i} = T(p, R_i)$$

$$T(p, R) = \frac{M}{R} \frac{K}{\sqrt{p}} \tag{3.1}$$

where

$$\begin{aligned}
T &= \text{Throughput of a TCP flow (in bits/sec)} \\
M &= \text{Maximum Segment Size or Packet Size} \\
R &= \text{Round Trip Time} \\
K &= \text{constant which varies between 1 and } \sqrt{8/3} \text{ [58]} \\
p &= \text{probability of packet loss}
\end{aligned}$$

To simplify matters even further we assume that all flows are uniform or they all have same round trip time $R$, same maximum segment or packet size $M$, and that maximum congestion window size advertised by TCP's receiver $W_{max}$ is large enough to not affect $T(p, R)$. This implies

$$\begin{aligned}
r_{t,i}(p, R) &= r_{t,j}(p, R), \quad 1 \leq i, j \leq N \text{ and hence} \\
&\leq \frac{C}{N}
\end{aligned}$$

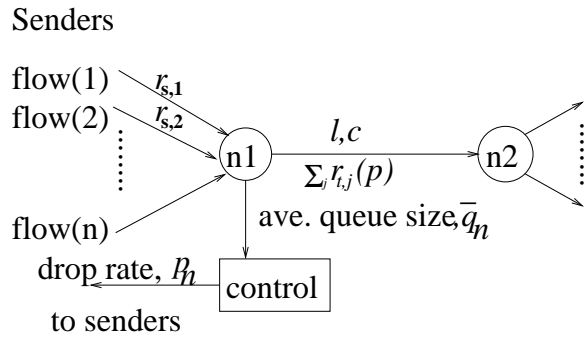

Figure 3.1: Simplified Network Diagram

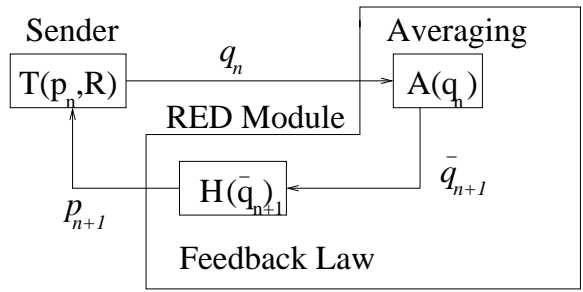

Figure 3.2: Feedback control system corresponding to the network shown

So this assumption enables us to reduce the $N$-flow system to a single flow system with feedback although it is important to keep in mind that feedback is based on the sending rate of all the flows since the router has no way of differentiating between them, at least in this set up.

To define this control system mathematically, we model the queue as a function of control variable $q = G(p)$, which acts as a plant in control system literature. To analyze this control system we also need the control function $p = H(\bar{q})$ implemented at the gateways. This control function $H$ is given by the policy implemented at the queue, such as Drop-Tail or RED [24]. Now following the procedure suggested in [21] we can define the plant function $G(p)$ as follows:

$$G(p) = \begin{cases} \min(B, \frac{C}{M}(T_R^{-1}(p, \frac{C}{N}) - R_0)) & : \quad p \leq p_0 \\ 0 & : \quad otherwise \end{cases} \quad (3.2)$$

where

$$R_o = \text{round-trip propagation and transmission time and}$$
$$p_o = T_p^{-1}(C/N, R_0)$$

Here, $T_R^{-1}(C/N, R_0)$ denotes the inverse of $T(p, R)$ in $R$, $T_p^{-1}(C/N, R_0)$ denotes the inverse of $T(p, R)$ in $p$, and $p_0$ is maximum probability for which the system is fully utilized i.e., for $p \geq p_0$ senders will have their rates too small to keep the link fully utilized. For $T(p, R)$ defined by eq. 3.1.

$$p_0 = \left(\frac{MK}{R_0 \frac{C}{N}}\right)^2 \quad (3.3)$$

$$G(p) = \begin{cases} \min\left(B, \frac{C}{M}(\frac{MK}{\frac{C}{N}\sqrt{p}} - R_0)\right) & , \quad \text{if } p \leq p_0 \\ 0 & , \quad \text{otherwise} \end{cases} \quad (3.4)$$

RED control law can be expressed as follows:

$$\begin{aligned} p &= H(\overline{q_e}) \\ &= \begin{cases} 0 & , \quad 0 \leq \overline{q_e} < q_{min} \\ \frac{\overline{q_e} - q_{min}}{q_{max} - q_{min}} p_{max} & , \quad q_{min} \leq \overline{q_e} < q_{max} \\ 1 & , \quad q_{max} \leq \overline{q_e} \leq B \end{cases} \end{aligned} \qquad (3.5)$$

where $\overline{q_e}$ is the exponential weighted moving average of queue size, $q_{min}, q_{max}, p_{max}$ are configurable RED parameters, and $B$ is buffer size.

## 3.3 Discrete Model for TCP-RED

It is argued in [21] that TCP adjusts its sending rate depending on whether it has detected a packet drop or not. Hence, this process can be modeled as a stroboscopic map where the instant of observation is one round trip time or RTT. This technique has been utilized before for different clocked systems in power electronics for modeling the dynamics of power converters [6]. Following similar arguments it seems reasonable to model TCP-RED dynamics as a discrete map. Although one would prefer that the sampling interval be regular, there are models where the dynamics is sampled at irregular intervals and the resulting maps are known as "impact maps" [18].

Let $p_k$ be the packet drop probability at $t_k$. At time $t_{k+1} = t_k + RTT$ the senders observe drop rate $p_k$ and in an average sense, adjust their transmission rates. This in turn forces the buffer to its new state $q_{k+1} = G(p_k)$ following the queue law in eq. 3.4. The RED mechanism now computes a new estimate of queue size $\overline{q}_{e,k+1} = A(\overline{q}_{e,k}, q_{k+1})$, following the exponential weighted moving average:

$$A(\overline{q}_{e,k}, q_{k+1}) = (1-w)\overline{q}_{e,k} + w \cdot q_{k+1} \qquad (3.6)$$

where $w$ is the weight used for averaging. After computing $\bar{q}_{e,k+1}$, the RED module adjusts it dropping rate to $p_{k+1} = H(\bar{q}_{e,k+1})$ given by its "feedback control law" in eq. 3.5. This results in the following mathematical relationships:

$$
\begin{aligned}
q_{k+1} &= G(p_k) \\
\bar{q}_{e,k+1} &= A(\bar{q}_{e,k}, q_{k+1}) \\
p_{k+1} &= H(\bar{q}_{e,k+1})
\end{aligned}
\tag{3.7}
$$

From eq. 3.7, we derive a simple one dimensional discrete time dynamical system representation. Since the maps $G(.)$ and $H(.)$ are static, the only dynamics that appear are from the map $A(.,.)$. Using substitution, we can easily derive the following equation for the exponential weighted moving average for the queue length at time $t_{k+1}$:

$$
\bar{q}_{e,k+1} = (1-w)\bar{q}_{e,k} + w \cdot G(H(\bar{q}_{e,k}))
\tag{3.8}
$$

Eq. 3.8 also provides a formula to compute the instantaneous queue occupancy $q_{k+1}$ at time $t_{k+1}$ from the exponentially averaged queue occupancy $\bar{q}_{e,k+1}$. This will be utilized later to plot both averaged and instantaneous queue occupancies.

$$
q_{k+1} = \frac{\bar{q}_{e,k+1} - (1-w)\bar{q}_{e,k}}{w}
\tag{3.9}
$$

Below, we illustrate some interesting dynamical behavior of eq. 3.8. This equation is rather simple in most of its domain of definition.

We know that $G(.)$ is identically 0 for all $p \geq p_0$. So we can find a corresponding value $b_1$ of $\bar{q}_{e,k}$ such that for any $\bar{q}_{e,k} \geq b_1$, $G(.)$ is identically 0 if we assume a monotone feedback law.

$$b_1 = \begin{cases} \dfrac{p_0(q_{max}-q_{min})}{p_{max}} + q_{min} &, \text{ if } p_{max} \geq p_0 \\ q_{max} &, \text{ otherwise} \end{cases} \quad (3.10)$$

This gives an explicit formula for the map in eq. 3.8 for all $\bar{q}_{e,k} \geq b_1$:

$$\bar{q}_{e,k+1} = (1-w)\bar{q}_{e,k}$$

Now consider the other boundary value $b_2$ of $\bar{q}_{e,k}$ such that for all $\bar{q}_{e,k} \leq b_2$ we have $G(.) = B$ or buffer is always full. This value can be computed from eq. 3.4 and eq. 3.5. $b_2$, and is given by:

$$b_2 = \dfrac{\left(\dfrac{NK}{B+\dfrac{R_0C}{M}}\right)^2}{p_{max}}(q_{max}-q_{min}) + q_{min} \quad (3.11)$$

This gives an explicit formula for map in eq. 3.8 for all $\bar{q}_{e,k} \leq b_2$:

$$\bar{q}_{e,k+1} = (1-w)\bar{q}_{e,k} + wB$$

It is clear that most of the interesting dynamics happens for $b_2 \leq \bar{q}_{e,k} \leq b_1$. Map in eq. 3.8 can be written for this region as follows:

$$\bar{q}_{e,k+1} = (1-w)\bar{q}_{e,k} + w\left(\dfrac{NK}{\sqrt{\dfrac{p_{max}(\bar{q}_{e,k}-q_{min})}{(q_{max}-q_{min})}}} - \dfrac{R_0C}{M}\right) \quad (3.12)$$

Finally combining the definition in all three regions, we get TCP-RED map:

$$q_{e,k+1} = \begin{cases} (1-w)q_{e,k} & \text{if } q_{e,k} > b_1 \\ (1-w)q_{e,k} + wB & \text{if } q_{e,k} < b_2 \\ (1-w)\bar{q}_{e,k} + w\left(\dfrac{NK}{\sqrt{\dfrac{p_{max}(\bar{q}_{e,k}-q_{min})}{(q_{max}-q_{min})}}} - \dfrac{R_0C}{M}\right) & \text{otherwise} \end{cases}$$
$$:= f(\bar{q}_{e,k}, \rho) \quad (3.13)$$

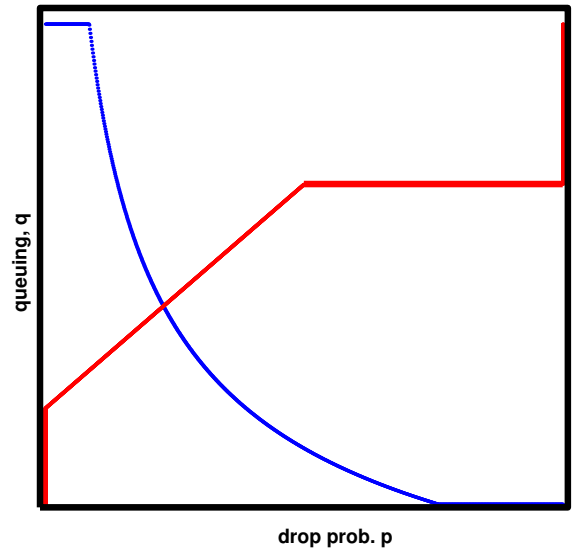

Figure 3.3: Feedback law $H(q)$ in red and queue law G(p) in blue

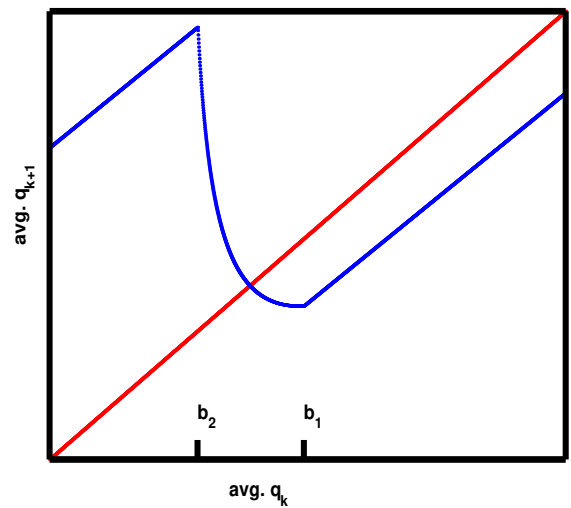

Figure 3.4: 1. Blue curve shows first return map, 2. $45^0$ degree line is shown in red.

where $\rho$ summarizes the parameters in the system. Fig. 3.4 shows the plot of first return map for TCP-RED map.

Existence of a fixed point in the interval $[b_2, b_1]$ for map given by 3.13 is proved next.

**Theorem 6** *Map given by 3.13 has a fixed point in the interval $[b_2, b_1]$ for all $w > 0$.*

**Proof:** Lets consider the function $g(x) = f(x, \rho) - x$ where $f(\cdot, \rho)$ is restriction of 3.13 on the interval $[b_2, b_1]$. Clearly, $g(x)$ continuous as $f(\cdot, \rho)$ and $-x$ are continuous. Evaluation of $g(b_2)$ gives $(1-w)*b_2 + wB - b_2 = w(B - b_2)$ which is positive for all $w > 0$ as $B > b_2$. $g(b_1)$ is $(1-w)b_1 - b_1 = -wb_1$ which is negative for all $w > 0$.

From Intermediate Value Theorem, there exist an $x^* \in [b_2, b_1]$ such that $g(x^*) = 0$ or $f(x^*, \rho) = x^*$. Hence, the proof. ∎

This result is very interesting in the sense that it does not make use of exact functional form of the middle segment in the TCP-RED map. It is true irrespective of which throughput function we use or presence of UDP type traffic in the model. More about genericity of these results for any throughput function will be discussed in section 3.8. More about UDP and TCP mixed traffic can be found in [75, 74].

Solving the TCP-RED map in eq. 3.13 leads to a third degree polynomial in fixed point $\overline{q}_e^*$ which interestingly does not depend on $w$ as should be expected since, both the "queue law" and the "feedback control law" are not functions of $w$. The polynomial is given below.

$$(\overline{q}_e^* - q_{min})(\overline{q}_e^* + \frac{R_0 C}{M})^2 = \frac{(NK)^2}{p_{max}}(q_{max} - q_{min}) \tag{3.14}$$

This equation tells the general trends of fixed point as a function of different parameters. In particular, it can be shown that fixed point is a strictly increasing function of number of active connections $N$.

**Lemma 2** *Fixed point $\overline{q}_e^*$ of TCP-RED system given by eq. 3.12 is a strictly increasing function of number of connection $N$.*

**Proof:** Differentiating eq. 3.12 with respect to $N$ gives:

$$(\overline{q}_e^* + \frac{R_0 C}{M})^2 \frac{d\overline{q}_e^*}{dN} + 2(\overline{q}_e^* + \frac{R_0 C}{M})(\overline{q}_e^* - q_{min})\frac{d\overline{q}_e^*}{dN} = \frac{2NK^2}{p_{max}}(q_{max} - q_{min}) \quad (3.15)$$

$$\frac{d\overline{q}_e^*}{dN} = \frac{\frac{2NK^2}{p_{max}}(q_{max} - q_{min})}{(\overline{q}_e^* + \frac{R_0 C}{M})^2 + 2(\overline{q}_e^* + \frac{R_0 C}{M})(\overline{q}_e^* - q_{min})} \quad (3.16)$$

This shows that the right hand side is strictly positive in the parameter region of interest. Hence, the result. ■

## 3.4 Normalization and Bifurcation Analysis

We begin this section with the following assumption, and proceed to the development of a normalized TCP-RED model and to the bifurcation analysis of the normalized model.

**Assumption 1** $p_{max} > p_0$.

This is relatively more interesting case from the application point of view as it avoids the effect of disconnected section of RED law where drop probability jumps from $p_{max}$ to 1.

### 3.4.1 Normalization scheme

$\bar{q}_{e,k}$ is scaled by buffer size $B$ to make this map over unit interval($[0,1] \to [0,1]$). This scaling modifies the system equations as follows:

$$\begin{aligned}
q_k^n &:= \frac{\bar{q}_{e,k}}{B} \\
\gamma &:= \frac{q_{max} - q_{min}}{p_{max} B} > 0 \\
q_{min}^n &:= \frac{q_{min}}{B} \\
b_2^n &:= \frac{b_2}{B} = \left(\frac{NK}{B + \frac{R_0 C}{M}}\right)^2 \gamma + q_{min}^n \\
b_1^n &:= \frac{b_1}{B} = p_0 \gamma + q_{min}^n
\end{aligned}$$

$$q_{k+1}^n = \begin{cases} (1-w)q_k^n & \text{if } q_k^n > b_1^n \\ (1-w)q_k^n + w & \text{if } q_k^n < b_2^n \\ (1-w)q_k^n + \\ \quad + w\left(\frac{NK}{B\sqrt{\frac{q_k^n - q_{min}^n}{\gamma}}} - \frac{R_0 C}{MB}\right) & \text{otherwise} \end{cases} \quad (3.17)$$

Eq. 3.17 maps unit interval into itself. The benefit of this normalization is in the reduction of parameters required to characterize the dynamical behavior of this system.

### 3.4.2 Linear stability analysis

Linear stability of a map can be analyzed by linearizing the map at fixed and computing the corresponding eigenvalue. For the normalized eq. 3.17 linearization and eigenvalue is given as:

$$\begin{aligned}
\left.\frac{\partial f(q_k^n, \rho)}{\partial q_k^n}\right|_{q_k^n = q^{n*}} &= 1 - w - \frac{wNK}{2B(q^{n*} - q_{min}^n)^{\frac{3}{2}}}\sqrt{\gamma} \\
&:= \lambda(\rho)
\end{aligned} \quad (3.18)$$

Note that the eigenvalue depends on the fixed point. Of special interest here are parameter settings that lead to loss of stability of the fixed point, giving rise to nonlinear instabilities through a system bifurcation. Numerical simulations of the system show the presence of oscillatory regimes as control and system parameters are varied, and indicate that a period doubling bifurcation occurs from the fixed point with the variation of any of the system or control parameters. Following theorem makes this result precise:

**Theorem 7** *TCP-RED system given by eq. 3.17 will go through a Period-Doubling bifurcation for a large enough $w$ such that eigenvalue*

$$\lambda(w) = 1 - w - \frac{wNK}{2B(q^{n*} - q^n_{min})^{\frac{3}{2}}} \sqrt{\gamma} = -1$$

.

**Proof:** Proof is direct application of Period-Doubling theorem 2 on the map given by eq. 3.17. In particular, eigenvalue condition is satisfied due to the assumption and $\lambda' = -1 - \frac{NK}{2B(q^{n*}-q^n_{min})^{\frac{3}{2}}} \sqrt{\gamma} < 0$ can be easily established. ■

This kind of results clearly hold for not only the averaging parameter $w$ but also for other parameters like $q^n_{min}$, $p_{max}$ etc., although the direction of bifurcation can be different. Further proof of this dynamical behavior can be seen in the numerical section 3.6 where eigenvalues are explicitly computed as bifurcation parameter is varied.

Thus, we are led to consider cases in which the eigenvalue given by (3.18) becomes $-1$, giving a period doubling bifurcation (PDB) leading to oscillatory behavior in the system. To demonstrate existence of such bifurcations, it is easiest to focus on the exponential averaging parameter $w$ as the distinguished bifurcation parameter. The critical value of $w$ is one for which the eigenvalue given by (3.18) is $-1$. The critical value can be expressed in closed form as follows:

$$w_{crit} = \frac{2}{1 + \frac{NK}{2B(q^{n*} - q^n_{min})^{\frac{3}{2}}}\sqrt{\gamma}} \quad (3.19)$$

Now we can formulate the linear stability criterion ($|\lambda(\rho)| < 1$) [37] as follows:

$$\left|1 - w - \frac{wNK}{2B(q^{n*} - q^n_{min})^{\frac{3}{2}}}\sqrt{\gamma}\right| < 1 \quad (3.20)$$

We look for parameter values for which the eigenvalue given by eq. 3.18 becomes $-1$ and gives rise to a period doubling bifurcation (PDB) leading to the first oscillatory behavior in the system. Further, we compute the second and third derivatives of the normalized map to analyze the nature of this bifurcation.

$$\left.\frac{\partial^2 f}{\partial q_k^{n2}}\right|_{q_n^k = q^{n*}} = \frac{3wNK}{4B(q^{n*} - q^n_{min})^{\frac{5}{2}}}\sqrt{\gamma} \quad (3.21)$$

$$\left.\frac{\partial^3 f}{\partial q_k^{n3}}\right|_{q_n^k = q^{n*}} = \frac{-15wNK}{8B(q^{n*} - q^n_{min})^{\frac{7}{2}}}\sqrt{\gamma} \quad (3.22)$$

A function $S = \left(\frac{1}{2}\left(\frac{\partial^2 f}{\partial q_k^{n2}}\right)^2 + \frac{1}{3}\left(\frac{\partial^3 f}{\partial q_k^{n3}}\right)\right)$ of the second and the third derivatives of a map are used to determine the nature of a period doubling bifurcation (see [29], pp.158). A positive $S$ implies that the bifurcation is supercritical, and a negative $S$ implies a subcritical bifurcation. A subcritical bifurcation is potentially dangerous due to the discontinuous appearance of the period two orbit. Function $S$ for the system given by eq. 3.17 reads

$$S = \frac{wnK\sqrt{\gamma}}{B(q^{n*} - q^n_{min})^{\frac{7}{2}}} \left[\frac{9}{32}\frac{wnK\sqrt{\gamma}}{B(q^{n*} - q^n_{min})^{\frac{3}{2}}} - \frac{5}{8}\right] \quad (3.23)$$

The expression for $S$ in eq. 3.23 shows that it may change sign giving rise to a subcritical bifurcation if the parameters are not chosen carefully. This should be kept in mind when designing a TCP-RED system to avoid any unexpected oscillations in router queues. It should be noted here that first instability in this system is not due to the discontinuous behavior of RED control law rather system's nonlinear throughput function leads the system to period doubling instability in the parameter region of interest. This result contrasts significantly with earlier results where discontinuous behavior in either RED control or plant function was believe to be responsible [21].

A close look at eq. 3.17 shows the most of the interesting behavior in the system happens due to middle nonlinear segment of this equation. It contains two parts: 1) from averaging, 2) due to RED's control action in the form of drop probability. If we further look at these two contributions in eq. 3.18 then they have different first derivative with respect to $\bar{q}_{e,k}$. Averaging term has a positive first derivative but RED-control term has a negative derivative. Hence, there is a possibility of very rich dynamical behavior in the presence of these two functions with dissimilar derivatives. Generally speaking, if the middle segment is decreasing uniformly then first period doubling will be immediately followed by border collision bifurcations. In case of middle segment having a minima there can be cascades of period doubling bifurcations before a border collision bifurcation. To characterize the monotonicity of the map we compute its minimizer $q^m$ explicitly by setting the expression for eigenvalue in eq. 3.18 to zero.

$$1 - w - \frac{wNK}{2B(q^m - q_{min}^n)^{\frac{3}{2}}}\sqrt{\gamma} = 0 \text{ implies}$$

$$q^m := q_{min}^n + \left(\frac{wNK\sqrt{\gamma}}{2B(1-w)}\right)^{\frac{2}{3}} \tag{3.24}$$

Clearly, this minima will not be present in the map if the border $b_1^n$ is smaller than the minimizer $q^m$. It will be present otherwise.

The rest of this section contains a discussion of possible routes to chaos in a TCP-RED map. We make the case for the possibility of a *border collision* bifurcation (BCB) by analytically proving some of the properties of the TCP-RED map given by eq. 3.17. We also provide some numerical evidence for the same.

**Lemma 3** : *The map described by Eq. 3.17 is piecewise monotone if $b_1^n < q^m$.*

**Proof:** : First, note that map is continuous by construction. The only two points where continuity needs to be checked is $b_1^n$ and $b_2^n$. For $0 \leq q_k^n < b_2^n$ and $B \geq q_k^n \geq b_1^n$, piecewise monotonicity can be shown by differentiating the map. Its differential $\forall q_k^n > b_1^n$ or $q_k^n < b_2^n$ is $1 - w$ which implies that map is strictly increasing.

If $b_1^n < q^m$, map in eq. 3.17 has a negative differential for $q_k^n = b_1^n$. Now, for any $q_k^n \in [b_2^n \; b_1^n]$, $q_k^n - q_{min}^n$ will be smaller than $b_1^n - q_{min}^n$. This implies that the third term in the expression of the differential of this map in eq. 3.18 will be getting more negative as $q_k^n$ moves towards the left of $b_1^n$ until $b_2^n$. Hence, eigenvalue stays negative for any $q_k^n \in [b_2^n \; b_1^n]$. This proves that the map is strictly decreasing in the interval $[b_2^n \; b_1^n]$. Hence, the map given by Eq. 3.17 is piecewise monotone. ∎

**Lemma 4** : *The map given by eq. 3.17 depends smoothly on $w$.*

**Proof:** : The differential of this map with respect to the parameter $w$ in the three regions is given as follows:

$$\frac{\partial f(q_k^n, w, \gamma)}{\partial w} = \begin{cases} -q_k^n & \text{if } q_k^n > b_1^n \\ 1 - q_k^n & \text{if } q_k^n < b_2^n \\ -q_k^n + \left(\frac{NK}{B\sqrt{\frac{q_k^n - q_{min}^n}{\gamma}}} - \frac{R_0 C}{MB}\right) & \text{otherwise} \end{cases} \quad (3.25)$$

Evaluating the partial differential with respect to $w$ as given in eq. 3.25 from the both sides near the borders $b_1^n$ and $b_2^n$, it can be easily seen that their values agree. Hence, this map is smooth in $w$. ∎

Based on the properties proven in these two lemmas we obtain following result which is based in lemma in [6]. It outlines the crucial role played by border collision and their eventuality.

**Lemma 5** *If a fixed point of map given by (3.17) undergoes a smooth (eigenvalue = $-1$) period doubling bifurcation at $w_1$ and the resulting period two orbit also goes through a smooth period doubling for $w_2 > w_1$ then the periodic orbit must collide with the border for some $w \in [w_1 \ w_2]$.*

**Proof:** The proof for above lemma directly follows from lemma 1 given in [6]. ∎

Now we are ready to use the *border collision bifurcation* theory [6, 8, 66, 65] to analyze the bifurcations due to the variation of parameter $w$.

## 3.5 Border Collision Bifurcation (BCB)

In this section we use the *border collision bifurcation* theory [6, 8, 66, 65] to analyze the bifurcations due to the variation of parameter $w$. BCBs occur in piecewise smooth maps, and involve a nonsmooth bifurcation when a parameter change results in a fixed point (or other operating condition) crossing a border between two regions of smoothly defined dynamics in state space. In the theory of border collision bifurcation, smooth dependence on parameters is required.

If a fixed point collides with the border(s) with a change in the parameters, there is a discontinuous change in the derivative of $\frac{\partial f}{\partial x}$, and the resulting phe-nomenon is called *border collision bifurcation*. This kind of bifurcation has been reported widely in economics [66], mechanical systems, and power electronic models [6, 8, 65].

Border collision is a local bifurcation and hence it can be studied by characterizing the local properties of a map in the neighborhood of the colliding border. It is shown in [65] that a normal form which is an affine approximation of $f$ in the border neighborhood is sufficient to quantify the possible border collision bifurcations. This normal form is

$$G(x,\mu) = \begin{cases} ax + \psi, & \text{if } x \leq 0 \\ bx + \psi, & \text{if } x \geq 0 \end{cases} \quad (3.26)$$

where

$$a = \lim_{x \to x_b^-} \frac{\partial}{\partial x} f(x, \psi^*) \quad (3.27)$$

$$b = \lim_{x \to x_b^+} \frac{\partial}{\partial x} f(x, \psi^*) \quad (3.28)$$

and $\psi^*$ is the parameter for which border collision happens. It can be assumed to be 0 without any loss of generality.

There are various types of bifurcation scenarios possible depending on the values of coefficients $a$ and $b$ in the normal form given by (3.27) and (3.28). For the sake of simplicity, we will discuss only the case relevant to the observed phenomena in our model and provide a numerical proof by computing the one sided coefficients (eigenvalues) for the same.

As we saw in the last section, piecewise monotonicity ensures that the fixed point of the map collides with the border before the sign of eigenvalue changes with the variation of parameters in interesting regimes. Again, under some technical conditions it will be shown that that TCP-RED map will go through border-collision type bifurcations in many general situations for periodic orbits rather than fixed points. The complicated nature of higher iterates of TCP-RED map in different part of unit interval makes the proof cumbersome. For simplicity sake we will analyze only second iterate. After a smooth period doubling bifurcation there will be two branches emerging around the original unstable fixed point. To check the border collision bifurcation we choose the bifurcated branch which lies above the original fixed point. Typically in useful parameter region, there is a smooth period doubling and then upper branch of the bifurcation diagram may collide with the border $b_1^n$ leading to chaos or other higher periodicities.

To analyze the border collision phenomena we have to look at the local behavior of second iterate or $G = F^2(\cdot, \cdot)$ where $F$ is the map given by eq. 3.17. Due to the evident nonlinearity of second iterate, it needs to be shown that it has *isolated fixed point crossing property* at the border collision value of exponential averaging weight parameter $w_{bc}$ and bifurcation point $b_1^n$. To prove this property we choose the neighborhood $(x^*, \frac{b_1^n}{1-w})$ around the border $b_1^n$ and a small neighborhood of critical parameter $w_{bc}$ as parameter interval $\triangle_0$. We note the following properties of second iterate of our map towards *isolated fixed point crossing property*.

(1) Curve $K_c = (b_1^n, w) \subset \triangle^2, w \in \triangle_0$ and for every $w \in \triangle_0$, the map $G$ is not differentiable at the point $(b_1^n, w)$. We note that border is independent of parameter $w$ by eq. 3.10.

(2) The set $K_1 = \{(x(w), w) : G(x(w), w) = x(w) \text{ and } w \in \triangle_0\}$ is a continuous curve and intersects the curve $K_c$ transversally at $(b_1^n, w_{bc})$, so that border $b_1^n$ is fixed point also. Transversality can be proved by showing that the locus of fixed points of period two map is not horizontal at the critical parameter value because period two fixed points depend on $w$ but border is independent of parameter $w$ by eq. 3.10. This is required to differentiate the border crossing from grazing-type bifurcation [17] due to tangentially approaching trajectories.

(3) For $w \in \;_0$, write $L_w = \{x \in \; : (x,w) \in N_0, x < b_1^n\}$ and $R_w = \{x \in \; : (x,w) \in N_0, x > b_1^n\}$. For all $w \in \;_0$, the map $G$ restricted to both $L_w$ and $R_w$ is a $C^1$-map, $\frac{\partial G}{\partial x}(x, w_{bc}) > 0$ for $x \in L_w$ because map is decreasing for both the iterates but $\frac{\partial G}{\partial x}(x, w_{bc}) < 0$ for $x \in R_w$ because it has one increasing and one decreasing iterate, and both the left and right limit do exist;

(4) $G(b_1^n, w)$ is a smooth function of $w$, and $\frac{\partial G}{\partial w} > 0$ for $w \in \;_0$ This is clear from the shape of bifurcation diagram where period two fixed point increases steadily with the increase in parameter $w$.

For our model, the case of interest is when

$$0 < a < 1 \text{ and } b < -1 \tag{3.29}$$

This is mentioned as case 8 in [65]. It is shown that in this case a fixed point attractor can bifurcate into a periodic attractor or a chaotic attractor as bifurcation parameter is varied beyond its critical value. This is the exact phenomenon we observe for our model when the bifurcation parameter $w$ is varied and a stable period two orbit transitions to chaos. Essentially, if we take the second iterate of our map, it exhibits a fixed point bifurcating into a chaotic orbit. Existence of chaos can be confirmed by computing the Lyapunov exponents [71].

A numerical example that provides evidence for our claim is given in the next section.

## 3.6 Numerical Example

The behavior of the map in (3.17) can be explored numerically in parameter space to look for interesting dynamical phenomena. When the eigenvalue exits the unit circle, the fixed point becomes unstable. Depending on the nature of the ensuing bifurcation, there can be new fixed points, higher period orbits, or chaos. There is also a possibility of an orbit (original fixed point or a bifurcated orbit) colliding with either border $b_1^n$ or $b_2^n$, leading to a rich set of possible bifurcations.

In this section we numerically validate our analysis using bifurcation diagrams. A bifurcation diagram shows the qualitative changes in the nature and the number of fixed points of a dynamical system as parameters are quasistatically varied. The horizontal axis is the parameter that is being varied, and the vertical axis represents a measure of the steady states (fixed points or higher period orbits). For generating the bifurcation diagrams, in each run we randomly select four initial average queue sizes, $q_1^{ave}(0), q_2^{ave}(0), q_3^{ave}(0)$ and $q_4^{ave}(0)$, and these average queue sizes evolve according to the map $f(\cdot)$ in (3.13), i.e.,

$$q_i^{ave}(k) = f(q_i^{ave}(k-1), \rho) \text{ , for } k = 1, \cdots, 1,000 \text{ and } i = 1, 2, 3, \text{ and } 4 \text{ .}$$

We plot $q_i^{ave}(k), k = 991, \cdots, 1,000$ and $i = 1, 2, 3,$ and 4. Hence, if there is a single stable fixed point or attractor $q^*$ of the system at some value of the parameter, all $q_i^{ave}(k)$ will converge to $q^*$ and there will be only one point along the vertical line at the value of the parameter. However, if there are two stable fixed points, $\tilde{q}_1^{ave}$ and $\tilde{q}_2^{ave}$, with a period of two, i.e., $f(\tilde{q}_i^{ave}, \rho) \neq \tilde{q}_i^{ave}$ and $f(f(\tilde{q}_i^{ave}, \rho)) = \tilde{q}_i^{ave}, i = 1, 2,$ then there will be two points along the vertical lines and the average queue size will alternate between $\tilde{q}_1^{ave}$ and $\tilde{q}_2^{ave}$.

In this section we study the effects of various system and control parameters on average and instantaneous queue behavior as each of these parameters is varied while the others are fixed. More specifically, we study how the averaging weight $w$, lower threshold $q_{min}$, the number of connections $N$, and the round-trip propagation delay $d$ affect system stability, queue behavior and their sensitivity to these parameters.

### 3.6.1 Effect of exponential averaging weight

We use the following parameters for the numerical examples presented in this subsection:

$$q_{max} = 750, \ q_{min} = 250, \ C = 75 \text{ Mbps}, \ K = \sqrt{3/2}$$

$$B = 3{,}750 \text{ packets}, \ d = 0.1 \text{ sec}, \ M = 4{,}000 \text{ bits}$$

$$N = 250, \ w = \text{bifurcation parameter}$$

The bifurcation plots in Fig. 3.5 and 3.6, show the effect of varying the averaging weight $w$ for different values of $p_{max}$, namely $p_{max} = 0.1$ and $p_{max} = 0.03$. Fig. 3.5(a) and Fig. 3.6(a) show the exponentially averaged queue sizes while Fig. 3.5(b) and Fig. 3.6(b) plot the actual queue sizes. For small $w$, these plots have a fixed point, which shows up as a straight line until some critical value of $w$ is reached, at which point the straight line splits into two. The emergence of two stable fixed points of period two is a consequence of a period doubling bifurcation. This is the first indication of oscillatory behavior appearing in the system due to the inherent nonlinearity of the interaction between RED mechanism and TCP, as opposed to a discontinuity in "queue or control law" which has been suggested in the past. This period two oscillation starts batching load at the router as shown in the plots.

Increasing $w$ further results in one of the period two fixed points colliding with the upper border of the map, giving a chaos type phenomenon. This is basically a bifurcation sequence expressed briefly as $1 \to 2 \to chaos$. This is a case of border collision bifurcation as shown in the analysis above.

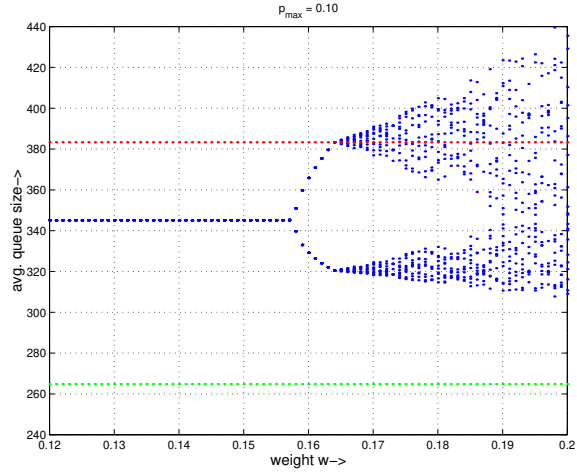

(a)

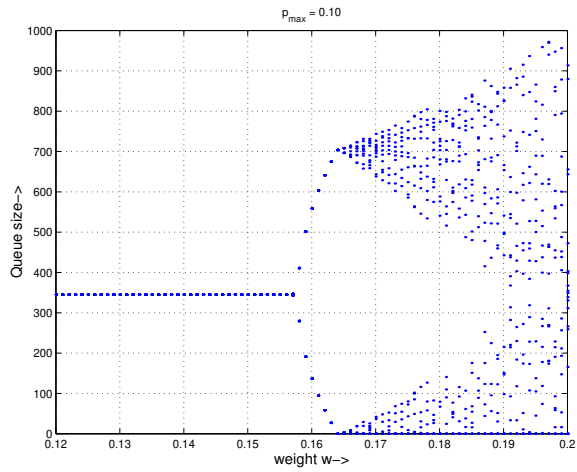

(b)

Figure 3.5: Bifurcation diagram of average and instantaneous queue length with respect to the averaging weight $w$ ($p_{max} = 0.1$).

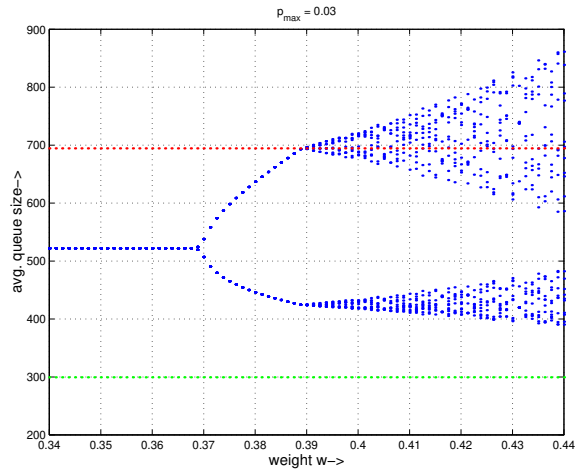

(a)

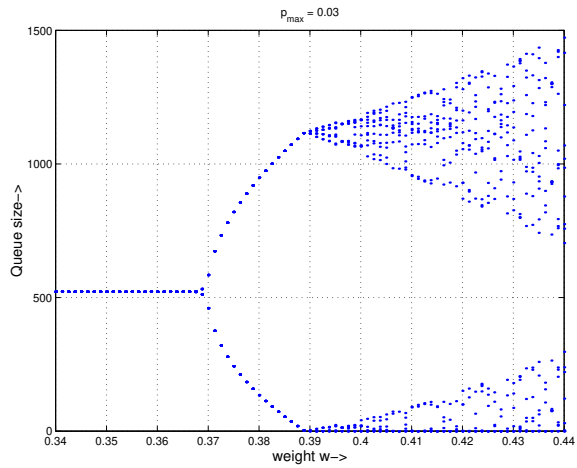

(b)

Figure 3.6: Bifurcation diagram of average and instantaneous queue length with respect to the averaging weight $w$ ($p_{max} = 0.03$).

It can be seen that when the bifurcation diagram for $q_k^{ave}$ collides with the border $b_1^n$, the queue empties, underutilizing the bottleneck link capacity. The implication of a relatively small oscillation in the average queue length is rather serious for the instantaneous queue length since the buffer starts getting empty and overly filled in every alternate cycle. This dynamical phenomenon is common to both plots in Figs. 3.5 and 3.6. We note that the distance between the initial period doubling bifurcation point and the border collision bifurcation point is short in both cases. This suggests that an effective way of controlling the instability may be to control the first period doubling bifurcation point, as demonstrated in [50, 73].

To illustrate the period doubling bifurcation in the system we compute the eigenvalue for the fixed point as $w$ is varied. It can be seen that it leaves unit circle along negative real line indicating a period doubling bifurcation. We also track the unstable fixed point and compute the corresponding eigenvalue to show that it indeed crosses the unit circle as shown in Table 3.1. We also notice that both stable and unstable fixed point ($q^{n*} = 0.092028$) is smaller than $q_1^n = 0.102222$ for the normalized model. Hence, it lies on the same side of the border even after smooth period doubling bifurcation.

To provide evidence to our claim for a BCB, we further compute the eigenvalue of a period two orbit of the map numerically and show that indeed one sided eigenvalues obey the condition given in (3.29). This computation is done for the set of parameters corresponding to Fig. 3.5. We define $\lambda(i, j) = \lambda(q_i^n) * \lambda(q_j^n)$.

In Table 3.2, the first and second rows show the four consecutive states (the exponentially averaged queue size at the router) corresponding to the parameter $w$ just before the BCB but after PDB. We note that all the states stay on the same side of the border with eigenvalue corresponding to a period-two orbit being less than unity. This implies the existence of stable period-2 orbit.

Table 3.1: Eigenvalues computed for different values of parameter $w$ to illustrate PDB, $b_1^n = 0.102222$.

| $w$ | $q_k^n$ | $\lambda(q_k^n, w)$ | Legend |
|---|---|---|---|
| 0.1561 | 0.092028 | -0.978111 | Close to PDB |
| 0.1572 | 0.092028 | -0.992051 | Closer to PDB |
| 0.1583 | 0.092028 | -1.005990 | After PDB |
| 0.1594 | 0.092028 | -1.019929 | After PDB |

Table 3.2: Eigenvalues computed for different values of parameter $w$ to illustrate BCB, $b_1^n = 0.1022$.

| $w$ | $q_k^n$ | $q_{k-1}^n$ | $q_{k-2}^n$ | $q_{k-3}^n$ | $\lambda(k, k-1)$ | $\lambda(k-2, k-3)$ |
|---|---|---|---|---|---|---|
| 0.1620 | 0.1001 | 0.0864 | 0.1001 | 0.0864 | 0.7864 | 0.7864 |
| 0.1631 | 0.0858 | 0.1013 | 0.0858 | 0.1013 | 0.7294 | 0.7294 |
| 0.1642 | 0.1021 | 0.0854 | *0.1022* | 0.0854 | 0.6921 | -1.8152 |

The third row depicts the same data just after a border collision bifurcation from a fixed point to chaos for the second iterate. Comparing the states with the border ($b_1^n$) reveals that $q_{k-2}^n$ and $q_{k-3}^n$ lie on different sides of the border. The eigenvalues corresponding to these two points, i.e., $\lambda(k-2, k-3)$, is negative. This eigenvalue $\lambda(k-2, k-3)$ can be used to approximate $b$ in (3.28) in this case. Similarly the eigenvalue corresponding to $q_{k-1}^n$ and $q_k^n$, i.e., $\lambda(k, k-1)$, can be used to approximate $a$ in (3.27). Since $a$ lies between 0 and 1 as shown in Table 3.2, and $b$ is smaller than -1, these values satisfy the condition given by (3.29).

Note the eigenvalues change discontinuously as $w$ is varied. This supports our contention that there is a *border collision* bifurcation in the system through which the system may become chaotic. It also stresses the role played by a border. We also note that there is a possibility of other rich nonlinear instabilities with different periodicity based on different parameter settings.

We also plot the Lyapunov exponents for the bifurcation scenario in Fig. 3.5 where $p_{max} = 0.1$. This is useful since a positive Lyapunov exponent indicates the presence of chaotic behavior (page 110, [2]). Fig. 3.7 shows that for small $w$ the exponent is negative, which corresponds to the single stable fixed point. It slowly increases to zero near the period doubling bifurcation, and then becomes negative again due to a stable period two orbit. Finally, it jumps to a positive value when one of the period two fixed points collides with one of the borders.

### 3.6.2 Effect of lower threshold value

In this subsection we explore how the lower threshold value $q_{min}$ affects system stability and behavior. The set of parameters used for the numerical example in this section is as follows:

$$q_{max} = 750, w = 0.15, \ C = 75 \text{ Mbps}, \ K = \sqrt{3/2}$$
$$B = 3{,}750 \text{ packets}, \ d = 0.1 \text{ sec}, \ M = 4{,}000 \text{ bits}$$
$$N = 250, \ p_{max} = 0.1, \ q_{min} = \text{bifurcation parameter}$$

Fig. 3.8 shows the fixed points of the system with varying $q_{min}$. As is clear from the figure, similar nonlinear phenomena are exhibited in this case as well. The system becomes less stable as $q_{min}$ is increased while other parameters are held fixed. Furthermore, this plot also exhibits similar sensitivity of the system behavior to $q_{min}$ as in the previous subsection.

We plot the Lyapunov exponent corresponding to the bifurcation scenario in Fig. 3.8. The Lyapunov exponent shown in Fig. 3.9 also stays negative in the for small $q_{min}$ like the other one plotted in Fig. 3.7. In a similar fashion, it increases to zero when the system goes through a period doubling bifurcation and again decreases when the system has a stable period two trajectory. Finally, it jumps to a positive value after a border collision bifurcation and stays positive.

### 3.6.3 Effect of system parameters on stability

A network manager may have control over the selection of control parameters such as $w, q_{min}, q_{max}$ and $p_{max}$. However, other system parameters such as the number of connections $N$ and the round-trip propagation delay $R$ are out of the network manager's control. Hence, it is important to understand how these system parameters affect the system stability and queue behavior in relation to the control parameters in order to understand how these control parameters should be set in practice.

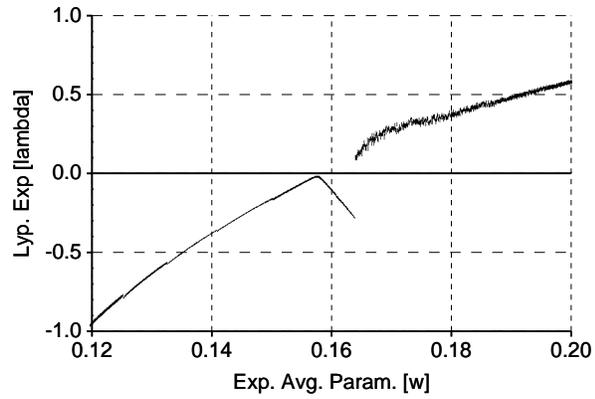

Figure 3.7: Lyapunov exponent computed for average queue length with respect to the averaging weight $w$ ($p_{max} = 0.1$).

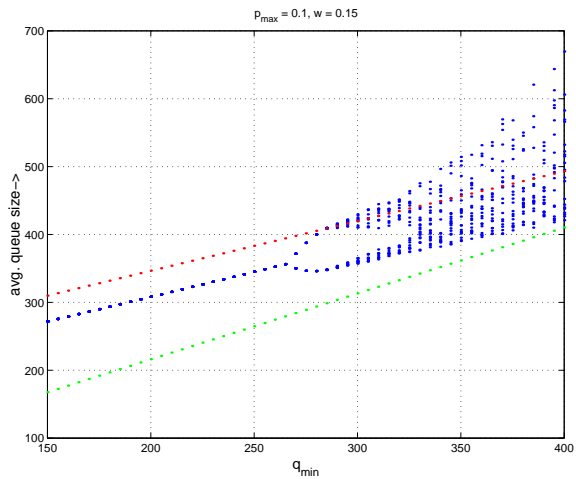

Figure 3.8: Bifurcation diagram of average and instantaneous queue length with respect to $q_{min}$ ($p_{max} = 0.1, w = 0.15$).

We first study the impact of the number of connections $N$ on the system behavior. The system parameters used for our example are as follows:

$$q_{max} = 750, \ q_{min} = 250, \ C = 75 \text{ Mbps}, \ K = \sqrt{3/2}$$

$$B = 3{,}750 \text{ packets}, \ R_0 = 0.1 \text{ sec}, \ M = 4{,}000 \text{ bits}$$

$$w = 0.15, \ p_{max} = 0.1, \ N = \text{bifurcation parameter}$$

The bifurcation diagram in Fig. 3.10 shows that the system stabilizes as the number of connections $N$ increases. In general, there is an agreement that a larger number of users tends to stabilize the system as stated in Lemma 2 [31, 56].

Another way of verifying that the system stability improves with $N$ is to compute the initial period doubling bifurcation point, i.e., the averaging weight $w^*$ at which the period doubling bifurcation occurs, as a function of $N$. This is shown in Fig. 3.11. The plot shows that the critical bifurcation parameter value $w^*(N)$ increases with the number of active TCP connections. Its implication for stability is that increasing the number of active TCP connections renders the queue length stable though at the expense of increased delay, since a larger value of $w$ is needed to induce the first period doubling bifurcation and consequently border collision bifurcation. This is in agreement with results in [31, 56], where under certain conditions a larger number of active TCP sessions stabilizes the system.

Similarly, we also plot a bifurcation diagram with respect to round-trip propagation delay $R_0$. Plot in Fig. 3.12 is in general agreement with the result in [31, 56] that larger delays cause instability. Parameters for this bifurcation diagram are as

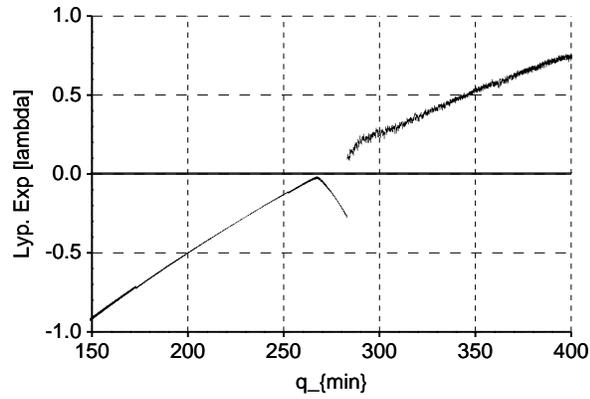

Figure 3.9: Lyapunov exponent computed for average queue length with respect to $q_{min}$

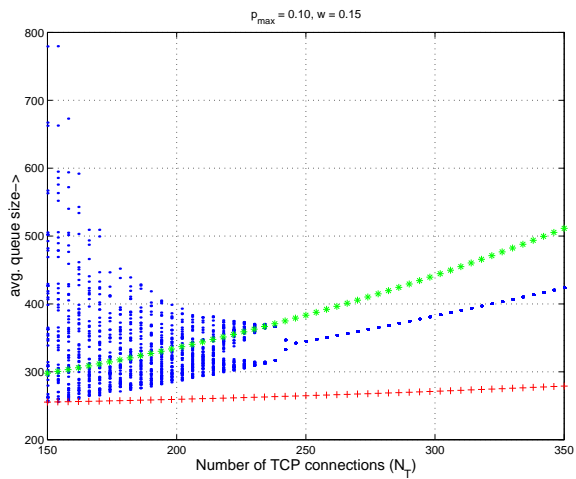

Figure 3.10: Bifurcation diagram of average queue length with respect to number of connections $N$ ($p_{max} = 0.1, w = 0.15$).

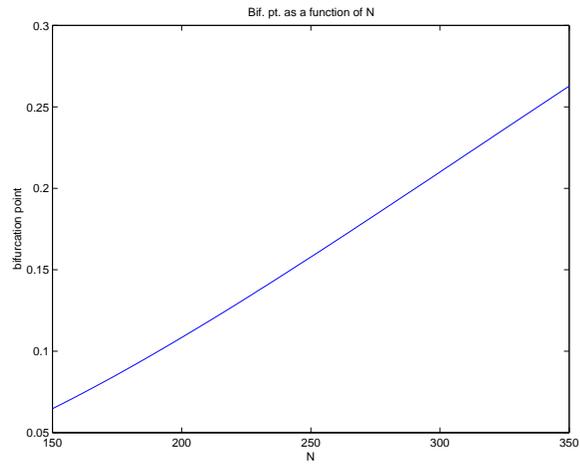

Figure 3.11: Initial PDB point as a function of number of active TCP-sessions.

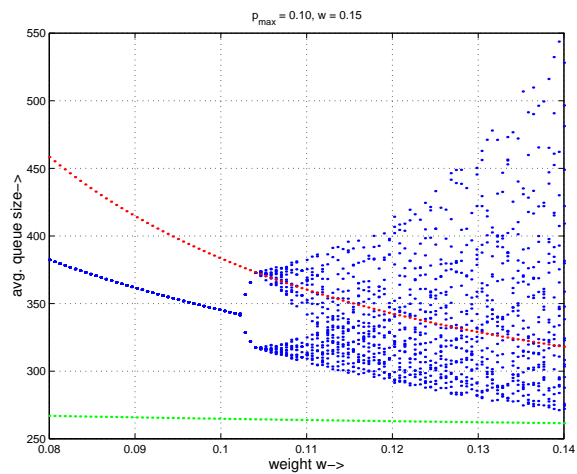

Figure 3.12: Bifurcation diagram of average and instantaneous queue length with respect to round-trip propagation delay ($R_0$)

follows:

$$q_{min} = 250, \ q_{max} = 750, \ C = 75 \text{ Mbps}, \ K = \sqrt{3/2}$$

$$B = 3{,}750 \text{ packets}, \ N = 250, \ M = 4{,}000 \text{ bits}$$

$$w = 0.15, \ p_{max} = 0.1, \ d = \text{bifurcation parameter}$$

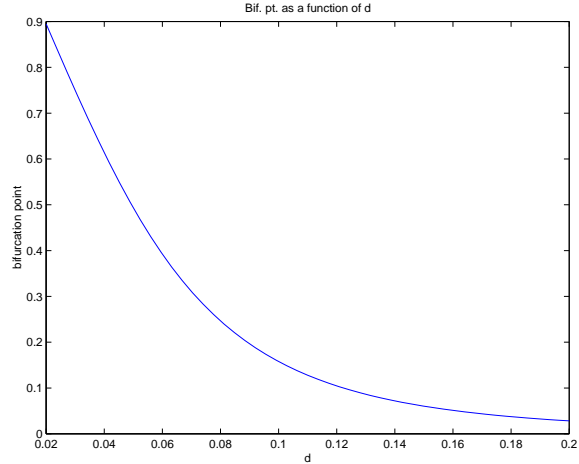

Figure 3.13: Initial PDB point as a function of round trip propagation delay.

The variation of the critical value of the bifurcation parameter $w^*(R_0)$ as a function of round-trip propagation delay $R_0$ is plotted in Fig. 3.13. It shows that the system is more stable for smaller values of round-trip propagation delay, as a larger averaging weight $w$ is needed to make it oscillate. This result again is in agreement with the general result of [31, 56] that smaller delays tend to keep the system stable. One should note that the system stability is rather sensitive to the variation in $R_0$.

## 3.7 Simulation Results

In this section we verify the existence of instabilities through the simulation results obtained using *ns*-2 simulator developed at UC Berkeley and Lawrence Berkeley Laboratory (LBL). We demonstrate the parametric sensitivity of the RED to different system parameters, such as the number of connections and round-trip delays. The simulated network topology is as shown in Fig. 3.1. TCP connections are Reno connections. The propagation delays of the edge links that connect the sources to node $r1$ or node $r2$ to the destinations are uniform random variables selected from [10 ms, 35 ms]. The capacity of these edge links are set to 30 Mbps. In congestion avoidance mode a TCP Reno connection increases its congestion window size by one during an RTT if there is no packet drop. When it detects a packet loss, it sets the window size to roughly half of what it was at the time of detection. Thus, although it is a very robust mechanism TCP leads to an oscillatory behavior due to this bandwidth estimation scheme in congestion avoidance mode. Hence, in practice, with a small number of TCP connections one would expect to see the average queue size fluctuate significantly. Such an oscillatory behavior will be mitigated with the increasing $N$ as demonstrated by Tinnakornsrisuphap and Makowski [84] and [51]. Thus, the oscillation in the average queue size per flow induced by this oscillatory behavior of TCP diminishes with the increasing number of flows. However, the oscillation induced by instability does not decrease with the number of flows as should be apparent from Section 3.3. Hence, we run the simulation with a reasonably large number of connections in order to reduce the fluctuation in the queue size due to the oscillatory behavior of TCP and set $N = 500$. The capacity and delay of the bottleneck link are set to $C = 150$ Mbps and 1.5 ms, respectively. Given these parameters the average round-trip propagation delay (without any queuing delay) is approximately 93 ms with the standard deviation of 20.4 ms. The packet size for TCP connections is 500 bytes, and the buffer size at node $r1$ and $r2$ is 7,500 packets. The threshold values $q_{min}$ and $q_{max}$ are set to 750 packets and 2,250 packets, respectively, and $p_{max}$ is set to 1/8. Simulation is run for 200 seconds.

### 3.7.1 Effects of exponential averaging weight

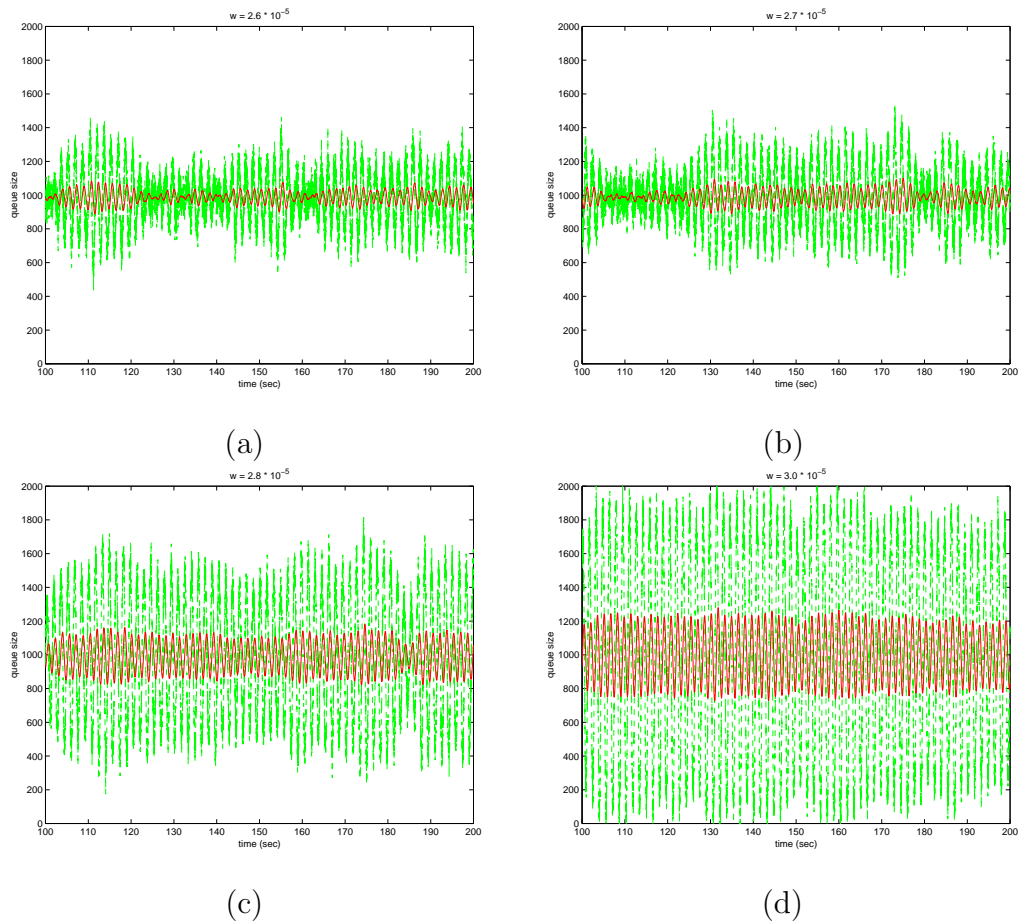

Figure 3.14: Average queue sizes with various $w$. (a) $w = 2.6 \cdot 10^{-5}$, (b) $w = 2.7 \cdot 10^{-5}$, (c) $w = 2.8 \cdot 10^{-5}$. (d) $w = 3.0 \cdot 10^{-5}$

In this section we vary the exponential averaging weight and study how it affects the queue behavior while other system and control parameters are fixed. Fig. 3.14 plots the evolution of the average queue size from $t = 100$ sec to $t = 200$ sec. In Fig. 3.14(a) and 3.14(b) the average queue size remains relatively stable around a single attractor or equilibrium, except for the fluctuations caused by the bandwidth estimation scheme of TCP. However, with increasing $w$ the average queue size begins to oscillate (Fig. 3.14(c)). Finally, when $w$ reaches some threshold value as predicted by our results, the average queue size shows no sign of any equilibrium point and constantly oscillates between the boundaries (Fig. 3.14(d)). This illustrates how one can expect very different behavior in queue sizes with small changes in the averaging weight $w$. One should note the exponential averaging weight is increased by only 15 percent from Fig. 3.14(a) to Fig. 3.14(d), clearly indicating a very sensitive nature of the RED to the parameter $w$.

### 3.7.2 Effects of number of TCP connections $N$

In this subsection we show how the number of TCP connections, $N$, affects the stability region through the simulation. We fix the exponential averaging weight at $w = 2.6 \cdot 10^{-5}$. Fig. 3.15 plots the queue evolution at $N = 450$, 475, and 500. As one can see the system becomes more stable with an increasing number of TCP connections in the simulation. This is consistent with our claim in Lemma 2 that the initial period doubling bifurcation point is an increasing function of $N$.

### 3.7.3 Effects of round-trip propagation delay $R_0$

This subsection demonstrates the sensitivity of the system stability to the round-trip delays of the connection. We have fixed the averaging weight at $2.6 \cdot 10^{-5}$, and increased the mean round-trip propagation delay of the connection by increasing the one-way propagation delay of the bottleneck link by 5 ms. Fig. 3.16 plots the queue evolution. Compared to Fig. 3.14(a), this plot shows much more pronounced oscillatory behavior although the delay is increased only by 10 ms.

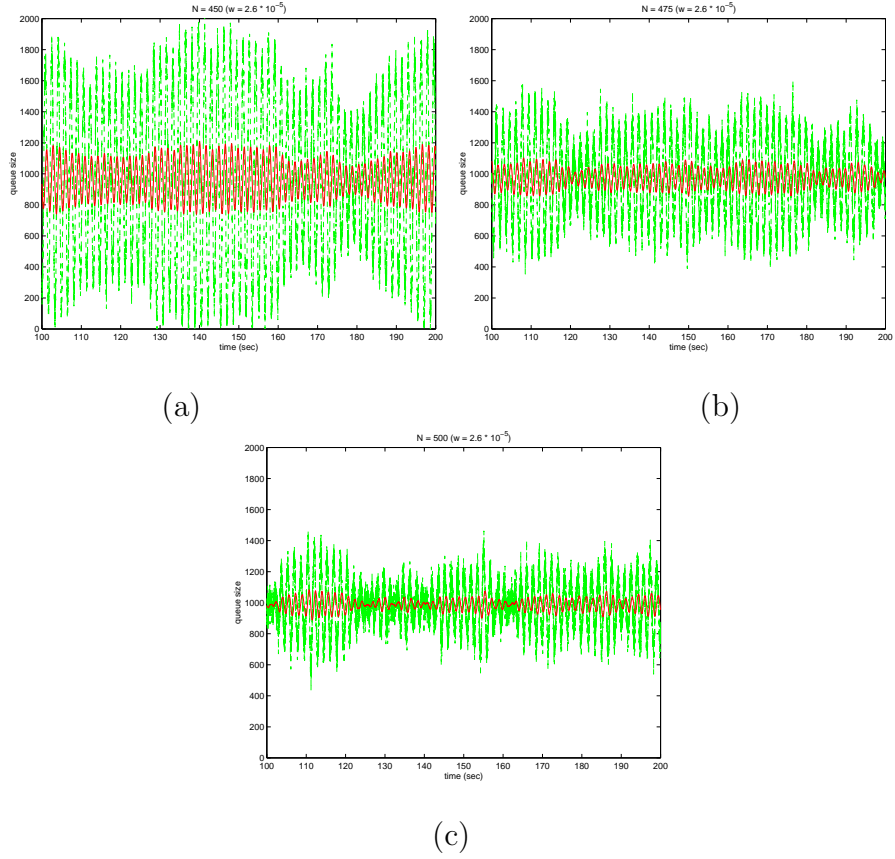

Figure 3.15: Queue evolution with different number of TCP connections ($w = 2.6 \cdot 10^{-5}$). (a) $N = 450$, (b) $N = 475$, (c) $N = 500$.

### 3.7.4 Effects of lower threshold value $q_{min}$

In this section we increase the lower threshold $q_{min}$ of the RED mechanism from 750 packets to 900 packets and plot the queue behavior in Fig. 3.17. The exponential averaging weight is set to $w = 2.6 \cdot 10^{-5}$. Fig. 3.17 shows much more unstable behavior compared to Fig. 3.14(a). This is consistent with the numerical results

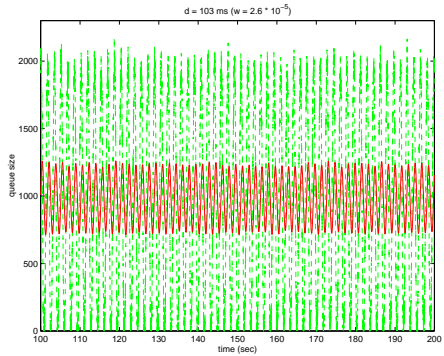

Figure 3.16: Queue evolution with $R_0 = 103$ ms ($w = 2.6 \cdot 10^{-5}$).

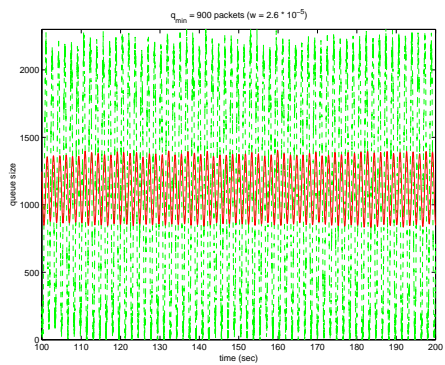

Figure 3.17: Queue evolution with $q_{min} = 900$ packets ($w = 2.6 \cdot 10^{-5}$).

plotted in Fig. 3.12 and 3.13 in section 3.6.3.

### 3.7.5 Short-lived flows

In this subsection we introduce short-lived flows and investigate how their presence affects system behavior. Here we reduce the number of long-lived TCP connections to 400 and introduce short-lived connections that arrive according to a Poisson process with $\lambda = 50$ connections/second. The duration of these short-lived connections is exponentially distributed with the mean of 2.0 seconds. This gives an average number of active connections of 500, as in the long-lived connection case, and approximately 80 percent of the traffic is generated by the long-lived connections.

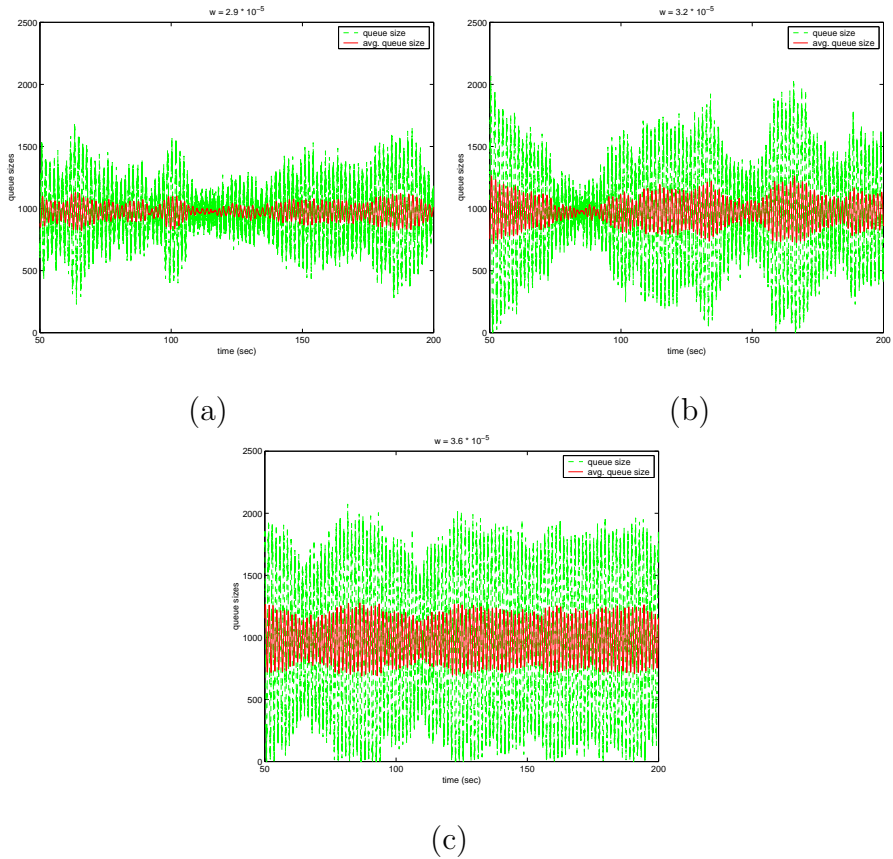

Figure 3.18: Queue evolution with short-lived and long-lived connections. (a) $w = 2.9 \cdot 10^{-5}$, (b) $w = 3.2 \cdot 10^{-5}$, (c) $w = 3.6 \cdot 10^{-5}$

As shown in Fig. 3.18 the queue behavior changes only slightly as the load on the network changes. However, even with short-lived connections the qualitative behavior of the queue size changes only marginally, and for sufficiently large averaging weight the queue size exhibits similar unstable behavior as in Fig. 3.14. This is consistent with the results presented in [51].

## 3.8 Modeling Using a Generic Throughput Function

In this section we extend the modeling formalism proposed in [71] to a whole general class of TCP-type throughput functions and their mix with UDP. This kind of models has only been possible due to a good understanding of the throughput functions $T(p, R)$ for different type of transport protocols. The basic idea behind this generalization is to first prove the existence of inverse functions of the throughput function individually in each of the variables $p$ (drop probability) and $R$ (return trip time). This is done for mixed traffic (TCP+UDP), and also for TCP-type traffic based on new congestion control algorithms proposed for multimedia [7]. The basic model, assuming RED-type active queue management, is the discrete-time nonlinear map [24]

$$\begin{aligned}\overline{q}_{e,k+1} &= \begin{cases} (1-w)\overline{q}_{e,k} & \text{if } \overline{q}_{e,k} > b_1 \\ (1-w)\overline{q}_{e,k} + wB & \text{if } \overline{q}_{e,k} < b_2 \\ (1-w)\overline{q}_{e,k}+ \\ \quad +\frac{wC}{M}(T_R^{-1}(p,\frac{C}{N}) - R_0) & \text{otherwise} \end{cases} \\ &:= f(\overline{q}_{e,k}, \rho) \end{aligned} \quad (3.30)$$

where $\rho$ represents the parameter vector of the system, the parameters being

$$\begin{aligned} N &= \text{Number of active TCP connections} \\ M &= \text{Maximum segment size or packet size} \\ R_0 &= \text{Round trip propagation delay} \end{aligned}$$

$$
\begin{aligned}
K &= \text{Modeling constant which varies between} \\
  &\quad \text{1 and } \sqrt{8/3} \text{ [58]} \\
w &= \text{Exponential averaging weight} \\
C &= \text{Bottleneck bandwidth} \\
T_R^{-1}(p, \tfrac{C}{N}) &= \text{Inverse of the TCP throughput function in } R. \\
B &= \text{buffer size}
\end{aligned}
$$

The parameters $q_{min}$, $q_{max}$ and $p_{max}$ are RED controller settings [24], and $b_1$ and $b_2$ are system borders to be defined later. Exponentially average queue occupancy is denoted by $\bar{q}_{e,k}$. Denote by $p_0$ the value of drop probability at which the link regime changes from under-utilized to fully-utilized; $p_0$ is given by the solution of following equation:

$$T(p_0, R_0) = \frac{C}{N}$$

The value $p_0$ is used to compute the border $b_1$, such that for any $\bar{q}_{e,k} \geq b_1$, the instantaneous buffer occupancy in the next return trip time remains 0 under a monotone AQM law. The border $b_1$ can be computed explicitly if $p_0$ and the RED-type control law $H(\bar{q}_{e,k})$ are known:

$$b_1 = \begin{cases} H^{-1}(p_0), & \text{if } p_{max} \geq p_0 \\ q_{max}, & \text{otherwise} \end{cases} \tag{3.31}$$

Similarly, denote by $p_1$ the maximum drop probability for which the buffer will be full in the next return trip time. Then the boundary value $b_2$ of $\bar{q}_{e,k}$ such that $\forall\, \bar{q}_{e,k} \leq b_2$, the buffer is full in the next return trip time can be expressed as:

$$b_2 = H^{-1}(p_1) \tag{3.32}$$

Let $R$ denote the return trip time which includes both propagation and queuing delay. As illustrated in [21], queue size $\bar{q}(p)$ for a given $p$ can be derived as a solution to the following equations:

$$T(p, R) = \frac{C}{N} \tag{3.33}$$

$$R = R_0 + \frac{\bar{q}(p)}{c} \tag{3.34}$$

The solution is given by $\bar{q}(p) = \frac{C}{M}(T_R^{-1}(p, \frac{C}{n}) - R_0)$, where $T_R^{-1}(p, \frac{C}{n})$ denotes the inverse of the TCP throughput function $T(\cdot, \cdot)$ in $R$. The Inverse Function Theorem is used to address the existence of such an inverse.

To proceed, we recall the *inverse function theorem* which is given as theorem 5 in chapter 2.

Inverse function theorem says that For a general function $T(\cdot, \cdot)$ the inverse in $R$ will exist if $\frac{\partial T(\cdot, \cdot)}{\partial R}$ does not vanish in the parameter region of interest. A closed form solution for the inverse may not be feasible due to a complex algebraic form of the TCP throughput function, for example the one given in [21]. However, it can be computed numerically. Computation of $T_R^{-1}(p, \frac{C}{N})$ is needed to compute the "plant function" $G(p)$ which plays a crucial role in the discrete time nonlinear model. The function $G(p)$ gives the next instantaneous queue occupancy in terms of the drop probability, and is expressed as follows:

$$G(p) = \begin{cases} \min(B, \frac{C}{M}(T_R^{-1}(p, \frac{C}{N}) - R_0)) & : \quad p \leq p_0 \\ 0 & : \quad otherwise \end{cases} \tag{3.35}$$

where

$$R_0 = \text{round-trip propagation and transmission time and}$$

$$p_0 = T_p^{-1}(C/N, R_0)$$

Finally, $p_1$ can be computed as the solution of the equation $G(p_1) = B$. This solution is guaranteed to exist under certain technical assumptions.

Hence, the solution of eq. 3.33, which is written as $p_0 = T_p^{-1}(\frac{C}{N}, R_0)$, will exist if $\frac{\partial T(\cdot,\cdot)}{\partial p}$ does not vanish in the parameter region of interest. Again, for a general TCP throughput function a closed form solution for $p_0$ may not be feasible, but it can be computed numerically if its Jacobian is invertible.

Assumption 1 below ensures applicability of the inverse function theorem guaranteeing existence of inverses in both $p$ and $R$. Assumption 2 is needed to reflect in the dynamic models the greedy behavior of TCP-type algorithms.

*Assumption 1: Let the given TCP throughput function $T(p, R)$ be smooth both in $p$ and $R$, and let its partial derivatives with respect to $p$ and $R$ be strictly negative ($\frac{\partial T(\cdot,\cdot)}{\partial R} < 0, \frac{\partial T(\cdot,\cdot)}{\partial p} < 0$) uniformly in the region of interest.*

*Assumption 2: The throughput function $T(p, R) \to \infty$ as $p \to 0$.*

These assumptions are very mild, and in general TCP throughput decreases as drop probability $p$ and return trip time $R$ increase. The second assumption is also generally satisfied due to the greedy nature of TCP where it tries to utilize the maximum amount of bandwidth available in low drop probability regimes. Although we have assumed smooth functions, similar results can be shown to hold for continuous but piecewise differentiable throughput functions as long as the negative derivative assumption holds.

**Lemma 6** *Under Assumptions 1 and 2, $\overline{q}(p)$, $p_0$ and $p_1$ exist and depend smoothly on data for general throughput function $T(p, R)$.*

**Proof:** Proof follows directly from inverse function theorem given in chapter 2 as theorem 5. ∎

*Example 0:* The assumptions above can be easily verified in the parameter regions of interest $(p > 0, T > 0)$ for the simplest TCP throughput function, given by [58, 30]:

$$T(p, R) = \frac{M}{R}\frac{K}{\sqrt{p}} \qquad (3.36)$$

where

$$\begin{aligned} T &= \text{Throughput of a TCP flow (in bits/sec)} \\ K &= \text{constant which varies between 1 and } \sqrt{8/3} \text{ [58]} \\ p &= \text{probability of packet loss} \end{aligned}$$

*Example 1:* The assumptions can again easily be verified in parameter regions of interest $(p > 0, T > 0)$ for the simplest TCP throughput function given above in the presence of some UDP traffic:

$$T(p, R) = \frac{\frac{NM}{R}\frac{K}{\sqrt{p}} + \lambda c(1-p)}{N} \qquad (3.37)$$

where $\lambda$ is the fraction of capacity as UDP traffic.

*Example 2:* The general class of TCP-type congestion control algorithms known as *Binomial Congestion Control algorithms* [7] can also be considered using the same framework. These algorithms have been developed to design TCP-friendly multimedia streams that cannot sustain window halving during congestion as prescribed in the multiplicative decrease phase of TCP. A theoretical derivation and empirical validation of throughput function for this class of algorithms can be found in [7]. Indeed, the empirical validation has been performed in the presence of RED gateways, which makes this a good candidate for analysis using the basic model (3.30). Briefly, the binomial congestion control algorithms can be described

as follows:

$$I : w_{t+R} \leftarrow w_t + \frac{\alpha}{w_t^k} \; ; \alpha > 0$$
$$D : w_{t+\delta t} \leftarrow w_t - \beta w_t^l \; ; 0 < \beta < 1$$

where $I$ denotes the increase in the window as a result of the receipt of one window of acknowledgements in a round-trip time (RTT) and $D$ denotes the decrease in window size on detection of congestion by the sender, $w_t$ denotes the window size at time t, $R$ is the $RTT$ of the flow, and $\alpha$ and $\beta$ are constants. Parameters $l$ and $k$ describe the functional dependence of increase and decrease respectively, on the current window size $w_t$. Binomial congestion control algorithms are characterized by $l + k = 1$. These algorithms are shown to possess the following approximate throughput profile [7]:

$$T(p, R) = \frac{M}{R} \frac{K}{p^{\frac{1}{k+l+1}}} \tag{3.38}$$

where

$$K = \frac{\alpha}{\beta}^{\frac{1}{k+l+1}}$$

Differentiating the throughput function with respect to $R$ and $p$, it can be easily shown that these derivatives are uniformly negative for the parameter region of interest.

$$\frac{\partial T(p, R)}{\partial R} = -\frac{M}{R^2} \frac{K}{p^{\frac{1}{k+l+1}}}$$
$$\frac{dT(p, R)}{dp} = -(k+l+1) \frac{M}{R} \frac{K}{p^{\frac{k+l+2}{k+l+1}}}$$

This class of congestion control algorithms are important due to their potential use for multimedia streams applications and TCP friendliness. This is to stress that nonlinear stability analysis framework proposed in [71] remains valid even for quickly growing presence of TCP-like non-TCP traffic as long as they react to the packets drops and have throughput functions similar to that of TCP. It can also be shown that these assumptions will remain valid under the limited presence of UDP traffic along with Binomial traffic.

*Example 3:* The assumptions for existence of an inverse can again be verified in parameter regions of interest $(p > 0, T > 0)$ for a general class of TCP and TCP-type binomial congestion control algorithms with timeouts considered in an average sense. Next, we analyze a more detailed TCP-throughput function [58, 7, 89] than in the examples above.

$$T(p,R) = \frac{M}{R\sqrt{\frac{\beta p}{\alpha}} + T_0 \min\left(1, 3\sqrt{\frac{\beta p}{\alpha}}\right)p(1+32p^2)} \qquad (3.39)$$

where

$$T_0 = \text{timeout period}$$

It is clear the average effect of timeout is to decrease the throughput. If we again consider this average effect of timeout opposed to the instantaneous effect, we can show that our modeling framework holds. First, analyzing the function $\min\left(1, 3\sqrt{\frac{\beta p}{\alpha}}\right)$ tells us that for drop probability $p < \frac{\alpha}{9\beta} = 0.222$ for $\alpha = 1$ and $\beta = 0.5$ (typical TCP implementations) which is reasonably high drop rate given that generally $p_{max}$ in RED approximately 0.1. Hence, throughput function can be reduced to:

$$T(p,R) = \frac{M}{R\sqrt{\frac{\beta p}{\alpha}} + 3T_0\sqrt{\frac{\beta p}{\alpha}}p(1+32p^2)} \qquad (3.40)$$

Again, by looking at the derivative of this function with respect to drop probability $p$ and return trip time $R$, it can be shown that both of these derivatives are negative in the parameter region of interest.

$$\frac{\partial T(p,R)}{\partial R} = -\frac{M(R\sqrt{\frac{\beta p}{\alpha}} + 3T_0\sqrt{\frac{\beta p}{\alpha}}p(1+32p^2))}{(R\sqrt{\frac{\beta p}{\alpha}} + 3T_0\sqrt{\frac{\beta p}{\alpha}}p(1+32p^2))^2}$$

$$\frac{\partial T(p,R)}{\partial p} = -\frac{M(0.5R\sqrt{\frac{\beta}{\alpha p}} + 3T_0(1.5\sqrt{\frac{\beta p}{\alpha}} + 112\sqrt{\frac{\beta p}{\alpha}}p^2))}{(R\sqrt{\frac{\beta p}{\alpha}} + 3T_0\sqrt{\frac{\beta p}{\alpha}}p(1+32p^2))^2}$$

Finally, even if the $p < \frac{\alpha}{9\beta}$ condition is not valid uniformly, we can show that throughput function is a piecewise smooth function and the inverse exists for separate segments. We have already shown that for small $p$ segment it satisfies the required properties to compute an inverse. For $p$ sufficiently large, throughput and corresponding derivatives can be written as:

$$T(p,R) = \frac{M}{R\sqrt{\frac{\beta p}{\alpha}} + T_0 p(1+32p^2)}$$

$$\frac{\partial T(p,R)}{\partial R} = -\frac{M(\sqrt{\frac{\beta p}{\alpha}} + T_0 p(1+32p^2))}{(R\sqrt{\frac{\beta p}{\alpha}} + T_0 p(1+32p^2))^2}$$

$$\frac{\partial T(p,R)}{\partial p} = -\frac{M(0.5R\sqrt{\frac{\beta}{\alpha p}} + T_0(1+96p^2))}{(R\sqrt{\frac{\beta p}{\alpha}} + T_0 p(1+32p^2))^2}$$

## 3.9 Deterministic First Order Model for TCP-RED

We start with discussing our deterministic first order model for TCP-RED presented in [71].

### 3.9.1 Approximate TCP throughput model

First model can be described by eq. 3.41 for a special case of TCP only traffic with its throughput function given by eq. 3.1 and $H(\overline{q}_{e,k})$ is classical RED [21].

$$\overline{q}_{e,k+1} = \begin{cases} (1-w)\overline{q}_{e,k} & \text{if } \overline{q}_{e,k} > b_1 \\ (1-w)\overline{q}_{e,k} + wB & \text{if } \overline{q}_{e,k} < b_2 \\ (1-w)\overline{q}_{e,k} + \\ +w\left(\dfrac{NK}{\sqrt{\dfrac{(\overline{q}_{e,k}-q_{min})p_{max}}{(q_{max}-q_{min})}}} - \dfrac{R_0 C}{M}\right) & \text{otherwise} \end{cases} \quad (3.41)$$

where

$$w = \text{Exp. averaging weight}$$

Moreover, $q_{min}$, $q_{max}$ and $p_{max}$ are RED controller settings [24]. Borders $b_1$ and $b_2$ in eq. (3.41) are given by eq. (3.42) and eq. (3.43) respectively.

$$b_1 = \begin{cases} \dfrac{p_0(q_{max}-q_{min})}{p_{max}} + q_{min}, & \text{if } p_{max} \geq p_0 \\ q_{max} & , \text{ otherwise} \end{cases} \quad (3.42)$$

$$b_2 = \dfrac{\left(\dfrac{NK}{B+\dfrac{R_0 C}{M}}\right)^2}{p_{max}}(q_{max}-q_{min}) + q_{min} \quad (3.43)$$

We remark that solving eq. 3.41 leads to a third degree polynomial in fixed point $\overline{q}_e^*$ which interestingly does not depend on $w$ as should be expected since, both the "queue law" and the "feedback control law" are not functions of $w$. The polynomial is given below.

$$(\overline{q}_e^* - q_{min})(\overline{q}_e^* + \dfrac{R_0 C}{M})^2 = \dfrac{(NK)^2}{p_{max}}(q_{max}-q_{min}) \quad (3.44)$$

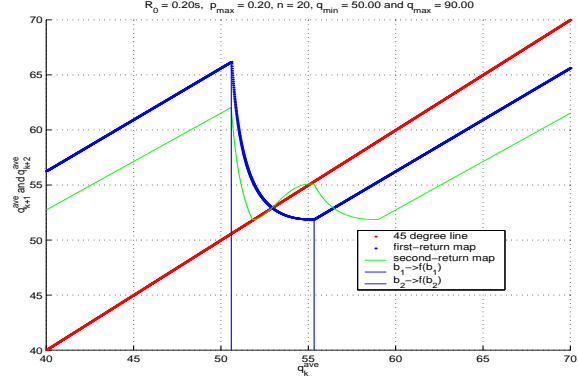

Figure 3.19: First and second return maps

### 3.9.2 Detailed TCP throughput model

In this section, to compare the validity of this modeling formalism we are going to use the detailed TCP throughput given by eq. (3.39) in example 3. Here, we notice the effect of timeout ($T_0 = 5R$ [21]) in the throughput function is to decrease it which will be have some interesting repercussions in bifurcation behavior. Using the formalism outlined in eq. (3.30) and classical RED [24], model can be described as:

$$\overline{q}_{e,k+1} = \begin{cases} (1-w)\overline{q}_{e,k} & \text{if } \overline{q}_{e,k} > b_1 \\ (1-w)\overline{q}_{e,k} + wB & \text{if } \overline{q}_{e,k} < b_2 \\ (1-w)\overline{q}_{e,k} + \\ \frac{wc}{M}\left(\frac{MN - cT_0 \min(1, 3\sqrt{\frac{3bp}{8}})p(1+32p^2)}{c\sqrt{\frac{2bp}{3}}} - R_0\right) & \text{otherwise} \end{cases} \quad (3.45)$$

where

$$p = p_{max}\frac{q_{e,k} - q_{min}}{q_{max} - q_{min}}$$

$b$ = average number of packets acknowledged by an ack (usually 2 [21])

$$T_0 = \text{timeout (which is typically } 5R_0 \text{ [21])}$$

To compute border $b_1$ in eq. (3.45), $p_0$ needs to be computed numerically as a solution of following equation:

$$\frac{M}{R_0\sqrt{\frac{2bp}{3}} + T_0 \min(1, 3\sqrt{\frac{3bp}{8}})p(1+32p^2)} = \frac{C}{N} \tag{3.46}$$

We use bisection method to compute this value in numerical simulation. Once $p_0$ is known, then border $b_1$ can be computed from the RED law.

$$b_1 = \begin{cases} \frac{p_0(q_{max}-q_{min})}{p_{max}} + q_{min}, & \text{if } p_{max} \geq p_0 \\ q_{max} & , \text{ otherwise} \end{cases} \tag{3.47}$$

Finally, $p_1$ is given as a solution to the following equation and bisection method can be again be used for numerical simulations:

$$\frac{C}{M}\left(\frac{MN - cT_0 \min(1, 3\sqrt{\frac{3bp}{8}})p(1+32p^2)}{c\sqrt{\frac{2bp}{3}}} - R_0\right) = B \tag{3.48}$$

This enables us to compute the second border $b_2$ from RED law.

$$b_2 = \frac{p_1}{p_{max}}(q_{max} - q_{min}) + q_{min} \tag{3.49}$$

## 3.10 Bifurcation Diagrams

In this section, bifurcation diagrams are given showing the qualitative changes in system behavior as parameters (e.g., $w, q_{min}$) vary. Using numerical simulation we show that both approximate and detailed models exhibit qualitatively similar behavior although critical parameter in both cases can be different.

### 3.10.1 Effect of exponential averaging weight $w$

Here we present the numerical simulation results for both approximate and detailed models. Parameters are chosen to reflect a realistic situation. Bifurcation parameters are chosen based on the fact that there is not much agreement still on how to set $w$ and $q_{min}$ in RED community. The following parameters are common to the next three bifurcation plots [21] for different values of $p_{max}$:

$$q_{max}=747, \quad q_{min}=249, \quad c=74.7 Mbps, k=\sqrt{3/2}$$
$$B=3735 \text{ packets}, \quad R_0=0.1 sec, \quad M=4kb$$
$$b=1 \quad T_0=5R_0 \quad N=249, \quad w=\text{bifurcation parameter}$$

Case 1. $p_{max} = 0.1$

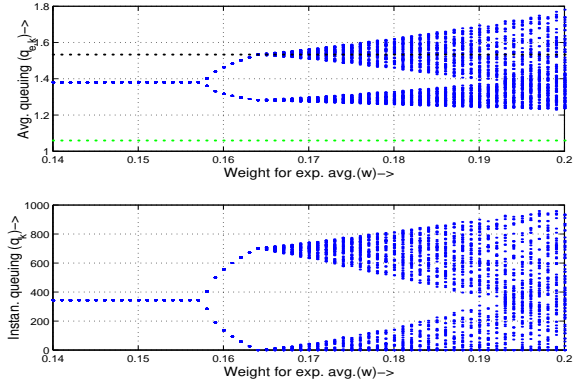

Figure 3.20: Bifurcation diagram of average and instantaneous queue length with respect to $w$, $p_{max} = 0.1$ for approximate model described by eq. (3.41)

The bifurcation diagrams in figs. (3.20, 3.22) show the effect of varying the exponential weight $w$ for different values of $p_{max}$ for approximate model and fig. (3.21) shows the same for detailed model given by eq. (3.45). For small $w$, these plots have a fixed point which looks like a straight line but after some critical value of $w$ this straight line splits into two and the map exhibits period-doubling bifurcation. This is the first indication of oscillatory behavior appearing in the system due to its inherent nonlinearity, as opposed to the discontinuities in "queue or control law" which have been proposed earlier. This period two oscillation reflects as a load batching at the router as shown in the plots. Increasing $w$ further shows that there are more period doubling bifurcations and finally, one of the branches collides with the upper border of the map giving birth to a chaos type phenomenon. This is basically a bifurcation sequence leading to chaos including non-smooth *border collision* bifurcation [6]. The border collision bifurcation is a well understood phenomenon in piecewise linear systems and has been shown responsible for chaos in several electric circuit and economic system models.

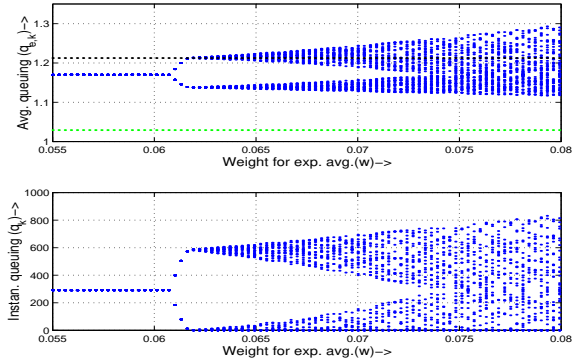

Figure 3.21: Bifurcation diagram of average and instantaneous queue length with respect to $w$, $p_{max} = 0.1$ for detailed model described by eq. (3.45)

Case 2. $p_{max} = 0.3$

### 3.10.2 Effect of RED parameter $q_{min}$

We do a numerical simulation of both approximate and detailed model using following parametric scenario:

$$q_{max} = 747, \quad w = 2^{-5}, \quad k = \sqrt{3/2}$$
$$R_0 = 0.1 \quad B = 3735 \text{ packets}, \quad M = 4kb \quad N = 249,$$

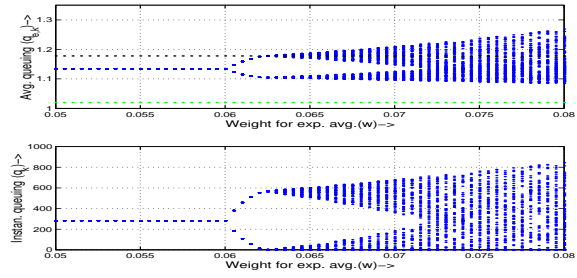

Figure 3.22: Bifurcation diagram of average and instantaneous queue length with respect to $w$, $p_{max} = 0.3$

$$b=1 \ \ T_0=5R_0 \ \ c=74.7 Mbps, \ \ q_{min}=\text{bifurcation parameter}$$

Bifurcation diagrams with respect to $q_{min}$ in fig.(3.23) also exhibits the bifurcation route to chaos for approximate model and fig. (3.24) shows the same for detailed model. TCP-RED exhibits this kind of dynamical variation with respect to many parameters like the number of active connections $n$, the round trip propagation delay $R_0$, and so on [71].

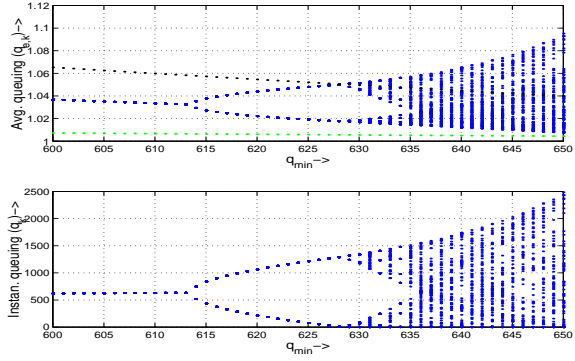

Figure 3.23: Bifurcation diagram of average and instantaneous queue length with respect to $q_{min}$, $p_{max} = 0.1$

In the foregoing section it was mentioned that average effect of timeout in detailed throughput function given by eq. (3.39) is to decrease the throughput of connections. This decrease reflects in bifurcation diagrams as a preponing of period doubling and border collision bifurcation. Comparing the bifurcation diagrams in fig. (3.23 and in fig. (3.24) confirms this. Similarly when exponential averaging weight $w$ is varied bifurcations are again preponed as evident from fig. (3.20) and fig. (3.21). It seems like any parameter change which decreases the throughput may lead to instability. This can be observed when $R$ or $p_{max}$ increases system approaches instability [71]. Similarly, decrease in the number of connection which decreases the total throughput also leads to instability.

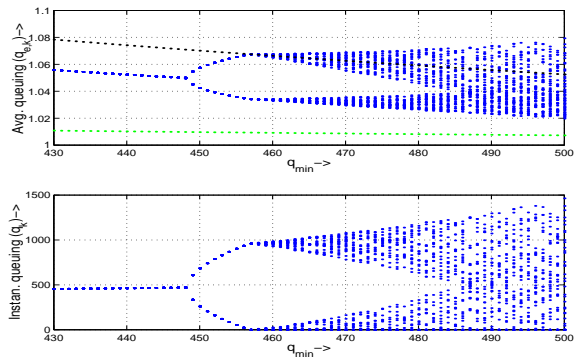

Figure 3.24: Bifurcation diagram of average and instantaneous queue length with respect to $q_{min}$, $p_{max} = 0.3$ and $R_0 = 0.1$ for detailed model given by eq. (3.45)

# Chapter 4

# Period Three Implies Chaos in TCP-RED

In this chapter, we continue the study beyond bifurcations and look for an analytic proof of chaos. In chapter 3, the emphasis was on modeling and numerical and simulation studies. We also developed the analysis further using bifurcation theory of both smooth and nonsmooth maps in chapter 2. Here, we appeal to a well known theorem to give a proof of the existence of chaos in the model studied numerically and through ns-2 simulation in the papers cited above. The occurrence of nonlinear instabilities in TCP-RED should not be surprising, considering that the protocols employ nonlinear rules and that the throughput function has a nonlinear dependence on drop probability during congestion.

The chapter proceeds as follows. In Section 4.1, we recall the first-order model for TCP-RED in a congested network from chapter 3. In Section 4.2, we recall the well known "period three implies chaos" result of Li and Yorke [53], and employ it to obtain sufficient conditions for the presence of chaotic behavior in the model. This theorem is a special case of a general theorem of Sarkovskii [16]. The results are illustrated numerically in Section 4.3.

## 4.1 First Order Model for TCP-RED

We start by discussing the deterministic first order model for TCP-RED presented in chapter 3. The model can be described as follows:

$$\overline{q}_{e,k+1} = \begin{cases} (1-w)\overline{q}_{e,k} & \text{if } \overline{q}_{e,k} > b_1 \\ (1-w)\overline{q}_{e,k} + wB & \text{if } \overline{q}_{e,k} < b_2 \\ (1-w)\overline{q}_{e,k} + \\ \quad +w\left(\dfrac{NK}{\sqrt{\dfrac{(\overline{q}_{e,k}-q_{min})}{(q_{max}-q_{min})/p_{max}}}} - \dfrac{R_0 C}{M}\right) & \text{otherwise} \end{cases}$$
$$:= f(\overline{q}_{e,k}, \rho) \tag{4.1}$$

where $\rho$ represents a vector of system parameters. The TCP parameters are

$$N = \text{Number of active TCP connections}$$
$$K = \text{Modeling constant between 1 and } \sqrt{8/3}$$
$$w = \text{Exp. averaging weight}, \ M = \text{Packet size}$$
$$R_0 = \text{Round trip time}, \ C = \text{Bottleneck bandwidth}$$

The parameters $q_{min}$, $q_{max}$ and $p_{max}$ are RED controller settings [24], and the borders $b_1$ and $b_2$ in eq. 4.1 are given by eq. 4.2 and eq. 4.3 respectively:

$$b_1 = \begin{cases} \dfrac{p_0(q_{max}-q_{min})}{p_{max}} + q_{min}, & \text{if } p_{max} \geq p_0 \\ q_{max} & , \text{ otherwise} \end{cases} \tag{4.2}$$

$$b_2 = \dfrac{\left(\dfrac{NK}{B+\frac{R_0 C}{M}}\right)^2}{p_{max}}(q_{max}-q_{min}) + q_{min} \tag{4.3}$$

Fig. 1 depicts the map $f$ in eq. (4.1).

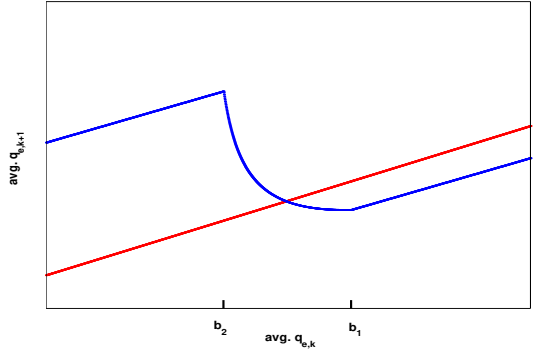

Figure 4.1: The map $f$ of Eq. (4.1) and its intersection with $45^o$ degree line shows the fixed point

## 4.2 Period Three Orbit and Chaos

We begin this section by recalling the following assumption, and proceed to the analysis of the chaotic region of the TCP-RED model.

*Assumption 1:* $p_{max} > p_0$

Here, $p_{max}$ is a RED parameter and $p_0$ is the maximum probability for which the system is fully utilized. That is, for $p \geq p_0$ senders will have their rates too small to keep the link fully utilized; $p_0$ is given by

$$p_0 = \left(\frac{MK}{R_0 \frac{C}{N}}\right)^2 \tag{4.4}$$

This is the parameter region of interest from the point of view of the application. The basic idea behind the application of the period three implies chaos theorem is the existence different monotone regions in the map given by eq.( 4.1). We have proved earlier [70] that the map eq. 4.1 is strictly increasing for $0 \leq \bar{q}_{e,k} \leq b_2$ and for $b_1 \leq \bar{q}_{e,k} \leq B$ but it can be strictly decreasing in the segment where $b_2 \leq \bar{q}_{e,k} \leq b_1$ under certain conditions. First, we recall the theorem from , a special case of the main result of [53] which is a special case of general result proved by Sharkovsky [77]:

**Theorem 8** : *Let  be an interval and let $F : \rightarrow$  be continuous. Assume that there is a point $a \in$  for which the points $b = F(a), c = F^2(a)$ and $d = F^3(a)$, satisfy*

$$d \leq a < b < c \text{ or } d \geq a > b > c$$

(4.5)

*Then*

*T1: for every $k = 1, 2, \ldots$ there is a periodic point in  having period $k$; and, furthermore,*

*T2: there is an uncountable set  $\subset$  (containing no periodic points), which satisfies the following conditions:*

*(A) For every $p, q \in$  with $p \neq q$,*

$$\limsup_{n \to \infty} |F^n(p) - F^n(q)| > 0$$

*and*

$$\liminf_{n \to \infty} |F^n(p) - F^n(q)| = 0$$

*(B) For every $p \in$  and periodic point $q \in$  ,*

$$\limsup_{n \to \infty} |F^n(p) - F^n(q)| > 0$$

A more general and detailed version of this result is due to Sharkovsky [77, 16]

which was interestingly proven ten years earlier than Li and Yorke's result. He proved the above result for the maps of real line rather than an interval.

**Theorem 9** *[77, 16, 9] Let $f : \quad \to \quad$ be continuous. Suppose f has a periodic point of period three. Then f has periodic point of all other periods.*

The more interesting part of this theorem is that it gives a complete accounting of a sequence of periods which imply other periods for continuous maps of $R$. Consider the following ordering of the natural numbers:

$$3 \to 5 \to 7 \to \cdots \to 2.3 \to 2.5 \to \ldots$$

$$\to 2^2.3 \to 2^2.5 \to \ldots \to 2^3.3 \to 2^3.5 \ldots$$

$$\to 2^3 \to 2^2 \to 2 \to 1$$

This sequence gives an ordering of natural numbers. The significance of this sequence is stated in the next theorem again due to Sharkovsky.

**Theorem 10** *[77, 16, 9] Let $f : \quad \to \quad$ be continuous. Suppose f has a periodic point of period k. if $k \to l$ in the above ordering, then f also has a periodic point of period l.*

This provides us with the theory for the computation of higher order fixed point and shows the existence of chaos due to the existence of any of the above mentioned orbits with odd periodicity.

Also, a great advantage in using this theorem is its lack of hypotheses and only needs the map to be continuous and hence is extremely useful for piece-wise smooth and continuous system.

In our case $\quad = [0 \ B]$, and $F$ is given by the function $f(\cdot, \cdot)$ which defines the TCP-RED map in eq. 4.1. It is also clear that the TCP-RED map is continuous by construction as long as Assumption 1 is in force. Also, note that existence of a period three orbit, i.e., $d = a > b > c$ or $d = a < b < c$ is a special case of the hypotheses of the theorem and proves the existence of chaos.

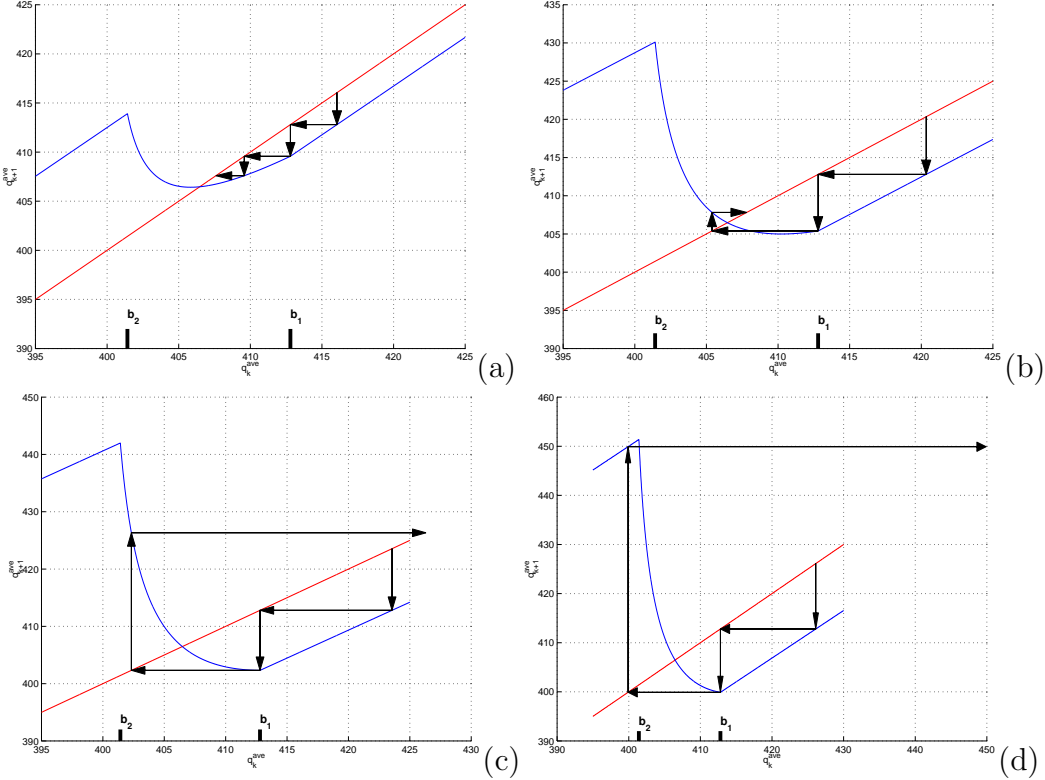

Figure 4.2: a. First return map and period three condition for $w = 2^7$, b. for $w = 2^{5.8}$, c. for $w = 2^{5.3}$ and d. for $w = 2^5$

To apply the theorem, we need to choose a starting point $a$ and iterate on it using the map $f$ three times and then apply the conditions stated in the theorem. We select $a = \frac{b_1}{(1-w)}$. This choice is made based on earlier numerical studies (Matlab and ns-2) which showed a strong tendency toward bifurcation and chaos when the system state $\bar{q}_{e,k}$ nears $b_1$. With this choice for $a$, we find that $b = f(a) = (1-w)a = b_1$ and $c = f^2(a) = f(b_1) = (1-w)b_1$. Looking at the Fig. 1, it is clear that there are two possible cases for the location of $c$: either $b_2 < c$ (Case I) or $b_2 > c$ (Case II). If $w$ is small, which is usually true, then $c = (1-w)b_1$ will be close to $b_1$ and therefore Case I will hold. However, conditions for Theorem 1 to apply will be found below for both Case I and Case II.

Case I: Let $b_2 < (1-w)b_1$ (this corresponds to $(1-w)b_1$ lying in the interval $((b_2, b_1))$. Then

$$f^3(a) = (1-w)^2 b_1 + w \left( \frac{NK}{\sqrt{\frac{p_{max}((1-w)b_1 - q_{min})}{(q_{max} - q_{min})}}} - \frac{R_0 C}{M} \right) \quad (4.6)$$

Hence the criterion $d \geq a > b > c$ of Theorem 1 ensuring existence of chaos gives:

$$(1-w)^2 b_1 + w \left( \frac{NK}{\sqrt{\frac{p_{max}((1-w)b_1 - q_{min})}{(q_{max} - q_{min})}}} - \frac{R_0 C}{M} \right) \geq a \quad (4.7)$$

Case II: Alternatively, suppose $b_2 \geq (1-w)b_1$. Then $f^3(a) = (1-w)^2 b_1 + wB$. Now the criterion $d \geq a > b > c$ of Theorem 1 ensuring existence of chaos in this case gives $f^3(a) - a \geq 0$, which reduces to the simple condition:

$$(1-w)^2 b_1 + wB \geq \frac{b_1}{(1-w)} \quad (4.8)$$

$$(1-w)^3 b_1 + w(1-w)B \geq b_1 \quad (4.9)$$

$$w(1-w)B - (1 - (1-w)^3) b_1 \geq 0 \quad (4.10)$$

$$w(1-w)B - (1 - 1 + 3w - 3w^2 + w^3) b_1 \geq 0 \quad (4.11)$$

$$b_1 \leq \frac{(1-w)B}{(3 - 3w + w^2)} \quad (4.12)$$

Summarizing, we have the following result.

**Theorem 11** *The TCP-RED system given by eq 4.1 is chaotic in the sense of Li and Yorke if either eq. (4.7) and $b_2 < (1-w)b_1$ hold, or eq. (4.12) and $b_2 > (1-w)b_1$ hold.*

Note that the condition above, given by eq. 4.7, is easy to check and parameter regions can be computed exactly where these inequalities are valid to guarantee chaos. Also, this condition does not depend on the fixed point of the map, hence it can be pre-computed if the system parameters are known. We will show empirically that this is indeed the case by showing the regions of bifurcation diagram where this condition holds.

## 4.3 Numerical Results

The behavior of the map can be explored numerically in parameter space to look for interesting dynamical phenomena. As the eigenvalue moves towards the unit circle, the fixed point will become unstable and depending on the nature of the ensuing bifurcation there can be new fixed points or chaos. There is also a possibility of the fixed point colliding with either border $b_1$ or $b_2$, leading to a rich set of possible bifurcations.

Here we use the same bifurcation diagrams to show that the period three criterion given by eq. 4.7 indeed implies the existence of chaos. To mark the region where this criterion is valid we have plotted the bifurcation diagram in red and it is blue otherwise. We consider the effect of various system and RED control parameters to study the chaotic phenomena with the help of bifurcation diagram. This is especially useful for $RED$ parameters ($q_{min}$, $q_{max}$, $p_{max}$ and $w$) and system parameters like the number of connections($N$), return trip time ($R_0$) etc.

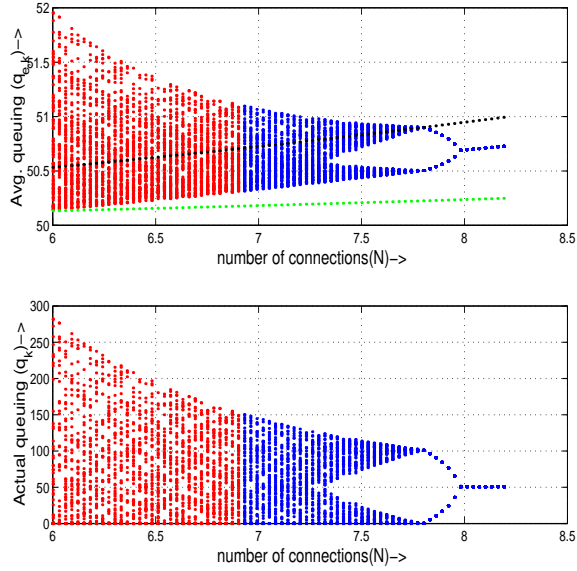

Figure 4.3: Bifurcation diagram of average and instantaneous queue length w.r.t. number of connections ($N$), Red region shows the validity of period three condition.

### 4.3.1 Effect of system parameters on dynamical behavior

To understand the effect of the number of connections on the system dynamical behavior, we have plotted the bifurcation diagram w.r.t. the number of connections ($N$). Other parameters for this bifurcation diagram are as follows:

$$q_{max}=100, \quad q_{min}=50, \quad c=1500 kbps, K=\sqrt{8/3}$$
$$B=300 \text{ packets}, \quad R_0=0.1 sec, \quad M=0.5 kb$$
$$w=2^{-7}, \quad p_{max}=0.1 \quad n=\text{bifurcation parameter}$$

The bifurcation diagram in Fig. 4.3 shows that the system stabilizes as the number of connections ($N$) increases. Also, it is clear that in the red region the system is chaotic. This chaotic region is also considerably separated from the period one and period two regions. This means the condition given by eq. 4.7 provides a robust criterion for chaos.

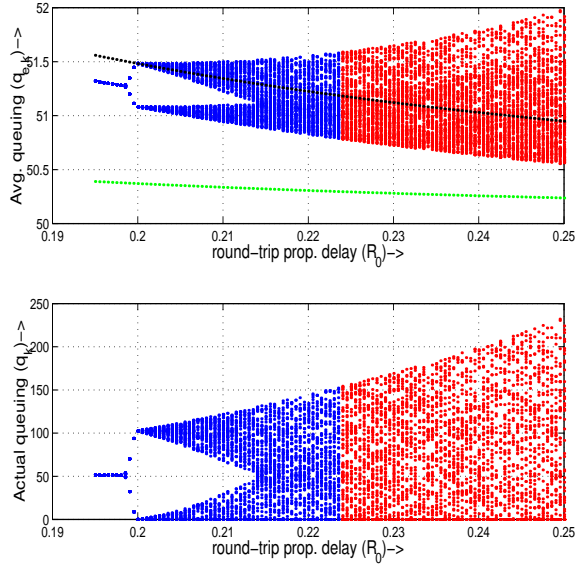

Figure 4.4: Bifurcation diagram of average and instantaneous queue length w.r.t. round-trip propagation delay ($R_0$). Red region shows the validity of period three condition.

Similarly, we also plot a bifurcation diagram w.r.t. round-trip propagation delay ($R_0$). Plot in Fig. 4.4 is in agreement with the result in [31] that larger delays cause instability. Also, we note the similar chaotic region in red which is significantly far from period behavior. Other parameters for this bifurcations diagram are as follows:

$$q_{max}=100, \quad q_{min}=50, \quad c=1500 kbps, K=\sqrt{8/3}$$
$$B=\tfrac{R_0 C}{M} \text{ packets}, \quad n=20, \quad M=0.5kb$$
$$w=2^{-7}, \quad p_{max}=0.1 \quad R_0=\text{bifurcation parameter}$$

We also plot the bifurcation diagram with respect to exponentially averaging parameter $w$ and $q_{min}$ to provide an empirical proof of the fact that indeed criterion given by eq. 4.7 is valid when system in chaotic and far from periodic operation as shown in the red part of the bifurcation diagrams in fig. 4.5 and in fig. 4.6.

Corresponding parameter scenarios are given by eq. 4.13 and eq. 4.16.

$$q_{max}=100, \quad q_{min}=50, \quad c=1500kbps, K=\sqrt{8/3} \qquad (4.13)$$

$$B=300 \text{ packets}, \quad R_0=0.1sec, \quad M=0.5kb \qquad (4.14)$$

$$N=20, \quad p_{max}=0.1, \quad w=\text{bifurcation parameter} \qquad (4.15)$$

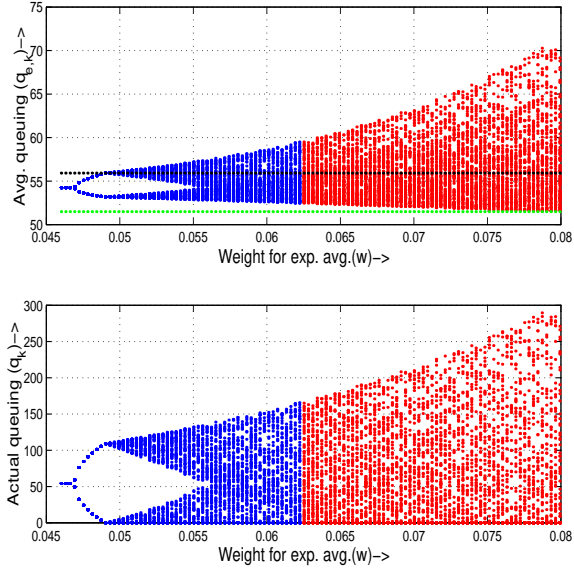

Figure 4.5: Bifurcation diagram of average and instantaneous queue length w.r.t. $w$, $p_{max} = 0.1$. Red region shows the validity of period three condition.

$$p_{max}=0.3, \quad q_{max}=100, \quad c=1500kbps, \quad K=\sqrt{8/3} \qquad (4.16)$$

$$B=300 \text{ packets}, \quad R_0=0.1sec, \quad M=0.5kb, \qquad (4.17)$$

$$N=20, \quad w=2^{-7} \quad q_{min}=\text{bifurcation parameter} \qquad (4.18)$$

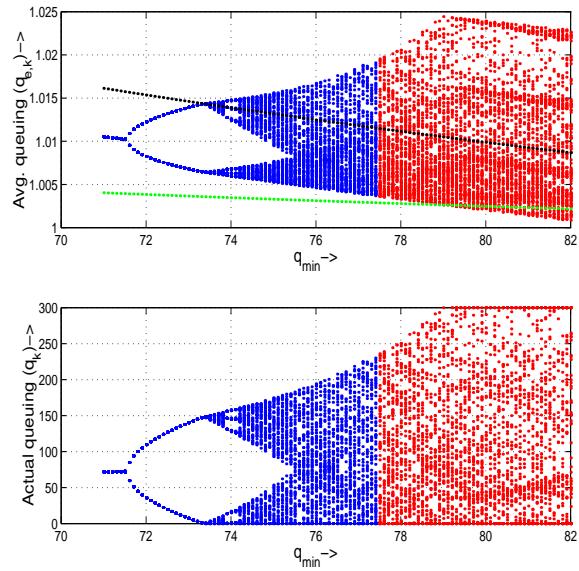

Figure 4.6: Bifurcation diagram of average and instantaneous queue length w.r.t. $q_{min}$, $w = 2^{-7}$. Here, the averaging queuing has been normalized by $q_{min}$ to see the structure of bifurcation diagram. Red region shows the validity of period three condition.

# Chapter 5

# Delay and Greed Trade-Offs for Stability

## 5.1 Introduction

With the unprecedented growth and popularity of the Internet the problem of rate/congestion control is emerging as a more crucial problem. Poor management of congestion can render one part of a network inaccessible to the rest and significantly degrade the performance of networking applications. Kelly has proposed an optimization framework for rate allocation in the Internet [41]. Using the proposed framework he has shown that the system optimum is achieved at the equilibrium between the end users and resources. Based on this observation researchers have proposed various rate-based algorithms that solve the system optimization problem or its relaxation [41, 47, 55] as described in section 1.4. However, the convergence of these algorithms has been established only in the absence of feedback delay, and the impact of feedback delay has been left open as well as any trade-off that may exist between stability and selected utility and cost functions. In particular the is-sue of stability has not been systematically investigated with large communication delays.

In this chapter we establish an intuitive global stability criterion for the system optimization problem in the presence of an *arbitrary* delay for a simple one resource problem with multiple flows. This stability condition is derived using invariance-based global stability results for nonlinear delay-differential equations [57, 36, 35, 34]. These global stability results are different from those based on Lyapunov or Razumikhin theorems in the sense that our approach also provides us with insight on the structure of emerging periodic orbits, *e.g.,* their periodicity and amplitude, in the case of loss of stability. This loss of stability and characterization of ensuing periodic orbit will also be studied in this chapter.

Generally speaking, our results tell us that if the user and resource curves have a stable market equilibrium, then corresponding dynamical equation for flow-optimization will converge to the optimal solution in the presence of arbitrary delay. This result essentially shows that stability is related to utility and price curves in a fundamental way. In particular, for a given price curve, it possible to design stable user utility functions such that the corresponding dynamical system converges to the optimal flow irrespective of communication delay. Conversely, if the underlying market equilibrium is unstable then it is possible to find a large enough delay for which the optimal point loses its stability and gives way to oscillations. This chapter contains the oscillatory orbit characterization by explicitly giving the bounds on their amplitude. It is also shown that these bounds are derived from an underlying discrete time map which goes through a period doubling bifurcation with the loss of stability.

These results provide an interesting perspective for designing end user algo-rithms and active queue management (AQM) mechanisms. It is also worth noting that in general characterizing the exact conditions for stability with a delay is difficult. Hence, our result provides a simple and yet robust way of dealing with the problem of widely varying feedback delay in communication networks through a clever choice of the user utility function and price functions.

This chapter is organized as follows. Section 5.2 describes the optimization problem for rate control. Relevant previous work on the stability criterion of a system given by a delayed differential equation is given in Section 5.3. Our main results on stability and instability are presented in Section 5.4. Results in the presence of non-responsive traffic are derived in section 5.6, which is followed by numerical examples in Section 5.7. Preliminary results on multi-user case are discussed in section 5.8. We conclude the chapter in Section 5.9.

## 5.2 Background

In this section we briefly describe the rate control problem in the proposed optimization framework. Consider a flow traversing a single resource. The rate control problem can be formulated as the following net utility optimization problem from the end user's point of view [41]:

$$\max_x \quad U(x) - x \cdot p(x) \tag{5.1}$$
$$\text{s. t.} \quad x \leq C$$

where $x$ is the rate, $U(x)$ is the utility of the user when it receives a rate of $x$, $p(x)$ is the price per unit flow the user has to pay when the rate is $x$, and $C$ is the capacity of the resource. The proposed end user algorithm in the absence of delay is given by the following differential equation [43].

$$\frac{d}{dt}x(t) = k\left(w(t) - x(t)\mu(t)\right) \qquad (5.2)$$

where $w(t)$ is the price per unit time user is willing to pay, $\mu(t) = p(x(t))$, and $k > 0$, is a gain parameter.

The case where $w(t)$ is a fixed constant, i.e., $U(x) = log(x)$, is studied in [42, 39]. In this work we assume that $w(t) = x(t) \cdot U'(x(t))$ with any monotonically decreasing concave family of utility functions [43] with their decrease rate smaller than $\frac{1}{x}$. Now, suppose that congestion signal generated at the shared resource, i.e., $p(x(t))$, is returned to the user after a fixed and common round trip time $T$. In the presence of delay the interaction is given by the following delayed differential equation

$$\frac{d}{dt}x(t) = k\left(w(t) - x(t-T)\mu(t-T)\right) \qquad (5.3)$$
$$= k\left(x(t)U'(x(t)) - x(t-T)p(x(t-T))\right) \qquad (5.4)$$

After normalizing time by $T$ and replacing $t = s \cdot T$, eq. 5.4 becomes:

$$\frac{1}{T}\frac{d}{ds}x(s) = k\left(x(s)U'(x(s)) - x(s-1)p(x(s-1))\right) \qquad (5.5)$$
$$\nu\frac{d}{ds}x(s) = x(s)U'(x(s)) - x(s-1)p(x(s-1)) \qquad (5.6)$$

where $\nu = \frac{1}{Tk}$. It is precisely eq. 5.6 we are interested in from stability point of view. For $T \gg 1$, this equation can be viewed as a singular perturbation

$$\nu\frac{d}{dt}x(t) = g(x(t)) - f(x(t-1)) \qquad (5.7)$$

of a general nonlinear difference equation with continuous argument given by

$$g(x(t)) = f(x(t-1)), \quad t \geq 0 \qquad (5.8)$$

where $g(x) = xU'(x)$ and $f(y) = yp(y)$ in the context of eq. 5.6. Under certain natural invertibility conditions on $g(\cdot)$, it leads to the much studied equation [79]

$$x(t) = F(x(t-1)), \quad t \geq 0 \tag{5.9}$$

where $F(\cdot) = g^{-1}(f(\cdot))$. For the solution of eq. 5.9 to be continuous for $t \geq -1$, along with the continuity of $F$ and $\phi(\cdot)$, which is the initial function, a so-called consistency condition $\lim_{t \to -0} \phi(t) = F(\phi(-1))$ is required [36, 79].

It turns out that a great deal about the asymptotic stability of eq. 5.7 can be learned from the asymptotic behavior of the following difference equation, with $Z_+$ denoting the set of positive integers:

$$x_{n+1} = F(x_n), \quad n \in Z_+ \tag{5.10}$$

Some relevant previous work on these equations is presented in the following section.

## 5.3 Previous Work

In this section we summarize some of relevant work presented in [35]. Consider a nonlinear delay differential equation of the following form:

$$\dot{x}(t) = f(x(t-\tau)) - g(x(t)) \tag{5.11}$$

where functions $f$ and $g$ are continuous for $\Re_+ = \{x : x \geq 0\}$ with the values in $\Re_+$. We make following additional assumptions these functions:

1. $g(x)$ is strictly increasing, $g(0) = 0$, and $\lim_{x \to \infty} g(x) = \infty$,

2. there is exactly one point $\bar{x} > 0$ such that $f(\bar{x}) = g(\bar{x})$; moreover, $f(x) > g(x)$ in $(0, \bar{x})$ and $f(x) < g(x)$ in $(\bar{x}, +\infty)$.

Eq. 5.11 can be written in a singular perturbation form by change of coordinates $t = \tau \cdot s$ and $\mu = \frac{1}{\tau}$.

$$\mu \frac{d}{dt}x(t) = f(x(t-1)) - g(x(t)) \tag{5.12}$$

Now define $F(x) := g^{-1}(f(x))$. Invariance and global stability of one dimensional map $F$ can be translated to those of eq. 5.11 for arbitrary time delay $\tau$ [36] as described here.

Let $I \subset \Re_+$ be a closed interval which is invariant under $F$. Also, let $X := C([-1,0], \Re_+)$ and $X_I := \{\phi \in X : \phi(s) \in I \ \forall s \in [-1,0]\}$.

**Theorem 12** *The set $X_I$ is invariant under eq. 5.12. For all $\phi \in X_I$ the corresponding solution $x(t; \phi)$ belongs to $I$ for all $\mu \geq 0$.*

Now suppose that $I_0 = \cap_{n \geq 0} F^n(I)$ degenerates into a single point, i.e., map $F$ has an asymptotically stable fixed point. Then, the following theorem holds.

**Theorem 13** *If $x = \bar{x}$ is the globally attracting fixed point of the map $F$, then for any initial function $\phi \in X$ and every $\mu > 0$ the corresponding solution $x(t)$ of eq. 5.12 approaches $\bar{x}$.*

## 5.4   Rate Control with Feedback Delay

We study the rate allocation problem in Kelly's optimization framework described in Section 5.2 [41] with the following class of price functions:

$$p(y) = \left(\frac{y}{C}\right)^b, \quad \text{where } b > 0 \tag{5.13}$$

This kind of marking function arises if the resource is modeled as an $M/M/1$ queue with a service rate $C$ packet per unit time and a packet receives a mark with a congestion indication signal if it arrives at the queue to find at least $b$ packets in the queue. Generally, any strictly increasing continuous function should suffice for our purpose.

The class of utility functions we consider here has the form

$$U_a(x) = -\frac{1}{a}\frac{1}{x^a}, \quad a > 0. \tag{5.14}$$

In particular, $a = 1$ has been found useful for modeling the utility function of Transmission Control Protocol (TCP) algorithms [45]. We say that a user $u_1$ with utility function $U_{a_1}(x)$ is greedier than another user $u_2$ with utility function $U_{a_2}(x)$ if $a_2 > a_1$. One can interpret the notion of greed here using the notion of elasticity of demand [85]. With the utility functions of the form in eq. 5.14 one can easily show that the elasticity of the demand decreases with increasing $a$ as follows. Given a price $p$, the optimal rate $x^*(p)$ of the user that maximizes the net utility $U_a(x) - x \cdot p$ is given by $p^{-\frac{1}{1+a}}$. The price elasticity of the demand, which measures how responsive the demand is to a change in price, is defined to be the percent change in demand divided by the percent change in price [85]. In our case the price elasticity of demand is given by

$$\frac{p}{x^*(p)}\frac{dx^*(p)}{dp} = \frac{p}{p^{-\frac{1}{1+a}}} \cdot \frac{-1}{1+a} p^{-\frac{1}{1+a}-1}$$
$$= \frac{-1}{1+a}. \tag{5.15}$$

Therefore, one can see that price elasticity of the demand decreases with $a$, i.e., the larger $a$ is, the less responsive the demand is.

The model used for design of the end-user rate control algorithm in [42] does not explicitly address the case where the total demand of the users exceeds the link capacity. In practice the total rate of the users is limited by the link capacity. In order to handle this shortcoming of the model we make following natural

assumption:

**Assumption 2** *Suppose that there are $N$, $N \geq 1$, homogeneous users, i.e., users with the same feedback delay and rate. We assume that the rate of each user is bounded above by $\frac{C}{N}$.*

Clearly, allowing more than $\frac{C}{N}$ may exceed the total throughput momentarily leading to severe penalties for all the users. This bound also makes sense due to the fact that networked users have a tendency to synchronize [39].

In the presence of time delay $T$, $N$ homogeneous end users with their utility function in eq. 5.14, each users rate dynamical equation is given by

$$\dot{x}(t) = k\left(\frac{1}{x(t)^a} - x(t-T)\left(\frac{Nx(t-T)}{C}\right)^b\right), \quad (5.16)$$

By substituting $t = T \cdot s$, we obtain

$$\nu \dot{x}(s) = \frac{1}{x(s)^a} - x(s-1)\left(\frac{Nx(s-1)}{C}\right)^b \quad (5.17)$$

where $\nu = \frac{1}{T \cdot k}$. In order to apply the theorems in Section 5.3, we can compare the forms in eq. 5.11 with that of eq. 5.16 where $g(x) = -k\frac{1}{x^a}$ and $f(x) = -kx\left(\frac{Nx}{C}\right)^b$. It is clear that although eq. 5.16 looks similar to eq. 5.11, it does not satisfy all the assumptions required to apply these theorems. In particular, these functions have their range in negative real numbers.

It turns out that by making a simple substitution we can make eq. 5.17 resemble the well studied eq. 5.11. Consider the following substitution:

$$g(x(t)) = x(t)U'(x(t)) := y(t), \quad \text{and} \quad (5.18)$$

$$f(x(t)) = x(t)\left(\frac{Nx(t)}{C}\right)^b \quad (5.19)$$

Although we give specific functional forms in eq. 5.18 and in eq. 5.19 results hold for any functional forms obeying following general assumptions on the functions $g(x)$ and $f(x)$.

**Assumption 3** *(i) The function $g(x)$ as given in eq. 5.18 is strictly decreasing with $-g'(x) > 0$ for all $x > 0$. (ii) The function $f(x)$ is increasing for all $x > 0$ (iii) Both $g(x)$ and $f(x)$ are Lipschitz continuous on strictly positive real axis.*

For our results in this work the exact form of users' utility functions or the resource price function needs not be given by eq. 5.14 and 5.13, respectively, but rather must satisfy Assumption 3. This allows us the following change of coordinate:

$$x(t) = g^{-1}(y(t)), \tag{5.20}$$

$$\dot{x}(t) = \frac{\dot{y}(t)}{g'(g^{-1}(y(t)))} \tag{5.21}$$

$$\nu \dot{y}(t) = g'(g^{-1}(y(t)))(y(t) - f(g^{-1}(y(t-1)))) \tag{5.22}$$

where the inverse $g^{-1}(\cdot)$ exists from assumption 3. Let $\kappa(y(t)) := -g'(g^{-1}(y(t)))$. Clearly, $\kappa(y(t)) > 0$ under assumption 3. Using this substitution in eq. 5.22 we get the following form which resembles eq. 5.11 closely, except for a multiplicative state-dependent gain $\kappa(y(t))$.

$$\nu \dot{y}(t) = \kappa(y(t)) \left( f(g^{-1}(y(t-1))) - y(t) \right) \tag{5.23}$$

It is eq. 5.23 which we wish to study and show that there is a close correspondence between invariance and global stability properties of map

$$y_{n+1} = f(g^{-1}(y_n)) := F(y_n) \tag{5.24}$$

and those of eq. 5.23 for fixed and periodic orbits. In particular, we wish to prove that if $y_{n+1} = F(y_n)$ has a period two fixed point then eq. 5.23 will have a periodic solution for large enough time delay $T$ if the initial function's range is contained in the immediate basin of attraction of this period two fixed point.

The proofs are based on the invariance property of the underlying map $F(\cdot)$ given in eq. 5.24 and the monotonicity of function $g(\cdot)$. The map $F(y)$ is decreasing because $g^{-1}(y)$ is strictly decreasing under assumption 3 and a composition of an increasing and a strictly decreasing function ($f$ is increasing from assumption 3) is a decreasing function. This is a much studied scenario for delay differential equations for oscillations where a typical requirement is $F' < 0$ or $xF(x) < 0$ around the origin.

**Assumption 4** *Suppose now that $I \subset \Re_+ := \{x : x > 0\}$ is a closed invariant interval under $F$. In particular let $I = [a\ b]$ be compact.*

Let $X := C([-1, 0], \Re_+)$, and $X_I := \{\phi \in X : \phi(s) \in I\ \forall s \in [-1, 0]\}$. Under this assumption, we have invariance for the solution of eq. 5.23 for all time $t \geq 0$ and for all $\nu \geq 0$ as will be shown in a moment.

### 5.4.1 Existence and uniqueness of solution

Since the functions $f, g$ involved in eq. 5.23 are Lipschitz continuous by assumption, solutions do exist for all $t \geq 0$ and are unique for any initial function $\phi \in X_I$, where $I$ is the assumed closed invariant interval under $F$. Furthermore, the invariance property of solutions, which is shown and stated below (Theorem 14), ensures that they stay positive and bounded by the initial set they start in, which is assumed to be invariant under map $F$.

**Theorem 14** *Invariance [72]: If $\phi \in X_I$, the corresponding solution $y(t) = y(t; \phi)$ satisfies $y(t) \in I$ for all $t \geq 0$. It means that set $I$ is invariant under eq. 5.23.*

**Proof:** Let $t_0$ be the first time when solution $y(t; \phi)$ leaves $I$ with $\phi \in X_I$. In particular, we can assume that $y(t_0) = b$ and every right hand neighborhood of $t_0$ will have a $t_1 > t_0$ such that $y(t_1) > b$. Then, we can find a point $t_2, t_0 < t_2 < t_0 + 1$, such that $y(t_2) > b$ and $\dot{y}(t_2) > 0$. Since $y(t_2 - 1) \leq b$, we have $\dot{y}(t_2) = \frac{\kappa(y(t_2))}{\nu} [f(g^{-1}(y(t_2 - 1))) - y(t_2)] < 0$ from eq. 5.23 and assumption 4 that $I$ is invariant under $F$, i.e., $f(g^{-1}(y(t_2-1))) \leq b$. This contradicts with the earlier assertion about the positivity of $\dot{y}(t_2)$.

Similarly, assume that $y(t_0) = a$ and the trajectory exits from left end of the interval. Then, every right hand neighborhood of $t_0$ will have a $t_1 > t_0$ such that $0 < y(t_1) < a$ due to the smoothness of solutions, and we can find $t_2, t_0 < t_2 < t_0+1$, such that $0 < y(t_2) < a$ and $\dot{y}(t_2) < 0$. From that $y(t_2 - 1) \geq a$, we have $\dot{y}(t_2) = \frac{\kappa(y(t_2))}{\nu} [f(g^{-1}(y(t_2 - 1))) - y(t_2)] > 0$ from eq. 5.23 and assumption 4. This contradicts with the negativity of $\dot{y}(t_2) < 0$. Hence, the theorem follows. ■

Here we note that uniform positivity of $\kappa(y(t))$ over the positive real line is crucial to the proof.

The next theorem considers the case when the map $F$ has an attracting fixed point $y^*$ with immediate basin of attraction $J_0 : F^n y_0 \to y^*$ for any $y_0 \in J_0$. Let $X_{J_0} = C([-1, 0], J_0)$ then following theorem holds.

**Theorem 15** *Stability [72]: For any $\nu > 0$ and $\phi \in X_{J_0}$, $\lim_{t \to \infty} y_\phi^\nu(t) = y^*$.*

Before proving the theorem we will state a lemma which is the key to the proof of theorem above.

**Lemma 7** *Suppose that an interval $J$ is mapped by $F$ into itself. If none of the endpoints of the interval $F(J)$ is fixed point then for every $\phi \in X_J = C([-1, 0], J)$ there exists a finite $t_0 = t_0(\phi, \nu, \kappa)$ such that $y_\phi^\nu(t) \in F(J)$ for all $t \geq t_0$.*

**Proof:** From last invariance theorem it is clear that $y_\phi^\nu \in J$ for all $t \geq 0$. The claim here is that after certain time $t_0$ it will be limited by $F(J) \subset J$.

First, assume that $\phi(0) \in \overline{F(J)}$. Then it can be shown that $y_\phi^\nu(t) \in F(J)$ for all $t \geq t_0$ by contradiction. Suppose that this is not true and let $t_0$ be the first time when $y_\phi^\nu(t)$ leaves the interval $\overline{F(J)}$. In particular assume that it leaves from the right end, i.e., every right-sided neighborhood of $t_0$ contains a point $t_1$ such that $y_\phi^\nu(t_1) > \sup \overline{F(J)}$. Then, the same neighborhood also contains a point $t_2$ such that $y_\phi^\nu(t_2) > \sup \overline{F(J)}$ and $\dot{y}_\phi^\nu(t_2) > 0$. As $y_\phi^\nu(t) \in J$ for all $t \in [t_0 - 1, t_0]$ and also $t_0 < t_2 < t_0 + 1$ can be assumed, we have $\dot{y}(t_2) = \frac{\kappa(y(t_2))}{\nu}[f(g^{-1}(y(t_2 - 1))) - y(t_2)] < 0$ from eq. 5.23. This contradicts the earlier assumption that $\dot{y}(t_2) > 0$. The other case where $y_\phi^\nu(t)$ leaves the interval from the left end can be handled similarly.

Now assume that $\phi(0) \notin \overline{F(J)}$. Particularly, let $\phi(0) > \sup \overline{F(J)}$. Claim here is that $y_\phi^\nu(t)$ is decreasing for all $t \in [0, t_0]$, where $t_0 \leq \infty$ is the first point with $y_\phi^\nu(t_0) = \sup \overline{F(J)}$. We first argue that $t_0 < \infty$ by contradiction. Suppose $t_0 = \infty$ and, hence, $y_\phi^\nu(t) > \sup \overline{F(J)}$ for all $t \geq 0$. From eq. 5.23 we have $\dot{y}(t_2) = \frac{\kappa(y(t_2))}{\nu}[f(g^{-1}(y(t_2 - 1))) - y(t_2)] < 0$ because $f(g^{-1}(y(t_2 - 1))) \leq \sup \overline{F(J)}$. Then there exists a limit $\overline{y} = \lim_{t \to \infty} y_\phi^\nu(t) > \sup \overline{F(J)}$ due to Bolzano-Weierstrass theorem argument, which states that every strictly decreasing sequence which is bounded from below has a limit [22]. As $\overline{y}$ is not a fixed point of map $F$, $\kappa(\overline{y})(\overline{y} - f(g^{-1}(\overline{y}))) := \delta > 0$. This tells us from eq. 5.23 that $\dot{y}(t) = \frac{\kappa(y(t))}{\nu}[f(g^{-1}(y(t - 1))) - y(t)] < -\frac{\delta}{2\nu}$ for large enough $t$. This implies that $y_\phi^\nu(t) \to -\infty$ as $t \to \infty$ which is a contradiction, for this means that $y_\phi^\nu(t)$ crosses $\sup \overline{F(J)}$ for some finite $t$. Hence, $t_0 < \infty$. Now, we invoke the first part of proof where system is restarted at time $t_0$ with $y(t_0) = \sup \overline{F(J)}$ and $y(t) \in J \; \forall t \in [t_o - 1 \; t_0]$. Using that argument $y(t) \in F(J)$ for all $t \geq t_0$. The other case can be handled similarly. ∎

Now we provide the proof for the theorem above using this lemma.

**Proof:** For any $\phi \in X_{J_0}$, define $m = \inf\{\phi(s), s \in [-1\ 0]\}$ and $M = \sup\{\phi(s), s \in [-1\ 0]\}$. Clearly, $[m\ M] \subset J_0$. Let $J'$ be the smallest closed invariant interval containing $[m\ M]$ which is a subset of $J_0$. Then, from existence of fixed point for the map $F$, $J' \supset F(J') \supset F(F(J')) \supset \ldots$ and $\cap_{i \geq 0} F^i(J') = x^*$. Using invariance and Lemma 7 repeatedly one can find arbitrarily small estimates for the range of trajectories with large enough time. Hence, the proof. ∎

Further, we are interested in the case where the map defined by eq. 5.24 goes through a period doubling bifurcation with its eigenvalue $\lambda := \frac{dF}{dx}\big|_{x=x^*} < -1$ where $x^*$ is an unstable fixed point of map $F$. In the following subsection we describe how the instability of underlying discrete-time map is translated to the instability of delay-different equation in eq. 5.23.

### 5.4.2 Linear instability analysis

Assuming that the map $F$ given by eq. 5.24 is locally smooth, it is possible to find conditions for linear instability of the fixed point of the map $y^*$ and that of constant function $y(t) = y^*$ for the delay-differential equation in eq. 5.23. In order for $y(t) = y^*$ to be locally asymptotically stable for all $T \geq 0$, following variational equation should have its zero solution stable.

$$\begin{aligned}
z'(t) &= \kappa(y(t))F'(y(t-T))\big|_{y=y^*} z(t-T) \\
&\quad + \left(\kappa'(y(t))[F(y(t-T)) - y] - \kappa(y(t))\right)\big|_{y=y^*} z(t) \\
&= \kappa(y(t))F'(y(t-T))\big|_{y=y^*} z(t-T) - \kappa(y(t))\big|_{y=y^*} z(t) \\
&\quad \text{(because } F(y^*) = y^* \text{ )} \\
&:= Bz(t-T) + Az(t) \quad (5.25)
\end{aligned}$$

where $B = \kappa(y^*)F'(y^*)$ and $A = -\kappa(y^*)$. Now to determine the stability of $y(t) = y^*$, we apply well known results from [63] as follows

(1) It is known that eq. 5.25 is stable for all $T \geq 0$ only if:

$$A \leq 0 \quad \text{and} -A \geq |B| \tag{5.26}$$

(2) In case the above is violated we have stability for some values of time delays:

$$-B > |A| \quad \text{and} \ T \leq T^* := \frac{\cos^{-1}(-\frac{A}{B})}{B^2 - A^2}. \tag{5.27}$$

For our case it follows from eq. 5.25 that $-\kappa(y^*)$ is always negative, which holds due to the fact that $\kappa(\cdot)$ is always positive. The second condition $\kappa \geq \kappa|F'|$ is crucial for the stability of eq. 5.23. Clearly, for the case when $F' < -1$ (period doubling condition for the map $F$) the linear stability condition given by eq. 5.26 is violated and for a large enough $T$ the constant solution $y(t) = y^*$ will not be stable. Thus, we know that in unstable situation solutions will be more complex than a constant function and will stay within the interval they initially start from due to the invariance results given by theorem 14. Next we show that in case of instability the bounds on the solution of delay-differential equation will be given by the period two solution of underlying discrete time map $F$.

**Theorem 16** *Let $I := [a, b]$ be a closed interval such that $F(I) := [a_1, b_1] \subset I$. Let the initial condition $\phi(t) \in Y_I$, where $Y_I = \{\phi^* \in Y \mid \phi^*(s) \in I \ \forall \ s \in [-1, 0]\}$ and $Y := C([-1, 0], \Re_+)$, be the solution of eq. 5.23. Now, if the points $a_1$ and $b_1$ are fixed points of $F$, then for all sufficiently small $\epsilon \geq 0$ there exists a finite $T = T(\phi, \epsilon, \kappa)$ such that $y(t) \in [a_1 - \epsilon, b_1 + \epsilon]$ for all $t \geq T$.*

**Proof:** The proof follows the same arguments used in the proof of Theorem 14, except that boundary considered this time will $b_1 + \epsilon$ from right. In the interval $[b_1 + \epsilon, b]$ solution will be strictly decreasing until it reaches the point $b_1 + \epsilon$. Afterwards from invariance theorem it stays bounded by $b_1 + \epsilon$ from above. Similar reasoning follows for lower bound. ∎

Above theorem essentially gives bounds for the interval which will contain the solution asymptotically. Now suppose $y(t)$ be a solution of eq. 5.23 under the instability condition that map $F'(y^*) < -1$. Due to invariance theorem we know that $0 < \liminf_{t \to \infty} y(t) = m \le \limsup_{t \to \infty} = M < +\infty$. Now based on the theory developed in [34] we make following observations:

(1) If the solution $y(t)$ is strictly monotone then $m = M = y^*$ because of the boundedness of solutions.

(2) If $m \ne M$ then the solution $y(t)$ is oscillating. In particular, the solutions will have a sequence of maxima at times $\{t_n\}$ and minima at times $\{s_n\}$. Clearly, $y'(t_n) = y'(s_n) = 0$ for all $t_n$ and $s_n$. This implies from eq. 5.23 that $y(t_n) = F(y(t_n - 1))$ and $y(s_n) = F(y(s_n - 1))$. This shows interesting discrete time map structures in the solution of delay-differential equation eq. 5.23.

(3) If $y(t)$ does not converge to $y^*$ then it oscillate around it. This holds due to the fact that image of the interval $[m, M]$ under $F$ contains $[m, M]$. Hence, it will have a fixed point $y^*$.

So, we conclude that solutions either converge to the fixed point or oscillate around it.

## 5.5 Applications

Suppose that there are $N \geq 1$, homogeneous users in the system. Since users are homogeneous, denote the rate of a user by $x(t)$. We assume that the utility function of the users is of the form in eq. 5.14 and the price function used at the resource is that of eq. 5.13. Then, the end user algorithm is given by

$$\dot{x}^{(N)}(t)$$
$$= k\left(x^{(N)}(t)U_a'(t) - x^{(N)}(t-T) \cdot p(N \cdot x^{(N)}(t-T))\right) \quad (5.28)$$
$$= k\left(\frac{1}{x^{(N)}(t)^a} - x^{(N)}(t-T)\left(\frac{N \cdot x^{(N)}(t-T)}{C}\right)^b\right), \quad (5.29)$$

where a superscript $(N)$ is used to denote the dependence on $N$. The underlying discrete time difference equation is given by

$$y_{n+1}^{(N)} = \left(\frac{N}{C}\right)^b y_n^{-\frac{b+1}{a}} := F^{(N)}(y_n^{(N)}) \quad (5.30)$$

$$\frac{1}{\left(x_{n+1}^{(N)}\right)^a} = x_n^{(N)}\left(\frac{N \cdot x_n^{(N)}}{C}\right)^b, \quad x_n^{(N)} > 0 \quad (5.31)$$

$$x_{n+1}^{(N)} = \left(\frac{(C/N)^b}{x_n^{(N)\,b+1}}\right)^{\frac{1}{a}} \quad (5.32)$$

Then, from eq. 5.32 the fixed point

$$x^{(N)*} = \left(\frac{C}{N}\right)^{\frac{b}{a+b+1}} \quad (5.33)$$

and the eigenvalue is given by

$$\lambda^{(N)}(x^{(N)*}) = -\frac{b+1}{a} \quad (5.34)$$

and is independent of $N$. Therefore, the stability of the system does not depend on the number of users in the system. This can also be explained using the price elasticity of demand. Since, given a utility function of the form in eq. 5.14 for some $a > 0$, the price elasticity of the demand is constant for all $x > 0$ from eq. 5.15, one would expect the stability of the system to be independent of the operating point, i.e., the fixed point, and capacity, but only on the choices of the utility and price functions that determine the responsiveness of the users and resource, respectively. Hence, $a > b + 1$ gives local stability. According to the Sharkovsky cycle coexistence ordering [78] the most general condition for the fixed point $x^*$ to be globally attracting is that the second iteration $F^2(\cdot)$ of the map $F(\cdot)$ does not have a fixed point in the relevant invariance set other than $x^*$, which is locally stable. These conditions hold in our example, and hence the fixed point $x^*$ is *globally* stable in the invariance set.

However, in the case of instability when $a < b + 1$, depending on the feedback delay $T$, it will oscillate. Clearly, the upper bound on the solution is given by the self-imposed limit $\frac{C}{N}$ of the users to avoid any capacity mismatch. According to Theorem 16 the lower bound will be given by $F(C/N) = \left(\frac{N}{C}\right)^{\frac{1}{a}}$. Hence, link will see a wide fluctuation from full to very low utilization irrespective of the number of users. Actually, having more users will reduce the oscillation amplitude due to lower bandwidth per user.

## 5.6  Presence of Non-Responsive Traffic

In this section we analyze the effect of the presence of non-responsive traffic in the traffic streams. As we know that non-responsive traffic has no dynamics and only adds up to the queuing. This traffic modifies the models given by eq. 5.31 and eq. 5.32 as follows:

$$\frac{1}{x_{n+1}^a} = x_n \left(\frac{Nx_n + p}{C}\right)^b, \quad x_n > 0 \tag{5.35}$$

$$x_{n+1} = \left(\frac{C^b}{(Nx_n + p)^b x_n}\right)^{\frac{1}{a}} \tag{5.36}$$

where $p < C$ is the non-responsive traffic. Fixed point for this system will be given as the solution of following equation:

$$x^{a+1}(Nx + p)^b = C^b \tag{5.37}$$

$$x^{\frac{a+1}{b}}(Nx + p) = C \tag{5.38}$$

$$Nx + p = \frac{C}{x^{\frac{a+1}{b}}} \tag{5.39}$$

It is clear from eq. 5.39 that as the amount of non-responsive traffic $p$ increases the fixed point solution $x^*$ of eq. 5.37 decreases. Now we will analyze the eigenvalue of eq. 5.36 to look into the effects of non-responsive traffic on stability. Computing the eigenvalue $\lambda^p(x^*)$ gives the following:

$$\lambda^p(x^*) = -\frac{C^{\frac{b}{a}}}{a} \left[\frac{1}{x^{\frac{a+1}{a}}(Nx+p)^{\frac{b}{a}}} + \frac{Nb}{x^{\frac{1}{a}}(Nx+p)^{\frac{a+b}{a}}}\right]\bigg|_{x=x^*} \tag{5.40}$$

Using the expression for $(Nx + p)$ from eq. 5.39, eq. 5.40 reads:

$$\lambda^p(x^*) = -\frac{C^{\frac{b}{a}}}{a}\left[\frac{1}{C^{\frac{b}{a}}} + \frac{Nbx^{\frac{a+b+1}{a}}}{C^{\frac{a+b}{a}}}\right]\bigg|_{x=x^*} \tag{5.41}$$

$$= -\frac{1}{a}\left[1 + \frac{bx^{\frac{a+b+1}{a}}}{\frac{C}{N}}\right]\bigg|_{x=x^*} \tag{5.42}$$

$$\tag{5.43}$$

This expression can be verified by using the expression for the fixed point in the absence of non-responsive traffic from eq. 5.33 which yields the same expression as the eigenvalue in the only responsive traffic case given by eq. 5.34. From the earlier observation that the fixed point decreases in the presence of non-responsive traffic $p$, it can be easily seen that the second term in the eigenvalue expression will be reduced and hence eigenvalue will be smaller in magnitude rendering the system stable. This confirms the earlier observations of stability due to non-responsive traffic.

## 5.7 Numerical Simulations

We take two homogeneous users with their utility functions of the form given by eq. 5.14 with $a = 3$ and price function as in eq. 5.13 with $b = 5$ and link capacity $C$ to be 5. It is clear that for these values of $a$ and $b$ rate control algorithm is unstable since $\frac{5+1}{3} = 2 > 1$. The optimal rates for both users in the absence of delay will be given by $x^* = \left(\left(\frac{2}{5}\right)^5\right)^{\frac{1}{9}} = 1.6637$. Their self imposed upper rate limit will be $C/2 = 2.5$. The lower limit on the solution according to the period two orbit of map $F$ will be given by $F(2/5) = \left(\frac{2}{5}\right)^{\frac{1}{3}} = 0.7368$.

Fig. 5.1(a) shows the first (red 'o') and second (green '+') return map $F$ for this case of parameters. Instability and period two orbits can be seen in the plot as $45^o$ line (blue '.') intersects with second return map. Fig. 5.1(b) has the rate waveform for a delay of one time unit which is not sufficient to send the system into the unstable mode, and hence both rates converge to their optimal value of 1.6637 . However, when the delay is increased to 10 time units, systems is already oscillating as shown in Fig. 5.1(c). The upper limit of 2.5 and lower limit of 0.7368 can be verified. Finally, in Fig. 5.1(d), which shows the same waveform for a delay of 50 time units, the waveform is more square-like compared to last figure. In the limit with increasing delay this waveform approaches a square waveform oscillating

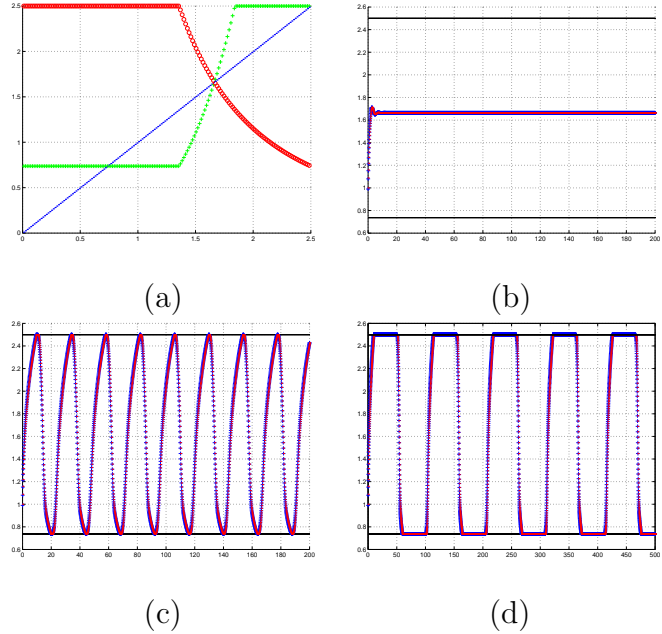

Figure 5.1: (a) Map given by eq. 5.24 for above scenario, (b) Rate waveform for the delay of 1 time unit, (c) for the delay of 10 time units, and (d) for the delay of 50 time units

between the period two orbit of corresponding map.

It is also evident that in both of the oscillating cases the period of the waveform is approximately twice of the delay and the interval between consecutive times when the waveforms cross $y(t) = y^* = 1.6637$ is more than the delay itself. Typically, these oscillating orbits are very difficult to describe as they vary from sinusoidal to square waves with increasing value of delay. This phenomena has been studied earlier in [12].

Clearly, these numerical solutions confirm the upper and lower limits for the trajectories for large enough communication delays. In particular, these periodic orbits remind of a particular periodic solution class devised specifically for delay-differential equations, namely *Slowly Oscillating Periodic (SOP)* orbits [57, 79].

Roughly, an SOP is a periodic orbit with its consecutive zeros (zero corresponds to the fixed point $y^*$ in our case) separated by more than one normalized time unit. The time unit used in our context corresponds to a round-trip time, which arises naturally as a measure for network performance and stability. This also supports the view that round-trip time may be the most useful time scale from the point of view of stability and oscillations [31]. For dynamical eq. 5.24 we have following conjecture regarding the existence of an SOP:

**Conjecture 1** *SOP: For all $0 < \nu < 1/T_0$, where $T_0$ is given by eq. 5.27 in linear stability context, eq. 5.23 has at least one slowly oscillating periodic solution with period $P(\nu) > 2$. Moreover, $T(\nu) \to 2$ as $\nu \to 0$.*

Although proving the existence of an SOP is technically complicated and its asymptotic behavior is even more challenging, we believe that these slowly oscillating periodic orbits are useful for the study of networks and networked control systems to understand the stability and oscillation behavior in the presence of non-negligible delays.

## 5.8 Multiple Users with One Bottleneck

In this section we extend the results for users with different utility functions of the form given by eq. 5.14 with different $a_i$. Following Kelly's rate control formulation, the rate of the $i$-th user evolves according to

$$\frac{d}{dt}x_i(t) = k_i \left(w_i(t) - x_i(t-T)\mu(t-T)\right) \tag{5.44}$$

$$= k_i \left(x_i(t)U'(x_i(t)) - x_i(t-T)p(\Sigma_j x_j(t-T))\right) \tag{5.45}$$

After normalizing time by $T$, and rescaling time as $t = s \cdot T$, and using the substitution $y_i = x_i U'(x_i) := g_i(x_i)$, the dynamic equation for the $i$−th user can be rewritten as

$$x_i(t) = g_i^{-1}(y_i(t)), \tag{5.46}$$

$$\dot{x}_i(t) = \frac{\dot{y}_i(t)}{g_i'(g_i^{-1}(y_i(t)))} \tag{5.47}$$

$$\nu \dot{y}_i(t) = k_i g_i'(g_i^{-1}(y_i(t)))(y_i(t) - f_i(g_1^{-1}(y_1(t-1)), ..., g_n^{-1}(y_n(t-1))))) \tag{5.48}$$

$$= k_i g_i'(g_i^{-1}(y_i(t)))(y_i(t) - f_i(\overline{g}^{-1}(\overline{y}(t-1)))) \tag{5.49}$$

where $\nu = \frac{1}{T}$, $f_i(\overline{g}^{-1}(\overline{y}(t-1))) = g_i^{-1}(y_i)p(\Sigma_j g_j^{-1}(y_j(t-1)))$, $n$ is the number of users in the system, and $\overline{y}(t-1) = (y_1(t-1), \cdots, y_n(t-1))$. Writing these equations in matrix form will show the striking structure they possess which is crucial to their convergence behavior:

$$\begin{bmatrix} \frac{1}{\nu}\dot{y}_1(t) = -k_1 g_1'(g_1^{-1}(y_1(t))) \left(f_1(\overline{g}^{-1}(\overline{y}(t-1))) - y_1(t)\right) \\ \frac{1}{\nu}\dot{y}_2(t) = -k_2 g_2'(g_2^{-1}(y_2(t))) \left(f_2(\overline{g}^{-1}(\overline{y}(t-1))) - y_2(t)\right) \\ \vdots \\ \frac{1}{\nu}\dot{y}_n(t) = -k_n g_n'(g_n^{-1}(y_n(t))) \left(f_n(\overline{g}^{-1}(\overline{y}(t-1))) - y_n(t)\right) \end{bmatrix} \tag{5.50}$$

Eq. 5.50 can be represented in following compact form:

$$\frac{1}{\nu}\dot{\overline{y}}(t) = \kappa(\overline{y}(t))\left(F(\overline{y}(t-1)) - \overline{y}(t)\right) \tag{5.51}$$

where $\kappa(\cdot)$ is state dependent diagonal gain matrix with $\kappa_{ii} = -k_i g_i'(g_i^{-1}(y_i(t)))$. Clearly, this decomposition is possible due to the fact that the utility of a user is a function only of its own rate and and does not depend on those of other users. It can be seen that in the wake of our assumptions $\kappa(\cdot)$ is strictly positive definite matrix, which turns out to be an important property to prove the convergence results for the system given by eq. 5.51. Again, $F(\cdot)$ given by eq. 5.52 is a multidimensional one step nonlinear map which is crucial for the understanding of the stability of the system. We note that this system of differential equations has a natural underlying difference map structure as seen earlier for uniform users case.

$$\overline{y}_{n+1} = F(\overline{y}_n), \ \ n \in Z_+, \ y_n \in R_+^n \text{ and } F_i(\overline{y}) = f_i(g_1^{-1}(y_1), ..., g_n^{-1}(y_n)) \quad (5.52)$$

The importance of this multidimensional one step map will be evident when it will be shown that stability of this map is a sufficient condition for the stability of the delya-differential system given by eq. 5.51. In addition to the stability, this map will also provide the insights in the instability behavior of this system which will be described by orbits of periods close to twice of underlying delay. This phenomena will generalize the existence of period doubling bifurcations in many familiar network setups.

### 5.8.1 Results

For multiple users with heterogenous utility function case we will need to prove the convergence in multidimensional space. This approach will be based again on the earlier approach given in a paper by Verriest and Ivanov [87]. The basic idea behind this approach is to use invariance and continuity of the underlying map for the differential equation and find bounds with intervals replaced by convex sets. Just like the one dimensional case one can construct multidimensional convex covers for the image of the convex set containing the initial functions. Dissipativity is derived from the dissipativity of the underlying map also, which provides the bounds for the trajectories of the delay-differential system. Following this plan, we will first prove the invariance of system given by eq. 5.51 if the underlying map given by eq. 5.52 has a convex invariance set. The assumption of existence of a convex invariance set is natural in wide parameter range for network rate control scenario as will be illustrated in next section. Our first result says that the set $C([-1, 0], D)$ is invariant under the action generated by eq. 5.51, given $D$ is closed, convex and invariant under $F$ given in eq. 5.52. Before we state the invariance theorem we need a proposition which establishes orthant invariance for a vector under the multiplication by a positive diagonal matrix.

**Proposition 1** *(Orthant-Invariance) For a diagonal positive matrix $K \in R^{nxn}$ and an arbitrary vector $v \in R^n$, $Rv$ remains in the same orthant as $v$.*

**Proof:** Any vector $v$ can be expressed as a linear combination of basis vectors $\{e_i\}$ or $v = \Sigma_i c_i e_i$. Clearly, due to the diagonal structure of positive matrix $K$, $Kv = \Sigma_i k_{ii} c_i e_i$ which means that all the coefficients $c_i$ retain their original sign even after the multiplication with $K$. This ensures the fact that they remain in the same orthant. ∎

Clearly, diagonal structure of positive gain matrix $\kappa(\cdot)$ is useful to determine the directions of right hand side of eq. 5.51. We note here that diagonal structure gives us more than required in the sense that a non-diagonal positive $\kappa(\cdot)$ is enough to ensure that the right hand side of eq. 5.51 stays directed towards the interior of a convex domain on any boundary point.

Next, we state the invariance theorem which establishes set bounds given that initial conditions lie in that particular set.

**Theorem 17** *(Invariance) If $D$ is a closed convex invariant domain under $F(\cdot)$, then for any initial function $\phi \in C([-1, 0], D) := X_D$ the solution of eq. 5.51 $\overline{y}^\nu(t)$ belongs to the domain $D$ for all $t \geq 0$ and $\nu \geq 0$.*

**Proof:** Let $t_0 \geq 0$ be the first time at which $\overline{y}^\nu(t_0) \in D$ and the solution $\overline{y}^\nu(t)$ leaves the domain $D$ for $t \geq t_0$. So, both $\dot{\overline{y}}^\nu(t_0)$ and $\nu \ddot{\overline{y}}^\nu(t_0)$ will be directed outside the domain. Now, since $t_0$ is first such point at which the trajectory leaves the domain $D$, the $F[\overline{y}^\nu(t_0 - 1)]$ lies within the domain $D$. Therefore, the vector $F[\overline{y}^\nu(t_0 - 1)] - \overline{y}^\nu(t_0)$ and $k(\overline{y}^\nu(t_0))[F[\overline{y}^\nu(t_0 - 1)] - \overline{y}^\nu(t_0)]$ will both be directed towards the inside of the domain $D$ because diagonal $k(\overline{y}^\nu(t_0)) > 0$ and $D$ is convex. This holds because of the orthant-invariance property of vectors under the multiplication with positive diagonal matrix as shown in prop. 1. But, $\nu \dot{\overline{y}}^\nu(t_0)$ and $k(\overline{y}^\nu(t_0))[F[\overline{y}^\nu(t_0 - 1)] - \overline{y}^\nu(t_0)]$ vectors are equal due to dynamical eq. 5.51 which is a contradiction. Hence, we have the proof. ∎

Again, we note that the monotone property of utility functions over the domain $D$ and individual optimization policies are crucial to the proof of this theorem. Now that invariance is established, we want to understand the asymptotic stability property of delay-differential equation under the natural assumption that the underlying map is asymptotically stable. The basic idea here is to look at the image $F(D)$ of initial convex set $D$ which may not be convex as will be shown through an example in Fig. 5.2. Note that the image is contained in a rectangular box.

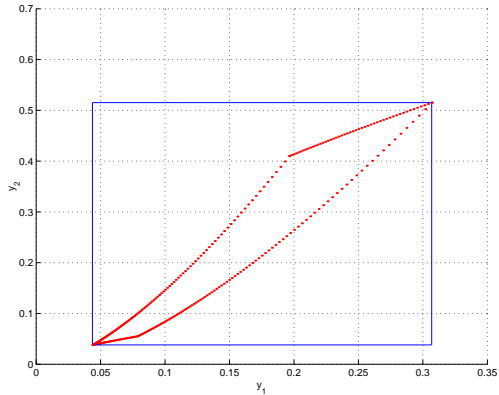

Figure 5.2: Non-convex image of a rectangle under map given by eq. 5.52 with $a_1 = 20, a_2 = 5, b = 3, C = 5$

This lack of convexity forbids the direct application of results developed by Verriest and Ivanov [87]. Instead our approach is to construct a series of convex coverings of image $F^i(D_i)$ and look at their asymptotic behavior. In particular, we would like to construct rectangular boxes which seem to be the most reasonable choice for the networking problems. These boxes are obviously convex, and if we can prove that they contain their own images then we have invariance. Also, due to the monotone property of map $F(\cdot)$ the co-ordinates of these constructed boxes and their images can be computed explicitly. So, again to recapitulate we want to construct a series of convex closed domains $\{D_i\}$ such that under certain stability conditions $F(D_i) \subseteq D_{i+1} \subseteq D_i$. Clearly, if $y^* = \cap_{n \geq 0} D_i$ is a point then all the solutions of the map given by eq. 5.52 with initial condition in $D_0$ will converge to $y^*$ asymptotically.

**Assumption 5** *Multidimensional map $F : R^n \to R^n$ has an arbitrary fixed point $y^*$ and there exists an open convex neighborhood $int(D_0)$ where $int(V)$ denotes the interior of set $V$. Also, assume that there is a sequence of closed convex domains $D_i, i \geq 0$ such that $F(D_i) \subseteq D_{i+1} \subseteq D_i$ and $y^* = \cup_{n \geq 0} D_i$.*

Let $Y_{D_0} = C([-1, 0], int(D_0))$ be a subset of initial functions and $\overline{y}_\phi^\nu$ a solution of eq. 5.51 constructed through $\phi \in Y_{D_0}$.

**Theorem 18** *(Dissipation) All solutions starting with initial functions $\phi \in Y_{D_0}$ converge to $y^*$ for all $\nu > 0$.*

This establishes that the attracting fixed point is stable in set $D_0$. We will first prove a lemma which will be used to prove the theorem.

**Lemma 8** *Let $V$ be an arbitrary open convex set containing a domain $D_1 \supset F(D_0)$ and contained in a domain $D_0$ and arbitrary initial data $\phi \in Y_{D_0}$. If $\phi(0)$ is in* *closure of the set $V$, $cl(V)$, then $\overline{y}_\phi^\nu$ is in the closure of $V$ for all $t \geq 0$.*

*If $\phi(0)$ is not in the closure of $V$ then there exists a time $t_0 = t_0(\phi, D_0, \kappa(\cdot))$ such that $\overline{y}_\phi^\nu \in \partial V$ and $\overline{y}_\phi^\nu \in cl(V)$ for all $t \geq t_0$ with $\partial V$ denoting the boundary of $V$.*

**Proof:** Assume $\phi(0) \in cl(V)$. As $cl(V)$ is a closed convex set and $V \supset F(D_0)$ implies $F(cl(V)) \subset cl(V)$. Hence, $\overline{y}_\phi^\nu \in cl(V)$ for all $t \geq t_0$ follows from the direct application of invariance theorem (Theorem 17).

Now suppose that $\phi(0) \notin cl(V)$ though $\overline{y}_\phi^\nu \in D_0$ from the last invariance theorem. If we have a first time $t_0$ such that $\overline{y}_\phi^\nu(t_0) \in \partial V$, then $\overline{y}_\phi^\nu \in cl(V)$ for all $t \geq t_0$. This is achieved essentially along the same last lines as last theorem.

Suppose that we begin with $\phi(0) \notin cl(V)$ and $\overline{y}_\phi^\nu \notin cl(V)$ for all time $t \geq 0$. Also, let $M$ be the maximal open convex set containing $V$ such that $M \cap cl(\{\overline{y}_\phi^\nu(t), t \geq 0\}) = \emptyset$ or intersection of $M$ and the closure of trajectories is an empty set. It is possible for $M$ to coincide with $V$ if the trajectory never enters $V$. Since $M$ is maximum such set there exists a sequence $t_i, i = 1, 2, ..$ such that $\overline{y}_\phi^\nu(t_i) \to y_0 \in \partial M$. Now there will be a sequence of finite vectors $\overline{r}_i = \frac{\kappa(\overline{y}_\phi^\nu(t_i))}{\nu}\left[F(\overline{y}_\phi^\nu(t_i - 1)) - \overline{y}_\phi^\nu(t_i)\right]$ which form the right hand side of eq. 5.51 at these time instants. They are bounded since $M \subset D_0$ and $F(M) \subset D_1$. Hence, all these $r_i$ are bounded away from zero and are directed strictly towards the inside of domain $M$. As the origins of the finite vectors $\overline{r}_i$ converge to the point $y_0 \in \partial M$ and $\overline{r}_i$ is tangent to the solution vectors $(\overline{y}_\phi^\nu, i = 1, \cdots, n)$ at $t = t_i$, and $|r_i| \geq \delta$ for some $\delta$, there exists $t_M$ such that $\overline{y}_\phi^\nu(t') \in M$ for some $t' \geq t_M$ or solutions are definitely going to enter the set $M$ which is a contradiction. Hence the proof. ∎

Now we can provide an easy proof for Theorem 18.

**Proof:** Since it is possible to construct a sequence of nested convex open neighborhoods $\mathcal{U}_{i+1} \subseteq \mathcal{U}_i$ such that $F(\mathcal{U}_i) \subseteq \mathcal{U}_{i+1}$ and $\cap_{i \geq 0} \mathcal{U}_i = y^*$. By repeated application of Lemma 8 there exists a sequence $\{t_i\} \to \infty$ such that $\overline{y}_\phi^\nu(t_i) \in cl(\mathcal{U}_i)$ and it stays there due to the invariance theorem for $t \geq t_i$. As $\cap_{i \geq 0} \mathcal{U}_i = y^*$, asymptotic convergence and implied stability for $y^*$ is achieved. ∎

Now we return to our networking problem with multiple users. We want to study the stability for the case when all users have their utility parameters $a_i > b+1$. First we will analyze the uniform case where bounds on the closed convex set can be computed analytically.

**Example 1** *We consider the case of two users with the same utility parameters $a > b+1$. From the one dimensional case of uniform users we know that this is a stable case. We want to analyze this example in light of Theorem 18. So we want to construct a sequence of nested interval $F(D_i) \subseteq D_{i+1} \subseteq D_i$ and $y^* = \cup_{n \geq 0} D_i$. We begin with a rectangular box $D_0$ with upper right corner at $d_0^u = \{\frac{c}{2}, \frac{c}{2}\}$ and lower left corner at $d_0^l = \{F_1(c/2), F_2(c/2)\}$. To compute the image of this box under $F$ we note that both $F_1$ and $F_2$ are strictly decreasing over convex set $D_0$. In particular it will have its maxima at $d_0^l$ and minima at $d_0^u$. This construction gives us ideas for $D_1$ which can be constructed with $d_1^u = \{F_1^2(c/2), F_2^2(c/2)\}$ and $d_1^l = d_0^l = \{F_1(c/2), F_2(c/2)\}$. Similarly, we can construct a series of nested boxes $D_i$ who vertices will have co-ordinates of the form $F^k(C/2)$ for some positive integer $k$. From the stability results of one dimension we know that $\cup_{n \geq 0} D_i = \{y^*, y^*\}$. Hence, Theorem 18 is applicable and we have the asymptotic stability in initial set $D_0$.*

Essentially what we see here is that the question of stability of a delay-differential equation has been reduced to the question of stability for a one step multi-dimensional map. Multi-dimensional difference equations get very complicated and there are very few tools to analyze them compared to one-dimensional maps which have been studied extensively as noted in [39]. As we will see in the next example where we take two users with diverse $a_i > b + 1$. Unfortunately it is hard to compute even the fixed points and Jacobian in closed form for this case.

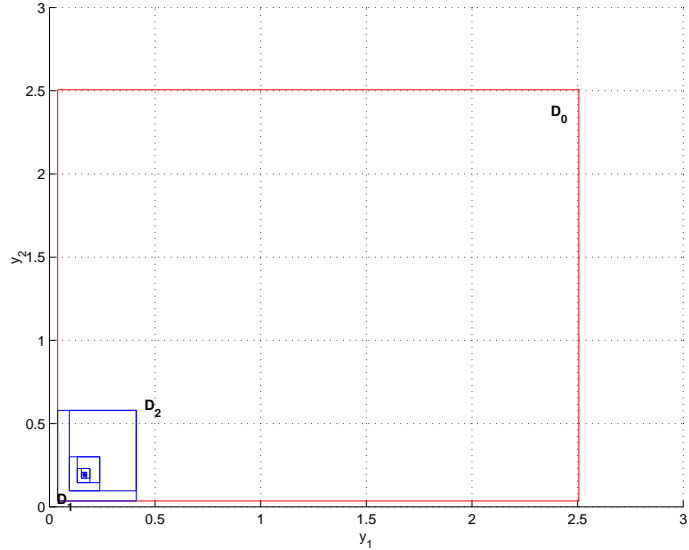

Figure 5.3: Nested convex domains constructed for example 2 for $a_1 = 10, a_2 = 5, b = 3, C = 5$

**Example 2** *Consider the situation as in example 1 but with different $a_i > b + 1$. Again we consider similar rectangular constructions but we do not have any analytic expressions to characterize them as compared to uniform case. Hence, we will plot these boxes numerically and illustrate that their intersection seems to be the fixed point of the system. A simple case for $a_1 = 10$ and $a_2 = 5$ with $b = 3$ is shown in Fig. 5.3.*

## 5.9 Conclusion

We show that dynamical stability of a rate control mechanism between users and a resource is determined by the interaction of underlying utility and price functions. In particular, we show that any mismatch in the users' utility and resource price function leads to network instability. We explicitly characterize this instability for the class of utility functions used.

Results in this chapter extend the earlier local stability results of delay differential equations in networking context for simple cases when user's weight is fixed in [39]. Compared to the results in [39], we treat the system with full nonlinearity and provide global results for one-dimensional cases. Global results for the multi-dimensional case seem intuitive but need more work.

We also note that the discrete-time framework arises as a natural tool to study the dynamics of delayed rate control schemes. It also indicates the structure of periodic trajectories and their bounds.

Models has been shown to give interesting but intuitive results for the presence of non-responsive traffic and confirm that indeed they stabilize the system.

Finally, we believe that SOP orbits may be very relevant to study of the structure of periodic orbits arising in engineering applications. These periodic orbits have been studied extensively in the mathematics community and also arise when the delay is state-dependent, which is a useful context in networking [31]. These SOP orbits support the earlier belief that the round-trip time may be the most relevant time scale for network stability studies.

# Chapter 6

# Wash-out Filter Enabled Control of RED

## 6.1 Introduction

The interaction of nonlinearities and delay in a network can have severe consequences for the network performance. For a robust operation of the network, it is important to understand its dynamical behavior beyond linear stability regime. In chapter 3 and chapter 5 we have shown that for a class of diverse looking models like that of Firoiu and Borden [21] and for Kelly's rate control framework [41] the natural mode of transition from fixed point operation to oscillations is through a period doubling bifurcation in naturally arising discrete-time maps. In this work we look for methods to control these oscillations by delaying and stabilizing the period doubling bifurcations locally in the parameter space.

Nonlinear instability and its control is a well studied area in control system literature [1, 13, 67]. In this chapter, we utilize the theory of bifurcation control using washout filters and other feedback based chaos control techniques to extend the stable parameter range of TCP-RED. The basic control paradigm which will be followed here is the stabilization of the fixed point that changes stability with parameter variation rather than introduction of new operating points. This essentially means that control becomes active only when the original fixed point of the system transitions into unstable regime. Both linear and nonlinear control formulations will be used to achieve this objective. Also, we view the problem as a bifurcation stabilization rather than stabilization at fixed parameters. This is important due to the fact that generally bifurcation reappears for different values of bifurcation parameter. For this purpose nonlinear control terms help in stabilization of reappearing bifurcation as parameters change.

Washout filter is a simple high-pass filter and has been shown to be very effective for robust control of nonlinear instabilities [1]. The basic idea behind this control scheme is to stabilize the original equilibrium point by feedback based, small parametric modulations introduced in RED parameters e.g. $p_{max}$ or $q_{max}$. It cancels the the instability effect introduced due to significant variation in other system parameters like number of connections ($N$) or round trip propagation delay ($R_0$) etc. Washout filter ensures the normal RED operation when system is operating in stable mode. Hence, washout filter augments the capability of RED by diluting its infamous parametric sensitivity. Linear control terms are used to delay the bifurcation whereas, nonlinear control terms are utilized to suppress the amplitude of oscillations in the case of instability [1]. Essentially, it makes TCP-RED interaction coherent in larger parametric region.

This chapter is organized as follows. In Section 6.2, we recall the first-order model for TCP-RED in a congested network from chapter 3 and its linear analysis. In Section 6.3, we outline the theory behind washout filter based control.

Section 6.4 contains the application of wash-out filter control theory in the context of TCP-RED. Section 6.5 contains the numerical simulations showing the bifurcation delay in parameter range and also ns-2 implementation of washout filter based control scheme.

## 6.2 Model

Before proceeding, let us recall the basic TCP-RED model from chapter 3:

$$\overline{q}_{e,k+1} = \begin{cases} (1-w)\overline{q}_{e,k} & \text{if } \overline{q}_{e,k} > b_1 \\ (1-w)\overline{q}_{e,k} + wB & \text{if } \overline{q}_{e,k} < b_2 \\ (1-w)\overline{q}_{e,k} + \\ \quad + w\left(\dfrac{NK}{\sqrt{\dfrac{(\overline{q}_{e,k}-q_{min})p_{max}}{(q_{max}-q_{min})}}} - \dfrac{R_0 C}{M}\right) & \text{otherwise} \end{cases}$$

$$= : f(\overline{q}_{e,k}, \rho) \qquad (6.1)$$

## 6.3 Feedback Control of Instabilities

In this section, we illustrate a simple delayed feedback control algorithm to control instabilities [69, 1]. The basic idea behind this control is to modulate system parameters by feeding back a function of the difference between the state and the desired fixed point [67]. Let the system model be given by the following equations with bifurcation parameter $\rho$ and $f : \quad \times \quad \rightarrow \quad$.

$$x_{n+1} = f(x_n, \rho) \qquad (6.2)$$

If a system described by eq.(6.2) is amenable to an additive control on the right hand side, then the modified system with additive linear feedback control term can be rewritten as:

$$x_{n+1} = f(x_n, \rho) + k(x_n - x^*(\rho)) \qquad (6.3)$$

where $x^*(\rho)$ is the fixed point of original system given by eq.(6.2) and $k$ is the linear feedback gain. It is clear that this kind of stabilization scheme leaves the fixed points of the system unchanged. Control is required only to enhance the stability of the naturally occurring fixed point $x^*(\rho)$ in case of instability. Also, the control effort is very small if the system starts in the vicinity of the original fixed point. The main challenge for this scheme is computation of the fixed point near or after bifurcation for a given system and bifurcation parameter. This problem is particularly difficult when the model may contain significant amount of uncertainty.

There are many ways to get around the computation of the fixed point by introducing moving average type filters to estimate the fixed point [13] or using washout filters [1] to maintain the fixed point despite uncertainty. The fixed point does not need to be computed exactly because with proper design of feedback control it is possible to drive the system closer to the approximate fixed point, which in turn produces more accurate estimates of the fixed point. In this setup one has to look into the stability of augmented system of fixed point estimator coupled with original plant process.

The second problem is the mode of actuation where the control can appear. Generally, this problem is solved by modulating a system parameter to cancel the effect of change in another system parameter.

Rewriting eq. 6.2 along with moving average fixed point estimation gives:

$$x_n^{*e} = \frac{1}{N}(\Sigma_{i=n-N+1}^{n} x_i^n) \tag{6.4}$$

$$x_{n+1} = f(x_n, \rho) + \gamma_0 + k(x_n - x_n^{*e}(\rho)) \tag{6.5}$$

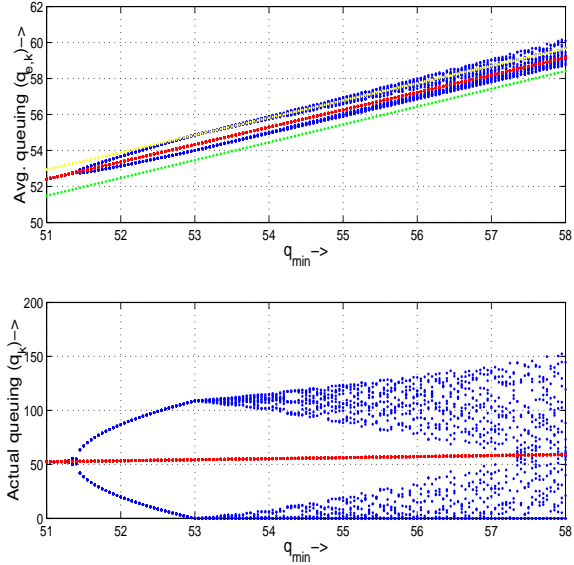

Figure 6.1: Bifurcation diagram with and without control by modulating $p_{max}$ with respect to $q_{min}$, $p_{max}^{nominal} = 0.3$, $N = 4$ and $k = 0.05$. Bifurcation diagrams in blue and red are plotted without and with control respectively.

Although control enters additively on the right side of the plant equation in most theoretical control settings, here control is administered by modulating system parameters. This reflects in eq. 6.5 as an addition of the feedback control signal to $\gamma_0$ where $\gamma_0$ is the parameter modulated to cancel the effect of changes in some other parameter $\rho$.

Control schemes based on these ideas are the motivation to modulate RED parameters like $p_{max}$, $q_{max}$ etc. in a TCP-RED model as described by eq.(6.1) to control the nonlinear instabilities leading to chaos. We take an indirect approach to extend the stable domain of operation by which we intend to control the first instability (period doubling bifurcation) to delay the bifurcation sequence in parameter space as much as possible.

We provide a numerical example to show how this kind of control renders the system stable and eliminates the load batching. We plot the bifurcation diagram in fig. 6.1 with and without control. Control is actuated by modulating $p_{max}$ in this scenario. It is clear from the plot that a significant bifurcation delay in the parameter can be achieved by this simple control mechanism. The basic idea behind this linear control mechanism is to delay the occurrence of bifurcation. Certain nonlinear controls described in the next section can be used to stabilize without changing the critical parameter value, and a combination of linear and nonlinear control terms achieve both delay in parameter space and stabilization of ensuing bifurcations and hence reduction in the amplitude of oscillations.

### 6.3.1 Washout filter based control

The washout filter mechanism has been successfully utilized to control a number of bifurcations in nonlinear models with uncertainty [1]. The basic idea behind using this mechanism for instability control is to preserve the equilibrium points of the given system. This approach for TCP-RED systems differs considerably from other schemes as wash-out filter based the control scheme tries to keep the operating point invariant under significant parametric variations [23, 3, 31]. Although the adaptive RED scheme also modulates $p_{max}$ to adapt to the changed operating conditions using a additive increase and multiplicative decrease (AIMD) algorithm, the thresholds for switching is fixed and tries to keep system operation independent of other parameter variations.

A simple discrete time high-pass filter can be used as an analogue of washout filter in continuous time. Consider the following high-pass filter discussed in [1]

$$G(z) = \frac{1 - z^{-1}}{1 - (1-d)z^{-1}} \tag{6.6}$$

This can have the following time domain implementation:

$$z_{k+1} = x_k + (1-d)z_k \tag{6.7}$$

$$y_k = x_k - dz_k \tag{6.8}$$

where $\{x_k\}$ is the input sequence to the washout filter, $\{y_k\}$ is the output sequence, and the washout filter constant $d$ should satisfy $0 < d < 2$. At steady state, $z_{k+1} = z_k$ so that by substitution into eq. 6.8 we have

$$0 = x_{eq} - dz_{eq} \tag{6.9}$$

Hence, from eq. 6.8 and eq. 6.8, $y_k \equiv 0$ at steady state. Thus, at steady state the output of the washout filter vanishes.

Now, we can consider a scalar nonlinear dynamical system with washout filter control:

$$x_{k+1} = f(x_k, u_k) \tag{6.10}$$

where $u_k$ is a scalar control input. If washout filter is put in the feedback loop with feedback function $h(\cdot)$, we have following modified system:

$$x_{k+1} = f(x_k, u_k) \tag{6.11}$$

$$z_{k+1} = x_k + (1-d)z_k \tag{6.12}$$

$$y_k = x_k - dz_k \tag{6.13}$$

$$u_k = h(y_k) \tag{6.14}$$

where $h : R \to R$ is any smooth function such that $h(0) = 0$). It can be shown that this type of feedback control does not modify the equilibrium point of the original system under no control ($u_k = 0$) [1]. However, with proper choice of feedback function $h(\cdot)$ and washout filter constant, it can enhance the stability of the original equilibrium point without need for accurate knowledge of the system model or equilibrium value.

## 6.4 Application to TCP-RED

In this section we look at the stabilization of map in eq. 6.2 with linear control terms in the neighborhood of fixed point $\overline{q}^*$. For this we need to compute the linearization of the map ($x_{n+1} = Ax_n + bu_n$) as a function of fixed point and system parameters.

$$\left.\frac{\partial f(\overline{q}_{e,k+1}, \rho)}{\partial \overline{q}_{e,k+1}}\right|_{\overline{q}_{e,k+1}=q^*} = 1 - w - \frac{0.5wNK}{(q^* - q_{min})^{\frac{3}{2}}}\sqrt{\frac{q_{max} - q_{min}}{p_{max}}}$$
$$:= \lambda_0(\rho) \qquad (6.15)$$

Depending on the RED parameter to be modulated, $b(p_{max})$ or $b(q_{max})$ can be computed. For one dimensional system with nonzero eigenvalue, both left ($l$) and right ($r$) eigenvectors are 1.

$$b(p_{max}) = \frac{\partial f}{\partial p_{max}} = -\frac{0.5wNK}{\sqrt{\frac{(\overline{q}_{e,k} - q_{min})}{(q_{max} - q_{min})}}p_{max}^{1.5}} \qquad (6.16)$$

$$b(q_{max}) = \frac{\partial f}{\partial q_{max}} = \frac{0.5wNK}{\sqrt{(\overline{q}_{e,k} - q_{min})(q_{max} - q_{min})p_{max}}} \qquad (6.17)$$

It is clear from the expressions of $b(\cdot)$ and nominal parameter ranges that $b(\cdot) \neq 0$.

From above two observations we conclude that $lb(\cdot) \neq 0$. This fact has consequences for linear stabilizability due to Popov-Belevitch-Hautus (PBH) eigenvector test for controllability of modes of linear time invariant systems [40]. Hence a linear stabilizing feedback exists in this case. This also means that a cubic feedback exists which we will see later in the nonlinear control section.

In the view of PBH test for controllability and wash out filter described above, we can view the RED exponentially averaged queue as input to the state estimation filter which provides the estimate $y_k$. This estimate can be used to construct the control depending on the functional form of $h$. In this section we consider only the linear control law because in linear analysis all the nonlinear terms vanish when system is linearized at the fixed point.

$$u_k = k_l y_k \text{ Linear Control Law}$$

In this framework, the TCP-RED system given by eq. 6.1 when augmented by washout filter, can be rewritten as follows:

$$z_{k+1} = \overline{q}_{e,k} + (1-d)z_k \tag{6.18}$$

$$u_k = h(\overline{q}_{e,k} - dz_k) \tag{6.19}$$

$$\overline{q}_{e,k+1} = \begin{cases} (1-w)\overline{q}_{e,k} & \text{if } \overline{q}_{e,k} > b_1 \\ (1-w)\overline{q}_{e,k} + wB & \text{if } \overline{q}_{e,k} < b_2 \\ (1-w)\overline{q}_{e,k} + \\ +w(\frac{NK}{\sqrt{\frac{(\overline{q}_{e,k}-q_{min})p_{max}}{(q_{max}-q_{min})}}} - \frac{R_0 C}{M}) & \text{otherwise} \end{cases} \tag{6.20}$$

$$\text{where } p_{max} = \min\{0.5, \max\{p_0, p_{max}^0 + u_k\}\}$$

where $p_{max}^0$ is original drop probability of the system and $u_k$ which is a function of the out of the washout filter given by eq.(6.14) depends on the particular functional form of the control employed which is linear and linear+cubic in this work. We also need to limit the drop probability by 0.5 from above due to the consideration of GENTLE_ mode of RED and to a small constant $p_0$.

Similarly, other parameter modulations like that of $q_{max}$ can be considered in eq. 6.21 with appropriate bounds on actuation.

### 6.4.1 Stability Analysis with Washout Filter

In this section we analyze the stability of washout enabled TCP-RED given by eq.( 6.21). Clearly, $[q^*/d, q^*]$ is the fixed point for the new system for $d \neq 0$. The Jacobian which will be evaluated at fixed point $[q^*/d, q^*]$ is given as follows:

$$A = \begin{pmatrix} 1-d & 1 \\ b\frac{\partial h(\bar{q}_{e,k}-dz_k)}{\partial z_k} & \frac{\partial f(\bar{q}_{e,k},\rho)}{\partial \bar{q}_{e,k}} + b\frac{\partial h(\bar{q}_{e,k}-dz_k)}{\partial \bar{q}_{e,k}} \end{pmatrix} \quad (6.21)$$

where $b$ is either $b(p_{max})$ or $b(q_{max})$ depending on the parameter to be modulated.

For a linear control, i. e. $u_k = k_l(\bar{q}_{e,k} - dz_k)$ and evaluation at fixed point $[q^*/d, q^*]$ eq. 6.21 reads:

$$A = \begin{pmatrix} 1-d & 1 \\ -dbk_l & \lambda_0 + bk_l \end{pmatrix} \quad (6.22)$$

where $\lambda_0 = \frac{\partial f(\bar{q}_{e,k},\rho)}{\partial \bar{q}_{e,k}}$ when evaluated at $q^*(\rho)$ as given by eq. 6.15

Next we recall Jury's stability test for second order discrete-time systems:

**Lemma 9** *(Jury's stability test for second order systems [54]).*

*A necessary and sufficient condition for the zeros of the polynomial*

$$p(\lambda) = a_2\lambda^2 + a_1\lambda + a_0$$

*($a_2 > 0$) to lie within unit circle is*

$$p(1) > 0,\ p(-1) > 0\ \ and\ \ |a_0| < a_2$$

The characteristics equation for matrix given by eq. 6.22 can be written as follows:

$$(\lambda - (1-d))(\lambda - (\lambda_0 + bk_l)) + dbk_l = 0$$
$$\Rightarrow \lambda^2 - \lambda((1-d) + \lambda_0 + bk_l) + (1-d)(\lambda_0 + bk_l) + dbk_l = 0$$
$$\Rightarrow \lambda^2 - \lambda((1-d) + \lambda_0 + bk_l) + (1-d)\lambda_0 + bk_l = 0$$

Using Jury's test for stability, the conditions for linear asymptotic stability can be given as follows.

$$d(1 - \lambda_0) > 0 \tag{6.23}$$

$$2 + 2bk_l + 2\lambda_0 - d(1 + \lambda_0) > 0$$
$$\Rightarrow k_l > \frac{(d-2)(1+\lambda_0)}{2b} \text{ for } b > 0. \tag{6.24}$$

$$|\lambda_0(1-d) + bk_l| < 1$$
$$\Rightarrow \frac{-1 - \lambda_0(1-d)}{b} < k_l < \frac{1 - \lambda_0(1-d)}{b} \text{ for } b > 0. \tag{6.25}$$

Similar inequalities can be formulated for linear stability in the case of $b < 0$. As we see here the stability region for pair $(d, k_l)$ is made up of three straight lines in $(d, k)$ plane which are described below:

$$(l_1) \qquad k = \frac{(1+\lambda_0)d}{2b} - \frac{1+\lambda_0}{b} \tag{6.26}$$

$$(l_2) \qquad k = \frac{\lambda_0 d}{b} - \frac{(1+\lambda_0)}{b} \tag{6.27}$$

$$(l_3) \qquad k = \frac{\lambda_0 d}{b} + \frac{(1-\lambda_0)}{b} \tag{6.28}$$

Under the generic assumption of $\lambda_0 < -1$ and $b > 0$, we can see that lines $(l_2)$ and $(l_3)$ are parallel as they have the same slope. Lines $(l_1)$ and $(l_3)$ intersect each other at $(d_0, k_0) = (\frac{4}{1-\lambda_0}, \frac{(1+\lambda_0)^2}{(1-\lambda_0)b})$. Also, lines $(l_1)$ and $(l_2)$ intersect each other at $(d_1, k_1) = (0, \frac{(1+\lambda_0)}{b})$. This essentially means that for a $(d, k)$ pair to be stabilizing, it should lie within the triangle made up by the vertices $(0, \frac{(1+\lambda_0)}{b})$, $(0, \frac{(1-\lambda_0)}{b})$, $(\frac{4}{1-\lambda_0}, \frac{(1+\lambda_0)^2}{(1-\lambda_0)b})$ as shown by shaded region in fig. 6.2.

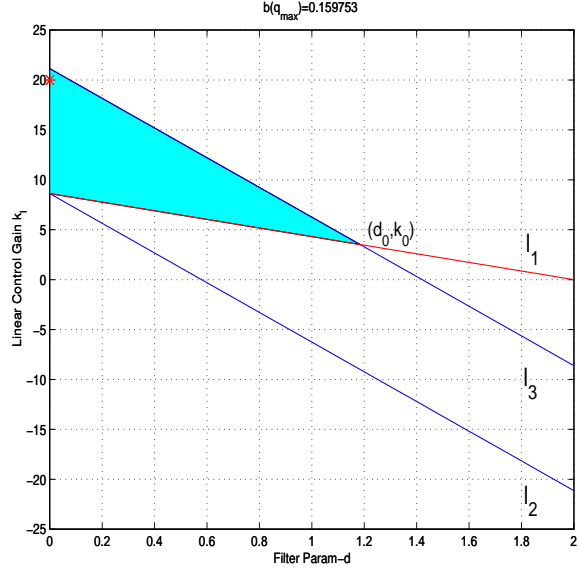

Figure 6.2: Control Region for washout augmented TCP-RED

**Proposition 2** *For a $(d,k)$ pair to be stabilizing, it should lie within the triangle made up by the vertices $(0, \frac{(1+\lambda_0)}{b})$, $(0, \frac{(1-\lambda_0)}{b})$, $(\frac{4}{1-\lambda_0}, \frac{(1+\lambda_0)^2}{(1-\lambda_0)b})$*

**Proof:** It can be easily verified that all the three conditions given by Jury's test in eq. 6.24, eq. 6.25 and eq. 6.25 are satisfied inside the above mentioned triangle. ∎

Parameter to be modulated for control will determine the value of $b$, e. g. $b < 0$ for $p_{max}$ but $b > 0$ for $q_{max}$. Gain $k_l$ needs to be chosen accordingly. Parameter $d$ is chosen such that $0 < d < 2$ and $\lambda_0 < -1$ in the regime after period doubling bifurcation. This shows that theoretically, it is possible to control the RED exponentially averaged queue locally near critical parameter value though allowable range for parameters likes $p_{max}$ or $q_{max}$ is severely limited by physical constraints in the real system. Also, these control gains need to be limited as to not cross the basin of attraction for the fixed point. Hence, though local stabilization near critical value of a parameter is possible, it may not be possible to stabilize in an arbitrary large parameter range. Next, using Jury's test we want to compute the parameter range where stabilization is possible for a fixed value of $k_l$ as different parameters like exponential averaging weight $w$, round trip time $R$ and number of active connection $N$ are varied.

### 6.4.2 Stabilization with respect to exponential averaging weight $w$

Stabilization with respect to exponential averaging weight $w$ is simpler to analyze since fixed point is independent of $w$ and eigenvalue $\lambda_0$ linearly decreases with respect to $w$. Hence, due to linear stabilizability of the original system ($lb \neq 0$) it is possible to stabilize the RED averaged queue by picking appropriate $k_l$ and $d$ which obey the conditions given by eqs. 6.24, 6.25 and 6.25. What we want to look for is the possibility of local linear stabilizing over all value of $0 < w < 1$. It turns out that due to some interesting properties of $\lambda_0$ and $b = b(q_{max})$ as given by eq. 6.17 it is possible to pick a $(d, k_l)$ pair to stabilize the system for all possible values of $w > w_{crit}$ where $w_{crit}$ is the value of $w$ at which first period doubling bifurcation happens in the uncontrolled system as given by eq. 3.19.

**Proposition 3** $\frac{(1-\lambda_0)}{b(q_{max})}$ *is independent of $w$.*

**Proof:** This can be verified by directly plugging the value of $\lambda_0$ from eq. 6.15 and $b(q_{max})$ from eq. 6.17. ∎

This provides important insight into the locus of triangular stability region as given by proposition 2. It shows that one of the vertices $(0, \frac{(1-\lambda_0)}{b})$ does not move as $w$ changes. Now, we need to understand the behavior of $\lambda_0$ and $b(q_{max})$ and that of $\frac{\lambda_0}{b(q_{max})}$ as $w$ is varied in unit interval. It is clear from eq. 6.15 that $\lambda_0$ decreases linearly as a function of $w$. Similarly, $b(q_{max})$ as given by eq. 6.17 increases linearly with $w$ but $\frac{\lambda_0}{b(q_{max})}$ is strictly decreasing with respect to $w$ which can be seen directly by differentiating the expression or using proposition 3. This means that all the three constraint lines given by eq. 6.27, eq. 6.28 and eq. 6.28 are going to get steeper with the increasing $w$. This leads to the squeeze of stability triangle. Finally, we use the fact that $w$ is bounded by one from above and evaluate the worst case stability region. Clearly, the eigenvalue remains finite even for $w=1$. Evaluating the vertices for $\lambda_0$ evaluated for $w = 1$ will provide the smallest triangle. Hence, if the stabilizing pair $(d, k_l)$ lies within this triangle then it will be stabilizing for all other values of $w > w_{crit}$.

**Theorem 19** *TCP-RED system along with washout filter for a given washout control parameter and linear control gain pair $(d, k_l)$ and all other parameters held fixed, will be stable for $w_{crit} < w < 1$ where $w_{crit}$ is value of $w$ corresponding to the first period doubling bifurcation, if $(d, k_l)$ lies within the triangle with vertices $(0, \frac{(1+\lambda_0(w=1))}{b}), (0, \frac{(1-\lambda_0(w=1))}{b}), (\frac{4}{1-\lambda_0(w=1)}, \frac{(1+\lambda_0(w=1))^2}{(1-\lambda_0(w=1))b})$ with $b = b(q_{max})$.*

**Proof:**   The proof follows from the simple application of Jury's test and proposition 2. ∎

### 6.4.3 Stabilization with respect to round trip time $R$

Round trip time $R$ is one of the crucial parameters in any networking scheme for stability purposes. In general $R$ tends to fluctuate a lot and stabilization with respect to the variation in $R$ is extremely important. TCP community has spent considerable amount of effort in the understanding the effect of round trip time and trying to design stable algorithms with respect to the variations in $R$. In washout filter control scheme, we achieve this goal by linear feedback modulation in RED parameters and thereby increasing the stable operation domain. Based on the Jury's criteria we state the following result:

**Theorem 20** *TCP-RED system along with washout filter for a given linear control gain $k_l > 0$ for $b(\cdot) > 0$ and all other parameters held fixed, will be linearly asymptotically stable for $R < R_0$ where $R_0$ is given as a solution of the following equation:*

$$\frac{(d-2)(1+\lambda_0(R_0))}{2b} = k_l \ \ for \ b > 0 \ and \ 0 < d < 2. \tag{6.29}$$

Solution to eq. 6.29 exists due to the fact that with increasing $R$ fixed point of the map decreases which makes the eigenvalue $\lambda_0(R)$ decrease monotonically in the parameter regime of interest.

### 6.4.4 Stabilization with respect to number of connections $N$

Number of connections $N$ another parameter which is beyond the control of network administrators. In general $N$ may vary wildly and stabilization with respect to the variation in $R$ is an important issue. In washout filter control scheme, we achieve this invariance by linear feedback modulation in RED parameters and thereby increasing the stable operation domain. Based on the Jury's criteria we state the following result:

**Theorem 21** *TCP-RED system along with washout filter for a given linear con-*

trol gain $k_l > 0$ for $b(\cdot) > 0$ and all other parameters held fixed, will be linearly asymptotically stable for $N > N_0$ where $N_0$ is given as a solution of the following equation:

$$\frac{(d-2)(1+\lambda_0(N_0))}{2b} = k_l \text{ for } b > 0 \text{ and } 0 < d < 2. \quad (6.30)$$

Solution to eq. 6.30 exists due to the fact that with decreasing $N$ fixed point of the map decreases which makes the eigenvalue $\lambda_0(N)$ decrease monotonically in the parameter regime of interest.

### 6.4.5 Nonlinear Control

It is possible to use small nonlinear control terms to further enhance the stability of a system going through period doubling bifurcation. Here we first recall the nonlinear control theorem given in [1] for local control of period doubling bifurcation. Before that we need the following hypothesis:

**Hypothesis 1** *Eq. 6.2 has a period-1 orbit at $x^*(\rho^*)$ where $x^*(\rho^*)$ is the fixed point at critical parameter value $\rho^*$. Furthermore, the linearization of the of 6.2 at $x^*\rho^*$ possesses a simple eigenvalue $\lambda_1(\rho)$ with $\lambda_1(\rho^*) = -1, \lambda_1'(\rho^*) \neq 0$.*

This hypothesis can be easily verified for TCP-RED map given by eq. 6.1. Next is the nonlinear stabilization theorem:

**Theorem 22** *[1] Under hypothesis ( 1) and for $lb \neq 0$ i.e. when the critical eigenvalue is controllable for linearized system, there is a feedback $u(x_k)$ with $u(x_k - x^*(\rho^*)) = 0$ which solves the local period doubling bifurcation control problem. Moreover, this can be accomplished with third order terms in $u(x_k)$, leaving the critical eigenvalue unaffected.*

Here, again $l$ is the left eigenvalue of linearized dynamics and $b$ is control vector. Again, for TCP-RED map $l = 1$ and $b \neq 0$. It can be easily verified that $lb \neq 0$.

Above theorem suggests a cubic control by itself can stabilize the system or a mixed control with linear terms can used to enhance the stability of bifurcation in extended parameter domain. Hence, we can consider different functional forms for the control in eq. (6.21). All these forms have been shown to enhance the stability of the fixed point, thus delaying the system bifurcations [1].

$$u_k = k_c y_k^3 \quad \text{Cubic Control Law}$$

$$u_k = k_l y_k + k_c y_k^3 \quad \text{Mixed Control Law}$$

The stability analysis done in [1] also suggests that $k_l$ and $k_c$ be based on the computation of $l$ and $b$. Clearly, we do not need a quadratic control due to critical eigendirection being linearly controllable. Cubic control can be used to change the nature of emerging period doubling orbit in the presence of uncertainty. According to the theoretical results in [1] it is possible to enhance the nonlinear stability terms by using just the cubic control terms. It is shown that stability co-efficient $\beta_2$ in the absence of any control equals:

$$\beta_2 = -2 \left( \frac{1}{2} \left( \frac{\partial^2 f}{\partial q_k^{n2}} \right)^2 + \frac{1}{3} \left( \frac{\partial^3 f}{\partial q_k^{n3}} \right) \right) \qquad (6.31)$$

It is clear that this co-efficient $\beta_2$ when evaluated for the linearized system decides if the bifurcation will be super ($\beta_2 < 0$) or subcritical ($\beta_2 > 0$). With the cubic control terms $\beta_2$ is changed by the following value:

$$\Delta = -4C_u(r, r, r) lb \qquad (6.32)$$

where $C_u(r,r,r)$ can be assigned any real value by appropriate choice of cubic feedback to stabilize the ensuing bifurcation.

Along with the linear feedback term there is a sound reason to use small cubic terms in order to stabilize. The theorem from [1] supports this idea due to the fact that by using cubic term one can achieve stable bifurcations with changes in parameters. There are indeed several reasons for using nonlinear feedback controls. First, the effect of linear feedback control designed to stabilize the linearized version of critical system on the original one parameter family of systems may be difficult to determine. Indeed, at least for small feedback control gains, one can expect that the bifurcation will reappear at a different value of the bifurcation parameter. The stability of this new bifurcation is not easily determined. Hence, simply using a linear stabilizing feedback may be unacceptable if the goal is to stabilize a bifurcation and not merely to stabilize an equilibrium point for a fixed parameter value. Second, it should not be surprising that in some situations a linear feedback which locally stabilizes an equilibrium point may result in globally unbounded behavior, whereas nonlinear feedback exists which stabilize the equilibrium both locally and globally [62].

## 6.5 Numerical and NS-2 Simulation of Washout Filter-Aided RED

Last section's results tell us that it is possible to control the loss of stability due to variation in any parameters locally. It also tells us that for fixed chosen control gain, there is a limit on the parameter range for guaranteed local stability and system may drift into instability if parameters move out of stable regime. In this section we illustrate the effect of washout filter-aided control on RED by numerical and NS-2 simulations by showing the stable allowable parameter region.

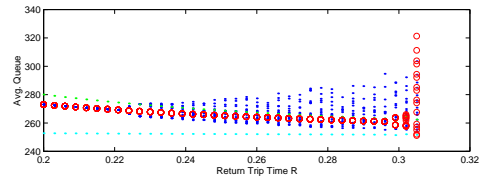

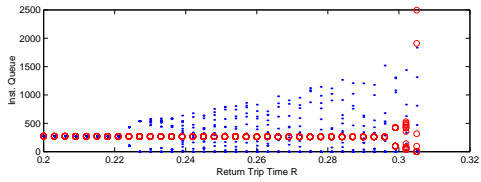

(a)

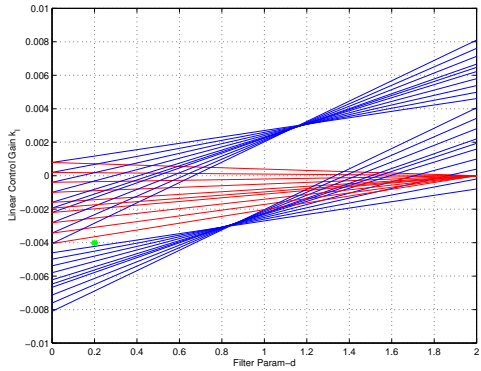

(b)

Figure 6.3: (a) Bifurcation diagram with and without control with respect to $R$ (with $p_{max}$ modulation). Bifurcation diagrams in blue and red are plotted without and with control, respectively. (b) Allowed $(d, k_l)$ region lies below the red line for stability.

We also note that wash-out filter stabilizes the queue by observing its dynamics and hence there is no need of per-flow datakeeping. This is an important issue when it comes to implementing a control scheme as methods which require per-flow data do not scale well with large number of flows traversing a router.

Fig. 6.3(a) plots the bifurcation diagram with respect to $R_0$ and the stability region of $(d, k_l)$. Here we modulate $p_{max}$ for feedback control, and only linear feedback control is used. The values of parameters used in the numerical example are as follows:

$$q_{max} = 747, \ q_{min} = 249, \ c=40 \text{ Mbps}, \ K = \sqrt{3/2},$$
$$B = 3,735, \ w = 2^{-5}, \ M = 4 \text{ kbits}, \ N = 129,$$
$$k_l = -15/b, \ d = 0.2, \ R_0 = \text{bifurcation parameter}$$

As shown in the figure the washout filter-aided control delays the bifurcation. However, once the bifurcation takes place with feedback control, the system becomes even more unstable than the system without feedback control. This demonstrates the need for nonlinear feedback control as explained in the previous section.

Fig. 6.3(b) shows the movement of allowable control region in $d - k_l$ space with changing $R_0$. It is clear from the diagram that for large $R_0$ system will become linearly unstable.

### 6.5.1  NS-2 Simulation

In this subsection we run the simulation with only long-lived TCP connections and compare the performance of RED with and without the feedback controller. The parameter $p_{max}$ is updated once every 500 ms. The gains for the first and third order terms, *i.e.*, $k_l$ and $k_c$, of the washout filter are set to $10^{-3}$ and $2.0 \times 10^{-8}$, respectively, and $d$ of the washout filter is set to 0.1. These parameters are not optimized, and the selection of robust parameters is left for future studies. We compare the performance of the controller with only linear term and both linear and third order terms as well.

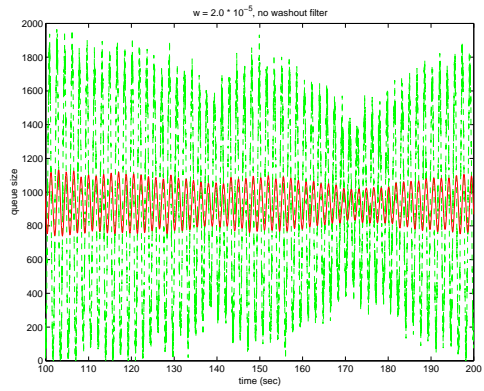

(a)

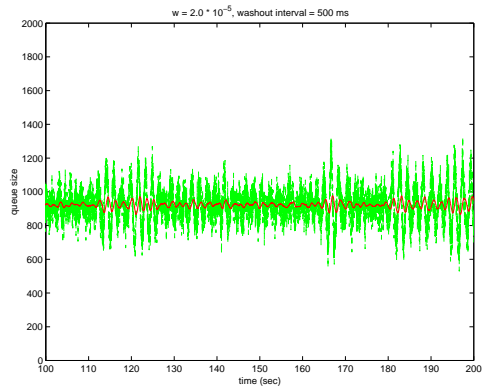

(b)

Figure 6.4: Long-lived connections. (a) RED without feedback controller ($w = 2.0 \times 10^{-5}$), (b) RED with feedback controller with both first and third order terms ($w = 2.0 \times 10^{-5}$).

Fig. 6.4 shows the evolution of the instantaneous and average queue sizes. As one can see the RED without any controller shows unstable behavior, while the RED with a feedback controller shows very stable behavior. Here we only show the performance of the controller with both first and third order terms. However, the controller with only the linear term shows similar improvement in the stability. This is because the third order term does not play a significant role in this example since the system is still stable with the controller.

Fig. 6.5 shows the queue evolution with $w = 4.0 \times 10^{-5}$. Unlike in the previous scenario, with a larger exponential averaging weight the difference in the performance is more visible. As one can see the linear controller is not able to control the average queue size as well as the controller with both terms, which still exhibits only small oscillations. This is consistent with the claim that the third order term (nonlinear term) reduces the amplitude of the oscillations in the presence of instability.

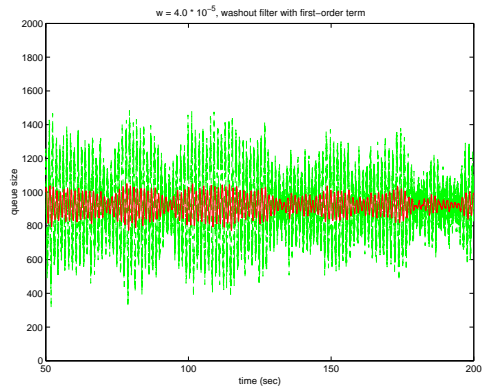

(a)

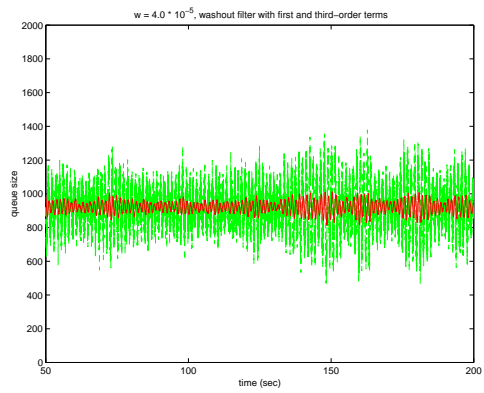

(b)

Figure 6.5: Linear controller vs. nonlinear controller. (a) RED with linear feedback controller ($w = 4.0 \times 10^{-5}$), (b) RED with feedback controller with both first and third order terms ($w = 4.0 \times 10^{-5}$).

## Chapter 7

# Conclusion and Future Research

In this work, we have shown the nonlinear dynamical behavior and parametric sensitivity of TCP-RED-type systems using a low dimensional model. The basic result which comes out of this work is that aggression of TCP-type flows when coupled with dynamic feedback strategy of RED may be responsible for the instability and parametric sensitivity observed in practise. As we have shown here, first instability is due to the square root nonlinearity in the throughput function and lack of damping with changing parameters of either RED or other system parameters like delay, or number of connections. The result is significant in the sense that a system administrator may have full control over the TCP/RED parameters but delay and number of users in a networked environment are highly volatile. This volatility potentially may lead the system towards nonlinear instabilities and into chaos. Identification of period doubling bifurcation is intuitive in the sense that when the RED queue is relatively empty, all the TCP flows show their aggression and send a lot of data but when RED queue occupancy increases it provides feedback by dropping/marking a lot of packets. Reacting to the feedback TCP streams recoil back but then again light occupancy in RED queue encourages the TCP streams into their aggressive more and the vicious cycle continues. The existence of different periodicities is reasonable depending on the time scales involved in aggression and recoiling of TCP-type streams and RED feedback parameters.

This work also proposes some preliminary ideas on controlling the instabilities by modulating RED parameters based on the feedback received from observing the average queue length. The basic idea is to control the period doubling bifurcation and increase the stable operating parameter range.

An interesting future direction along this work will be connecting two bottleneck routers and looking at their dynamical behavior at both modeling and NS-2 simulation level as discussed in [49]. Another future direction will be to try to model the relatively quick traffic with small packet load. It is not clear how traffic statistics is going to affect a network system near instability. The variety of dynamical behavior in TCP-RED will also be interesting to explore in the context of traffic patterns and anomaly detection systems. When traffic is unstable or chaotic then anomaly detection systems which are generally trained for certain preconceived patterns may activate a false alarm. The rich dynamical behavior exhibited by TCP-RED traffic may make the anomaly detection harder once the nonlinear behavior is taken into consideration.

It will also be interesting to explore the distributions induced on the buffer size due to chaotic behavior of TCP-RED map. It may provide some interesting QoS clues about the utilization of the buffer.

A second class of models studied in this work deals with optimal rate control in networks and are based on the rate-control framework proposed by Kelly. Using the results on delay-differential equation stability, the stability and its lack thereof is studied through an underlying map which arises naturally in time delay systems.

An invariance property of this map is used to prove delay-independent stability and to compute bounds on periodic oscillations. In particular, it is shown that the sender's utility and the receiver's price functions, both are responsible for the dynamical behavior of the system. Instability results indicate that the system again loses stability through a period doubling bifurcation resulting in oscillations. These oscillations are well understood in the Mathematics community and are known as "slowly oscillating periodic (SOP)" orbits.

It is also argued that these periodic orbits will be observed in networking and network-based control systems more often.

An interesting future direction will be to look at the intereaction of rate controlled senders with AQMs. Preliminary results show very rich dynamical behavior even in the absence of the delay.

Finally, this works is intricately related to Network Weather Service [88]. The Network Weather Service is a distributed system that periodically monitors and dynamically forecasts the performance that various network and computational resources can deliver over a given time interval. The service operates a distributed set of performance sensors (network monitors, CPU monitors, etc.) from which it gathers readings of the instantaneous conditions. It then uses numerical models to generate forecasts of what the conditions will be for a given time frame. One can think of this functionality as being analogous to weather forecasting, and it is from this analogy that the system inherits its name. Incidentally, in weather prediction chaos theory has a significant role as weather models can get chaotic in certain parameter regions. Some of the basic ideas and models which exhibit chaos can be found in [82, 80]. The models proposed in this thesis and their analysis illustrate the way a simple two node network can transition into an unpredictable regime. The understanding of the underlying dynamics can significantly enhance our prediction capabilities leading to better QoS and better utilization of networking/computing resources.